\documentclass[arxiv]{melba}
\usepackage{subcaption}

\usepackage{mwe} 

\usepackage{amsmath,amsfonts}

\usepackage[draft]{changes}

\melbaid{2025:026}  
\doi{https://doi.org/10.59275/j.melba.2025-ag5g}
\melbaauthors{Chen, Gu, Chen, Yang, Dong, Cao, Camarena, Mantyh, Colglazier, and Mazurowski}  
\email{yaqian.chen@duke.edu}
\volume{3}
\firstpageno{581}  
\melbayear{2025}  
\datesubmitted{2025-06-14}  
\datepublished{2025-11-13}  

\melbaspecialissue{Medical Imaging with Deep Learning (MIDL) 2020}
\melbaspecialissueeditors{Marleen de Bruijne, Tal Arbel, Ismail Ben Ayed, Hervé Lombaert}

\ShortHeadings{Automated Muscle and Fat Segmentation in Computed Tomography for Comprehensive Body Composition Analysis}{Chen, Gu, Chen, Yang, Dong, Cao, Camarena, Mantyh, Colglazier, and Mazurowski}

\title{Automated Muscle and Fat Segmentation in Computed Tomography for Comprehensive Body Composition Analysis}

\author{
	\firstname Yaqian \surname Chen\aff{1},
	\name Hanxue \surname Gu\aff{1},
        \name Yuwen \surname Chen\aff{1},
        \name Jichen \surname Yang\aff{1},
        \name Haoyu \surname Dong\aff{1},
        \name Joseph Y. \surname Cao\aff{2},
        \name Adrian \surname Camarena\aff{3},
        \name Christopher \surname Mantyh\aff{3},
        \name Roy \surname Colglazier\aff{2},
        \name Maciej A. \surname Mazurowski\aff{1,2,4,5},
}
\affiliations{
	\num 1 \addr Department of Electrical and Computer Engineering, Duke University, Durham, NC 27708 \\
	\num 2 \addr Department of Radiology, Duke University, Durham, NC 27708\\
	\num 3 \addr Department of Surgery Duke University School of Medicine, Durham, NC 27708\\
        \num 4 \addr Department of Biostatistics \& Bioinformatics, Duke University, Durham, NC 27708\\
        \num 5 \addr Department of Computer Science, Duke University, Durham, NC 27708\\
}

\abstract{
    Body composition assessment using CT images can potentially be used for a number of clinical applications, including the prognostication of cardiovascular outcomes, evaluation of metabolic health, monitoring of disease progression, assessment of nutritional status, prediction of treatment response in oncology, and risk stratification for surgical and critical care outcomes. While multiple groups have developed in-house segmentation tools for this analysis, there are very limited publicly available tools that could be consistently used across different applications. To mitigate this gap, we present a publicly accessible, end-to-end segmentation and feature calculation model specifically for CT body composition analysis.
    Our model performs segmentation of skeletal muscle, subcutaneous adipose tissue (SAT), and visceral adipose tissue (VAT) across the chest, abdomen, and pelvis area in axial CT images. It also provides various body composition metrics, including muscle density, visceral-to-subcutaneous fat (VAT/SAT) ratio, muscle area/volume, and skeletal muscle index (SMI), supporting both 2D and 3D assessments. To evaluate the model, the segmentation was applied to both internal and external datasets, with body composition metrics analyzed across different age, sex, and race groups. The model achieved high dice coefficients on both internal and external datasets, exceeding 89\% for skeletal muscle, SAT, and VAT segmentation. The model outperforms the benchmark by 2.10\% on skeletal muscle and 8.6\% on SAT compared to the manual annotations given by the publicly available dataset. Body composition metrics show mean relative absolute errors (MRAEs) under 10\% for all measures. Furthermore, the model provided muscular fat segmentation with a Dice coefficient of 56.27\%, which can be utilized for additional analyses as needed. 
    Our model with weights is publicly available at~\url{https://github.com/mazurowski-lab/CT-Muscle-and-Fat-Segmentation.git}.
    }


\begin{document}

\twocolumn[\maketitle]

\section{Introduction}\label{sec:intro}
Correlating body composition metrics based on computed tomography (CT) images with disease and clinical variables, such as cancer \citep{rutten2016loss, kumar2016muscle, deluche2018impact, yoshikawa2020sarcopenia, iwase2016impact, boer2020impact}, cachexia \citep{ali2014sarcopenia, fearon1990body, al2023body, baracos2010body} and frailty \citep{falsarella2015body, reinders2017body, villareal2004physical}, is becoming a widely adopted approach to leverage medical imaging data for real-world clinical applications \citep{tolonen2021methodology}. By measuring body composition, such as quantity and location of fat as well as quantity and quality of muscle, clinicians are able to gain valuable insights into a patient’s physiological status \citep{prado2014lean}. This information enables them to assess disease progression \citep{baracos2013clinical}, evaluate treatment efficacy \citep{bates2022ct}, and predict clinical outcomes \citep{weston2019automated}. 

Several key metrics are frequently utilized in body composition analysis, including muscle density, the visceral-to-subcutaneous fat (VAT/SAT) ratio, muscle area or volume, and the skeletal muscle index (SMI). Most studies in this field measure these metrics from a single slice, most commonly at the third lumbar vertebra (L3) \citep{arayne2023comparison}, while others employ volumetric analysis \citep{connelly2013volumetric}. However, regardless of the approach, extracting these metrics relies on effective segmentation models to accurately identify and quantify various tissues within the body. Traditional methods, such as pixel thresholding based on Hounsfield units (HU) \citep{wang2020artificial} and fuzzy c-means clustering \citep{wang2020artificial, christ2011fuzzy}, often require manual adjustments and are time-intensive \citep{wang2020artificial}. Furthermore, pixel thresholding algorithms cannot differentiate between visceral fat, subcutaneous fat, and intramuscular fat—an essential distinction when measuring the VAT/SAT ratio. Deep learning segmentation is a response to these limitations, and some groups have developed in-house segmentation models customized for their private datasets \citep{fu2020automatic, lee2017pixel, weston2019automated, wang2017two, hemke2020deep, koitka2021fully}. These models typically lack public accessibility and are designed for specific tasks. Furthermore, we observed inconsistency in how muscular fat (both intra-muscular and inter-muscular fat) is utilized in research. While some studies include muscular fat as part of skeletal muscle measurements \citep{hou2024enhanced, blanc2020abdominal, weston2019automated}, others classify it under VAT \citep{camus2014prognostic, wirtz2021ct, connelly2013volumetric}, and a smaller subset considers it part of SAT \citep{ozturk2020relationship}. 

In order to advance the research on the relationship of imaging-based body composition with disease and clinical variables, a robust, thoroughly validated, and publicly available tissue segmentation model and body composition variable calculation is needed. This model will allow different research groups to test the model with their data and correlate the unified body composition measurements with the clinical outcomes of their interest, building consistent scientific evidence of the importance of body composition in human health. 

Toward this goal, we developed and evaluated segmentation models based on nine publicly available network architectures that were recently published and have demonstrated state-of-the-art performance across various medical segmentation tasks, including nnU-Net and its variants \citep{isensee2021nnu, isensee2024nnu, roy2023mednext}, as well as foundation model finetuning approaches \citep{kirillov2023segment, ma2024segment, cheng2023sam}. For the model training and evaluation, we collected 813 CT volumes of chest, abdomen, and pelvis for 483 patients from Duke University Health System. In both training and test datasets, we included volumes from different years (from 2016 to 2019), various scanners, and diverse patient demographics to ensure the model's generalizability. Additionally, we incorporated the publicly available Sparsely Annotated Region and Organ Segmentation (SAROS) dataset \citep{koitka2023saros, clark2013cancer} for segmentation evaluation, demonstrating the generalizability of our proposed model. Based on the performance comparison, we selected the best-performing model from the nine. Using the automatic segmentation masks generated by this model, we analyzed the relationships between four body composition metrics (muscle density, VAT/SAT ratio, muscle area/volume, and SMI) and three key demographic variables: age, sex, and race (as shown in Sec. \ref{sec:Body composition analysis}). Notably, to facilitate broad adoption, we have made the model publicly available.

The main contributions of this work are summarized as follows:
\begin{itemize}
    \item To the best of our knowledge, this is the first publicly available deep-learning model designed to segment skeletal muscle, SAT, VAT, and muscular fat across the chest, abdomen, and pelvis on CT.
    \item We provide end-to-end standardized and publicly available measurements for four common body composition metrics, including muscle density, visceral-to-subcutaneous fat (VAT/SAT) ratio, muscle area/volume, and SMI on both L3  for 2D and T12 to L4 for 3D measurement.
    \item We evaluated nine publicly available network architectures based on both segmentation performance and clinically relevant body composition measurements to provide users with insights for model selection.
    \item Our model outperforms TotalSegmentator \citep{wasserthal2023totalsegmentator} and Enhanced segmentation \citep{hou2024enhanced} by 2.10\% on skeletal muscle and 8.6\% on SAT compared to the manual annotations given by publicly available dataset SAROS.
    \item We perform the statistical analysis to correlate four metrics (muscle density, VAT/SAT ratio, muscle area/volume, and SMI) with three patient demographic variables: age, sex, and race in both 2D and 3D settings. 
\end{itemize}

\section{Related Works}
\subsection{Body composition analysis using CT}
Body composition plays a crucial role in influencing physical performance \citep{hernandez2024relationship, falsarella2015body}, metabolic health \citep{trouwborst2024body, kakinami2022body}, and disease outcomes \citep{rutten2016loss}. Imaging offers an objective, quantitative approach to its analysis through various techniques, including CT, magnetic resonance imaging (MRI), and ultrasonography \citep{hou2024enhanced, tan2024quantitative, sharafi2024quantitative, xu2024value}. Among all the modalities, CT offers high spatial resolution, faster acquisition times, and superior contrast between tissues \citep{zhang2021improving}, making it particularly suitable for assessing visceral and subcutaneous fat, skeletal muscle, and organ-specific fat deposits \citep{wathen2013vivo}. 

During body composition calculation on CT, several key metrics are frequently utilized, including muscle density, the VAT/SAT ratio, muscle area/volume, and the SMI. Muscle density in CT provides insights into muscle quality, which is linearly influenced by muscular fat content \citep{engelke2018quantitative}. A reduction in muscle density is often associated with increased fat infiltration within the muscle, known as myosteatosis \citep{chang2024prognostic}, which compromises muscle function and structural integrity. This reduction serves as a critical indicator of sarcopenia, a condition characterized by the progressive loss of skeletal muscle mass and strength, as well as frailty and diminished physical performance, particularly in aging populations \citep{cawthon2015assessment}. 

The VAT/SAT ratio, on the other hand, is a key metric for assessing metabolic risk \citep{kaess2012ratio, oh2017visceral}. While visceral adipose tissue (VAT) is strongly associated with metabolic disturbances and cardiovascular risk \citep{vasamsetti2023regulation}, its volume alone may reflect both overall fat mass and an individual's tendency to store fat viscerally \citep{kaess2012ratio}. In contrast, the VAT/SAT ratio offers a more precise assessment, as it accounts for the balance between visceral and subcutaneous fat, providing insight that is independent of total body fat percentage \citep{kaess2012ratio}.

Muscle area/volume and SMI are essential measurements of total muscle quantity and its proportionality to body size. These metrics provide critical information about an individual’s muscle reserves, which are vital for mobility, metabolic function, and overall health status \citep{chen2023really}. Studies highlight them as significant markers of nutritional status \citep{risch2022assessment}, which are crucial for recovery from illness, mortality, and treatment-related complications, such as the length of hospital stays and the rate of readmissions \citep{schuetz2021management, kaegi2021evaluation, guenter2021malnutrition}. Furthermore, they also serve as important factors in assessing metabolic health \citep{cruz2019sarcopenia, prado2014lean, martin2013cancer, dodds2015epidemiology}, as lower muscle mass is associated with insulin resistance and impaired glucose metabolism.

The collection of these metrics pictures the clear body condition of patients, showcasing a comprehensive overview of their muscle composition, fat distribution, and overall physiological status.

\subsection{Traditional methods for body composition segmentation}
Most early studies on body composition analysis rely on semi-automated threshold-based segmentation using predefined Hounsfield unit (HU) ranges to differentiate lean muscle mass from adipose tissue \citep{lee2017pixel, ji2022thresholds}. Despite its simplicity, threshold-based segmentation presents significant challenges due to the overlapping HU values between different tissue types, such as SAT and skin, as well as muscle and adjacent organs \citep{lee2017pixel}. The method is also highly susceptible to image noise \citep{sehgal2022ct, diwakar2020ct}, which can significantly compromise tissue classification accuracy, particularly in low-quality or artifact-prone scans. As a result, the method typically requires manual correction based on visual analysis by highly skilled radiologists and is impractical on large datasets due to the expense and time required.

To overcome these limitations, researchers have developed various advanced segmentation algorithms, including rule-based \citep{kamiya2009automated, kamiya2011automated}, clustering-based \citep{positano2009accurate, positano2004accurate, christ2011fuzzy}, and finite-element-method-based \citep{popuri2015body} approaches. Kamiya et al. proposed a rule-based expert system for segmenting the psoas major and rectus abdominis muscles from CT images, approximating muscle shapes with simple quadratic functions \citep{kamiya2009automated, kamiya2011automated}. Positano et al. utilize a fuzzy c-mean algorithm to make unsupervised classification of image pixels on MRI \citep{positano2009accurate, positano2004accurate}. Karteek and the team developed a novel FEM deformable model for muscle and fat segmentation from CT \citep{popuri2015body}.

However, these methods primarily focus on extracting specific muscle groups from CT or MRI scans and are unable to differentiate between visceral fat, subcutaneous fat, and intramuscular fat—an essential distinction in many body composition analysis tasks \citep{staley2019visceral, torres2013nutritional, iwase2016impact}. A potential approach to address these challenges is the use of deep learning-based segmentation algorithms.

\subsection{Deep learning-based models for body composition segmentation}
Deep learning-based segmentation has been proven to be a reliable technique in various clinical applications \citep{gu2024segmentanybone, dong2024segment, wasserthal2023totalsegmentator, mazurowski2023segment}. While networks offer high accuracy, reduce human labor, and provide greater generalizability compared to traditional segmentation algorithms, it is straightforward to apply deep learning-based segmentation algorithms for body composition analysis.

The majority of current deep learning-based segmentation models for body composition are still based on convolutional neural networks (CNNs) \citep{nowak2020fully}. U-Net and its variants are among the most widely used architectures in this domain, providing precise segmentation of body composition components such as skeletal muscle, SAT, and VAT \citep{paris2020body, weston2019automated}. However, these models are typically not publicly accessible and are often designed for specific tasks \citep{mai2023systematic}. While a few commercial models are available \citep{cespedes2020evaluation, mai2023systematic, lee2021deep}, they are often associated with high costs and limited customization options. TotalSegmentator \citep{wasserthal2023totalsegmentator}, a recently published general CT segmentation model based on nnU-Net, also supports muscle and fat segmentation. However, studies have shown that its performance in segmenting muscle, SAT, and VAT can be further improved, and its non-commercial license restricts broader usage. Therefore, there remains a significant need for publicly accessible, transparent, and generalizable segmentation models for body composition analysis.
\section{Methods}
Our model provides four-label segmentation from chest to pelvis CT and automatic body composition measurement for four common metrics, including muscle density, VAT/SAT ratio, muscle area/volume, and SMI. To achieve this, the development of our algorithm has three parts: vertebrae detection, muscle and fat segmentation, and body composition metrics measurement. Vertebrae detection is used to identify the body region for 2D and 3D body composition measurements. Muscle and fat segmentation provides direct four-label segmentation results that can be utilized for other purposes and also serves as input for body composition measurement. Detailed information on muscle and fat segmentation is provided in Sec. \ref{Method:Segmentation algorithms}, while Sec. \ref{Method:Body composition metrics} describes vertebrae detection and body composition measurement.

\label{sec:method}
\subsection{Datasets} \label{sec:Datasets}
In this study, we utilize two datasets: an internal dataset collected from Duke Hospital and the publicly available SAROS dataset \citep{koitka2023saros, clark2013cancer} that integrates four publicly available CT datasets from TCIA with sparse annotations. The internal dataset is split into three non-overlapping subsets: \textbf{Internal Training} (used to train the segmentation model), \textbf{Internal Test} (used to evaluate segmentation performance), and \textbf{Demographic Analysis} (used to analyze body composition metrics across demographic groups). The \textbf{Internal Training} dataset is used exclusively for model development to ensure compliance with licensing constraints, while both the \textbf{Internal Test} and SAROS datasets are used for evaluating segmentation performance. \textbf{Demographic Analysis} dataset is utilized for investigating how body composition varies across age, sex, and race groups shown in Sec. \ref{sec:Body composition analysis}; this is aimed to provide insights for readers on how different body composition metrics correlate with patient demographics. The datasets span multiple institutions, acquisition years, and patient demographics, supporting the robustness and generalizability of our evaluation and analysis.

\subsubsection{Dataset 1: Internal dataset}
\label{dataset:internal}
For this project, we collected a total of 8,948 CT volumes from the Duke University Health System, spanning January 2016 to November 2019, including chest, abdomen, and pelvis exams. These volumes represent all CT images that had been previously downloaded and stored in our system. The downloads were carried out in two phases: (1) we randomly sampled 500 CT exams from 500 unique patients imaged between 2016 and 2019, primarily focusing on the two most frequent clinical protocols—CT chest abdomen pelvis with contrast w MIPS and CT chest with contrast; (2) additionally, we retrieved all CT exams (chest, abdomen, and pelvis) acquired between 2017 and 2018 that had corresponding MRIs of the same anatomical region, enabling potential future multi-modal analyses.

From the initial collection, we further identified 1927 volumes from 854 patients based on two criteria: (1) axial view exam was available (2) the volumes were original axial acquisitions, not derived from multiplanar reconstructions (MPR) or reformatted from other planes. These criteria were selected to align the model with real-world clinical scenarios. These 854 patients were further assigned into three non-overlapping subsets. All the annotations are generated through 3D slicer and the CT acquisition parameters for our internal dataset are available at Appendix \ref{appendix:CT parameters}.

\textbf{Internal Training}
A total of 453 patients were assigned to the Internal Training set, from which 1,863 slices were randomly sampled and annotated for model development. The annotation process was carried out by four Duke students under the direct supervision of a radiologist and a surgeon. This cohort was determined through an iterative development process, where slice sampling continued until performance on a fixed evaluation set stabilized. Ultimately, 1,863 slices (including those from the evaluation set) were used for model training and thus became our final training set.

\textbf{Internal Test}
Thirty patients were allocated to the Internal Test set, with one volume randomly selected per patient to avoid sampling bias. For each volume, slices were sequentially selected and annotated at uniform intervals (one slice every 2.5$cm$) to avoid redundant annotation of highly similar adjacent slices while still providing consistent spatial sampling for quantitative assessment. All annotations in this subset were independently reviewed, modified, and approved by a radiologist who had no involvement in the training dataset annotation process. Notably, selecting one slice every 2.5$cm$ only applied for evaluating muscle and fat segmentation performance and annotation. Since vertebrae detection heavily depends on spatial information as discussed in Sec. \ref{totalsegmentator eval}, all the vertebrae detection is launched on the full volume.

\textbf{Demographic Analysis}
The remaining 371 patients were included in the Demographic Analysis set. This dataset was designed to assess the model’s ability to analyze body composition across demographic groups such as age, sex, and race. To prevent any potential bias in the segmentation results, we excluded all patients who were involved in any part of the model development process. All masks in this cohort were generated entirely by the trained segmentation model without any manual editing, allowing for large-scale analysis.

\subsubsection{Dataset 2: SAROS dataset}
\label{dataset:external}
The SAROS dataset \citep{koitka2023saros, clark2013cancer} is a comprehensive collection of CT imaging volumes available on TCIA, featuring sparse annotations for 13 body region labels and six body part labels. The 13 annotations for body regions include the abdominal cavity, thoracic cavity, bones, brain, breast implants, mediastinum, muscles, parotid glands, submandibular glands, thyroid glands, pericardium, spinal cord, and subcutaneous tissue. The six body parts are the left arm, right arm, left leg, right leg, head, and torso, comprising a total of 900 CT volumes from 882 unique patients.

Given the torso label overlap with chest, abdomen, and pelvis regions, which are the focus body regions for our study, we utilize the slices with torso labels for our model evaluation by comparing the model segmented results with annotated skeletal muscle and SAT annotations. Furthermore, to ensure the flexible use of our model, we only selected a subset of the SAROS dataset covered under a commercial license for evaluation, more details for collection selection are shown in Appendix Sec \ref{sec:Data collections from SAROS}. As a result, in total, 650 CT volumes from 632 unique patients with  CT slices are selected for segmentation model evaluation. The detailed data collections for SAROS are shown in Table \ref{tab:SAROS_data_collection}.

\subsubsection{Patient demographics}
Table \ref{tab:demographic} provides a demographic overview of patients from the five dataset collections, derived from two sources: the internal dataset (including Internal Training, Internal Test, and Demographic Analysis collections, shown in Sec \ref{dataset:internal}) and the external dataset (SAROS collections, in Sec \ref{dataset:external}). Patients' ages and races are unknown for the SAROS dataset due to de-identification.
\begin{table*}[ht!]
\resizebox{\textwidth}{!}{%
\centering
\renewcommand{\arraystretch}{1.2} 
\resizebox{\textwidth}{!}{%
\begin{tabular}{|c|c|c|c|c|c|c|}
\hline
\multicolumn{3}{|c|}{} & \textbf{Internal Training} & \textbf{Internal Test} & \textbf{Demographic Analysis} & \textbf{SAROS Dataset} \\ \hline
\multicolumn{3}{|c|}{Number of patients} & 453 & 30 & 371 & 632 \\ \hline
\multicolumn{3}{|c|}{Number of studies} & 783 & 30 & 371 & 650 \\ \hline
\multicolumn{3}{|c|}{Number of slices} & 1863 & 636 & 183972 & 10038 \\ \hline
\multirow{7}{*}{Demographics} & \multicolumn{2}{|c|}{Age} & 60.1 (3 - 89) & 58.5 (5 - 84)  & 58.3 (0.25-7) & - \\ \cline{2-7} 
 & \multirow{2}{*}{\centering Sex} & Female & 51.88\% (235) & 53.33\% (16) & 52.02\% (193) & 54.75\% (346) \\ \cline{3-7} 
 &  & Male & 48.12\% (218) & 46.67\% (14) & 47.98\% (178) & 45.25\% (286) \\ \cline{2-7} 
 & \multirow{5}{*}{\centering Race} & White & 67.11\% (304) & 63.33\% (19) & 68.19\% (253) & - \\ \cline{3-7} 
 &  & Black/ African American & 25.83\% (117) & 30.00\% (9) & 23.99\% (89) & - \\ \cline{3-7} 
 &  & Asian & 1.99\% (9) & 0\% (0) & 1.62\% (6) & - \\ \cline{3-7} 
 &  & American Indian & 0.66\% (3) & 0\% (0) & 1.08\% (4) & - \\ \cline{3-7} 
 &  & Other & 4.42\% (20) & 3.33\% (1) & 5.12\% (19) & - \\ \hline
\end{tabular}%
}}
\caption{\textbf{Patient demographics for four collections:} Table presents demographic details for four collections, including Internal Training, Internal Test, Demographic Analysis, and SAROS Dataset Test. For sex and race, the absolute number of patients is shown in parentheses alongside the percentages. For age, the mean values for ages are provided in years along with the minimum and maximum age in parentheses. Notably, the youngest patient, recorded as 3 months old, is consistently represented as 0.25 years.}
\label{tab:demographic}
\end{table*}

\subsection{Segmentation algorithms}\label{Method:Segmentation algorithms}
For this study, we implemented \textbf{nine} 2D segmentation models that were recently published and have demonstrated state-of-the-art performance on medical segmentation tasks. 2D segmentation was chosen because training on randomly selected slices from a large, diverse patient cohort provides better exposure to inter-patient anatomical variability compared to annotating full CT volumes for 3D segmentation with the same annotation effort. This strategy has also been shown to improve generalization in large-scale medical imaging tasks \citep{zhang2024convolutional, sha2023enhancing, crespi20223d}.

The nine segmentation models can generally be split into two categories: nnU-Net \citep{isensee2021nnu, isensee2024nnu, roy2023mednext} and foundation model finetuning \citep{kirillov2023segment, ma2024segment, cheng2023sam}. In the following subsections, we provide detailed information for each of these model categories.

\textbf{nnU-Net} nnU-Net \citep{isensee2021nnu, isensee2024nnu} is a highly adaptable semantic segmentation method designed to automatically configure an optimized U-Net-based pipeline for any given dataset. Recent updates to the nnU-Net methodology have introduced enhancements to the U-Net baseline, emphasizing the importance of using advanced CNN architectures like ResNet \citep{isensee2024nnu} and ConvNeXt \citep{roy2023mednext} variants, leveraging the robust nnU-Net framework, and employing model scaling for improved performance. Therefore, in this study, we tested the segmentation performance on five variants including nnU-Net ResEnc (residual encoder blocks) M, nnU-Net ResEnc L, nnU-Net ResEnc XL, MedNeXt M k5, and MedNeXt L k5. These selections represent recent improvements and scaling strategies within the nnU-Net and MedNeXt families.

\textbf{Foundation Model Finetuning} Recently, an increasing number of studies have demonstrated the strong capability of vision foundation models in medical segmentation tasks mainly due to its strong generalization \citep{gu2024segmentanybone, dong2024segment}. In this study, we also evaluated three representative foundation model finetuning approaches: Segment Anything Model (SAM) finetuning \cite{kirillov2023segment}, MedSAM finetuning \citep{ma2024segment}, and SAM-Med2D finetuning \citep{cheng2023sam}. MedSAM and SAM-Med2D methods build upon the SAM by adapting it to the medical domain through supervised finetuning on task-specific annotations.


\subsubsection{Segmentation metrics} \label{Method:Segmentation metrics}
To evaluate our model's segmentation performance, we utilized three metrics: the Dice coefficient, the mean relative absolute errors (MRAE), and coefficient of determination ($R^2$). The details of these two metrics are introduced in this section.

\textbf{Dice coefficient} is commonly used for image segmentation tasks to evaluate segmentation accuracy by measuring the overlap between predicted and ground truth regions. It ranges from 0 to 1, where a value of 1 indicates perfect overlap and 0 signifies no overlap. The mathematic formula for the Dice coefficient is Eq. \eqref{dice}
\begin{equation}\label{dice}
    \text{Dice} = \frac{2 |A \cap B|}{|A| + |B|}
\end{equation}
where A represents the model predicted mask and the B represents the set of pixels in the ground truth. 

\textbf{MRAE} measures the absolute mean for relative errors across all data points (shown in Eq. \eqref{MRAE}, which is normally applied to measure the relative error between prediction and ground truth. Lower MRAE values indicate better model performance, reflecting smaller deviations between predicted and actual values. 
\begin{equation}\label{MRAE}
\text{MRAE} = \frac{1}{n} \sum_{i=1}^{n} \left| \frac{A_i - B_i}{A_i} \right|
\end{equation}
where $A_i$ represents the ground truth values, $B_i$ represents the predicted values, and $n$ is the total number of data points.

Compared with Dice, MRAE directly quantify how well the derived clinical measurements agree which offers complementary insight especially for clinical downstream decision-making. MRAE has also been adopted in previous studies for evaluating segmentation-derived quantitative outcomes, including organ volume estimation and fat quantification \citep{akhavanallaf2021whole, akhavanallaf2021personalized}.

\textbf{Coefficient of Determination (\(\mathbf{R}^2\))} indicates the proportion of the variance in the dependent variable that is explained by the independent variable in the model. The mathematical definition of $R^2$ is shown in Eq. \eqref{R2}, where $y_i$ represents the observed values and $\hat{y}_i$ shows the predicted values. The $R^2$ value ranges from 0 to 1, where $R^2 = 1$ indicates perfect prediction, while $R^2 = 0$ indicates a complete failure to explain the variance.

\begin{equation}\label{R2}
R^2 = 1 - \frac{\sum (y_i - \hat{y}_i)^2}{\sum (y_i - \bar{y})^2}
\end{equation}

\subsection{Body composition metrics}\label{Method:Body composition metrics}
Our model is capable of measuring four commonly used body composition metrics: muscle density, VAT/SAT ratio, muscle area/volume, and SMI in both 2D and 3D settings. The detailed descriptions for these four body composition including their calculation methods, clinical significance, and associations with diseases are summarized in Table \ref{tab:body_composition_summary} in Appendix \ref{sec:Summary of Body Composition Metrics}. By analyzing previous studies on body composition analysis, we selected the third lumbar vertebral level (L3) for 2D body composition measurements and the region spanning the twelfth thoracic vertebral level (T12) to the fourth lumbar vertebral level (L4) for 3D measurements. L3 is considered the most commonly used standard for body composition assessment in multiple clinical applications, including rectal cancer assessment \citep{han2020association, arayne2023comparison}, sarcopenia evaluation \citep{amini2019approaches, pickhardt2020automated, pickhardt2020automate}, and obesity research \citep{liu2023fully, malietzis2015role}. For the 3D body composition measurement, T12 is selected as the beginning of the 3D measurement region following the approach of previous studies \citep{demerath2007approximation, tong2014optimization}. This focus is particularly relevant for assessing visceral adipose tissue (VAT) and subcutaneous adipose tissue (SAT); however, recent studies related to sarcopenia and rectal cancer also pay increasing attention to T12 \citep{fernandez2024ia, arayne2023comparison, soh2024prognostic}. L4 is selected as the ending point since 92.62\% of our internal abdominal CT volumes include L4, while only 78.07\% include L5.

TotalSegmentator \citep{wasserthal2023totalsegmentator} is utilized for automatically extracting, T12, L3, and L4. We select the slice with the largest L3 label among all slices with L3 mask for our 2D body composition measurement. For 3D measurement, we extract the portion between the largest T12 label and the largest L4 since selecting the largest slice of the label demonstrate higher robustness on TotalSegmentator. In the subsequent sections, we detail the calculations for muscle density, VAT/SAT ratio, muscle area/volume, and SMI.

\textbf{Muscle density} measures the average Hounsfield Unit (HU) values within the segmented skeletal muscle area (SMA) with higher values indicating leaner muscle and lower values (typically from -29 to 29 HU \citep{salam2023opportunistic}) suggesting fat infiltration. Muscle density is the crucial biomarker for muscle quality \citep{looijaard2016skeletal, cleary2015does, wang2021muscle} and is frequently associated with evaluations of sarcopenia and myosteatosis \citep{cawthon2015assessment, sergi2016imaging, tagliafico2022sarcopenia}.

\textbf{VAT/SAT ratio} is more commonly related to obesity-related health risks, such as diabetes, cardiovascular disease, and metabolic syndrome \citep{piche2018overview, frank2019determinants, goossens2017metabolic, ladeiras2017ratio, tanaka2021distinct}. A higher ratio indicates a predominance of visceral adipose tissue (VAT) over subcutaneous adipose tissue (SAT), reflecting an unfavorable fat distribution pattern \citep{ladeiras2017ratio}. Visceral fat is metabolically active and associated with chronic inflammation, insulin resistance, and dyslipidemia, which contribute to the development and progression of these conditions \citep{hardy2012causes, chait2020adipose, bansal2023visceral}. 

\textbf{Muscle area/volume} assesses the total skeletal muscle within the region of interest (ROI). Specifically, for 2D measurement, this metric, also referred to as skeletal muscle area (SMA), is calculated by multiplying the number of pixels within the segmented skeletal muscle mask by the area ($m^2$) represented by each pixel. For 3D measurement, the segmented skeletal muscle area/volume is determined by multiplying both the pixel size and the slice thickness ($m^3$). The SMA is one of the standard metrics for muscle quantity evaluation \citep{goodpaster2000composition, sinelnikov2016measurement, vella2020skeletal} and has been demonstrated to be highly correlated with patients' post-operative recovery and survival rates \citep{antoniou2019effect, bradley2022relationship, antoniou2019effect}. 3D muscle area/volume provides better representation of the entire muscle \citep{momose2017ct}. 

\textbf{SMI} is another commonly used metrics for muscle quantity measurement. This metrics normalizes the muscle cross-sectional area (CSA) by dividing it by the individual's height squared ($m^2$).

\section{Experiments}
This section demonstrates the data preprocessing and implementation detail for nine segmentation models that are previously mentioned in Sec. \ref{Method:Segmentation algorithms}. Consistent with the structure in Sec. \ref{Method:Segmentation algorithms}, the models are grouped into two categories—nnU-Net and foundation model finetuning—as models within each category share similar implementation strategies.

\subsection{Data preprocessing}
We utilize DICOM header metadata, specifically the Rescale-Slope and RescaleIntercept parameters, to convert raw pixel values to standardized Hounsfield Units (HU). CT slices are resampled to a fixed size of 512 $\times$ 512 before being fed into segmentation models. For nnU-Net and its variants, data normalization was handled automatically by the framework. For foundation model finetuning algorithms, we applied the same normalization strategy shown in nnU-Net to align the preprocessing: intensity values were first clipped to the [0.5, 99.5] percentile range, followed by z-score normalization.

\subsection{Segmentation model implementation detail}
\textbf{nnU-Net:} All nnU-Net and MedNeXt models \citep{isensee2021nnu, isensee2024nnu, roy2023mednext} were trained using the default training pipeline provided by nnU-Net official implementations. For all models, we adopted the standard 5-fold validation strategy with shared training and validation data split. Training was conducted using the default hyperparameters: a learning rate of 0.01 with the SGD optimizer, a total of 1,000 epochs, and automatic adjustment of patch size and batch size based on available GPU memory. The best model was selected based on the Dice score achieved on the validation set to avoid potential overfitting.

\textbf{Foundation Model Finetuning:} For SAM and MedSAM, we fine-tune the foundation models following the test strategy described in \citep{gu2024build}, which identifies that fine-tuning using parameter-efficient learning with Adapter blocks on both the image encoder and the mask decoder can achieve the best task-specific performance for automatic segmentation. For PEFT with Adapter blocks, we implemented it based on \cite{gu2024build}, which added MLP layers to SAM’s transformer blocks. In the image encoder, adapter blocks were added into the transformer blocks on two locations: (1) right after the multi-head attention, and (2) within the MLP residual block using a scalable parameter. We added them to the first 2 and last 2 transformer blocks of the image encoder. In the mask decoder, adapters were added after the multi-head attention in the two-way transformer block. We also followed the same normalization strategy as the original inputs for each foundation model to ensure the inputs remain in scale. For all experiments, we used a base learning rate of 1e-4 with 200 warmup iterations, and applied AdamW with a weight decay of 0.1 for the remaining iterations, running a total of 1000 epochs.
 
 For SAM-Med2D, we follow the finetuning strategy provided by their official implementation. The model is trained using the Adam optimizer with a learning rate of 1e-4, a batch size of 2, and maximum epoch equals to 1000 with the best model selected by the performance on validation set. Notably, unlike our previous training setup where a single multi-label ground-truth mask was used for each CT slice, SAM-Med2D requires four separate binary masks—one per class—for both training and evaluation.
 
All models used the same training and validation dataset split as provided by the nnU-Net model.

All the algorithms was trained and evaluated on an NVIDIA RTX 3090 GPU, ensuring efficient computation and high performance.

\section{Segmentation results}
This section evaluates TotalSegmentator’s performance in detecting the body region for body composition measurement and presents the segmentation performance of our model on three selected labels: skeletal muscle, SAT, and VAT, as well as four key body composition metrics: muscle density, muscle area/volume, SMI, and VAT/SAT ratio. The evaluation is provided both qualitatively through visual comparisons, and quantitatively, using the Dice coefficient and MRAE to assess the overlap between the manually annotated ground truth and the model’s segmentation.



\subsection{Vertebra detection evaluation}\label{totalsegmentator eval}
This section represents both the quantitative and qualitative evaluations of TotalSegmentator’s vertebra detection ability : (1) we asked a human reader to select slice to measure for L3, and calculated the mean absolute difference (in $cm^2$) between the human-selected slices and those extracted by TotalSegmentator; (2) we computed the absolute slice distance between the human-selected and TotalSegmentator-selected slices, as well as the percentage differences in four body composition metrics (muscle density, muscle area, subcutaneous fat area, and visceral fat area) measured on those slice pairs. This evaluation was conducted across all volumes in the Internal Test dataset using L3 to assess the impact of slice selection discrepancies. Notably, the whole experiment on this evaluation is executed on the original dicom file without any data preprocessing and resampling.

\begin{figure*}[htbp]
    \centering
    \begin{subfigure}[b]{0.48\textwidth}
        \centering
        \includegraphics[width=\textwidth]{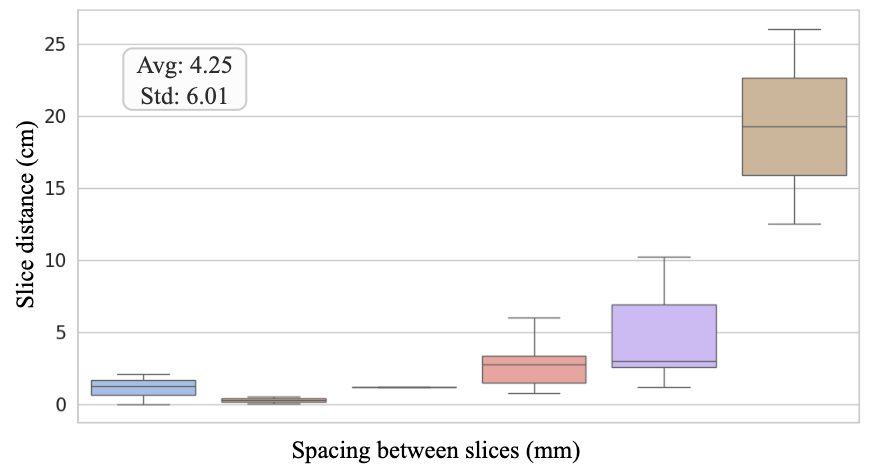}
    \end{subfigure}
    \hfill
    \begin{subfigure}[b]{0.48\textwidth}
        \centering
        \includegraphics[width=\textwidth]{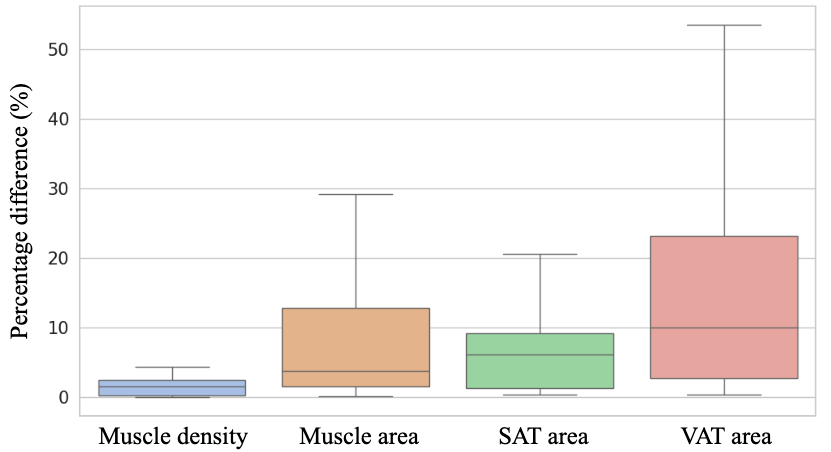} 
    \end{subfigure}
    \caption{\textbf{Quantitative Evaluation of Vertebrae Detection by TotalSegmentator:} The box plot in the \textbf{left} panel shows the slice distance between human-selected and automatically selected slices across different slice spacings. The \textbf{right} panel displays the percentage difference in body composition metrics caused by slice selection mismatch.}
    \label{fig:slice_selection_analysis}
\end{figure*}

The absolute slice distance error is shown in the left panel of Fig. \ref{fig:slice_selection_analysis}. As demonstrated by the box plot, on average, there is a slice position error of 4.25 $\pm$ 6.01$cm$ between the automated and manual selections. One interesting finding is, the figure demonstrates a clear trend where the error increases with the spacing between slices. In particular, cases with 5.0$mm$ spacing exhibit the largest variation, with some errors exceeding 20$cm$, while cases with small image spacing (e.g., 0.6$mm$ or 0.625$mm$) show minimal differences.

The right panel of Fig. \ref{fig:slice_selection_analysis} presents the percentage difference in body composition measurements caused by this slice selection mismatch. Among the four metrics, muscle density is the most stable with a median difference below 5\%, whereas visceral fat shows the highest variability with a median difference above 20\%. Muscle area and subcutaneous fat show moderate variation, with median differences around 10\% and 7\%, respectively.

For qualitative evaluation, we randomly selected four examples and visualized them in the sagittal view, showing the automatically segmented vertebrae, the automated selected slices, manually selected slices, and the slice distance between the automated and manual selections. The visualization is shown in Fig. \ref{fig:slice_selection_analysis_qual} in Appendix \ref{appendix:vertebra}.

\subsection{Qualitative evaluation}
\begin{figure*}[t]
    \centering
    \includegraphics[width=0.9\textwidth]{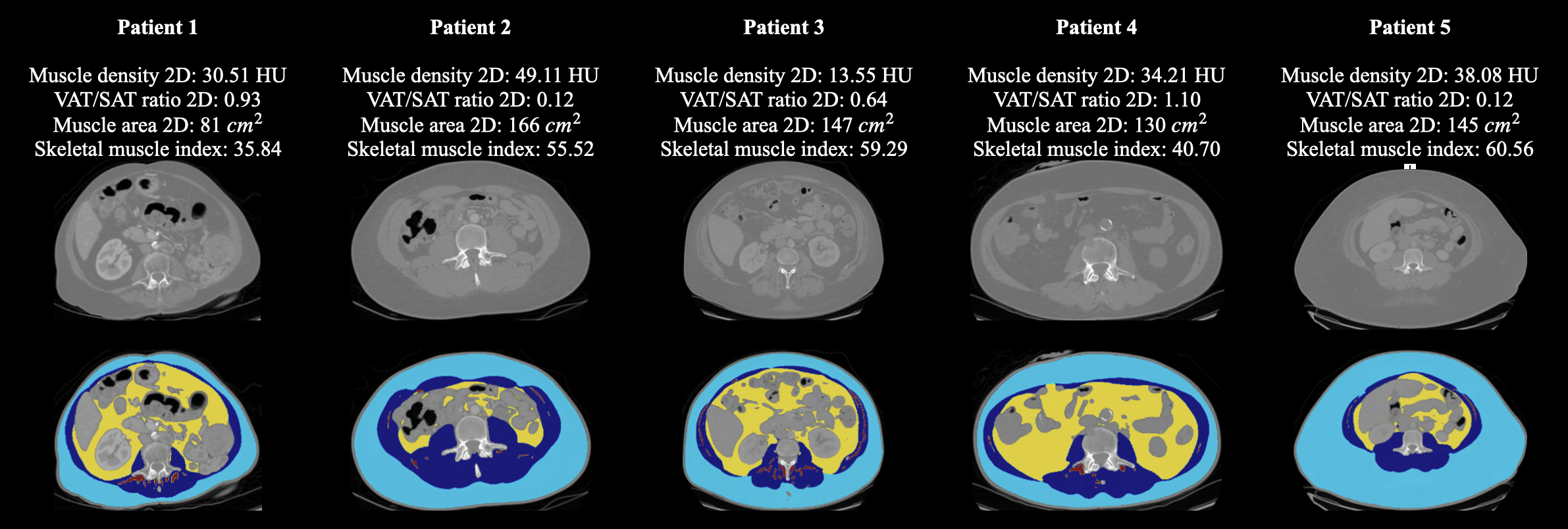}
    \caption{\textbf{Qualitative evaluation of our segmentation model:} Figure shows segmentation results of the abdominal L3 slice. In the segmentation, dark blue shows skeletal muscle, light blue SAT, yellow VAT, and maroon muscular fat.}
    \label{fig:2D_qualitative_evaluation_small}
\end{figure*}

Fig. \ref{fig:2D_qualitative_evaluation_small} demonstrates the L3 segmentation results and their corresponding body composition metrics (muscle density, VAT/SAT ratio, muscle area/volume, and SMI) for five selected patients. These examples were carefully selected to reflect extreme variations in body composition within the Demographic Analysis dataset, including exceptionally low and high values of muscle density (Patient 3 and Patient 2, respectively), VAT/SAT ratio (Patient 2/5 and Patient 4), muscle area (Patient 1 and Patient 2), and SMI (Patient 1 and Patient 5).

For a more comprehensive qualitative evaluation across patients with varying body composition levels, including all four metrics in both 2D and 3D visualizations, the figures are presented in Appendix \ref{appendix:comprehensive_vis}.

Notably, as shown in figure \ref{fig:2D_qualitative_evaluation_small} result, there's no simple correlation between the four body composition metrics. For example, Patient 2 with the highest muscle density does not exhibit the highest skeletal muscle index. More precise relationship analysis based on Pearson Correlation coefficient for each body composition metrics pair is shown in Appendix \ref{sec:Body composition metrics relationship}.

\subsection{Quantitative evaluation} \label{sec:Quantitative evaluatio}
By launching three experiments, our quantitative evaluation aims to demonstrate four main points: (1) the rationale for selecting nnU-Net ResEnc XL for our performance comparison and body composition analysis; (2) the validity of our data annotation; (3) the final model’s segmentation performance across different body regions; and (4) the final model’s accuracy in measuring body composition metrics.

The first point is addressed by comparing the nine segmentation models introduced in Sec. \ref{Method:Segmentation algorithms} on all slices from the Internal Test and SAROS datasets, as described in Sec. \ref{sec:Datasets}. Details of this experiment are provided in Sec. \ref{sec:External dataset evaluation}.

Sec. \ref{sec:External dataset evaluation} also includes a comparison between our segmentation model and two external models \citep{wasserthal2023totalsegmentator,hou2024enhanced} on the publicly available SAROS dataset \citep{koitka2023saros, clark2013cancer}. Since these external models were trained on their own datasets and we evaluated all three models on the same external dataset, this comparison supports point 2, demonstrating the reliability of our data annotation and segmentation performance.

With the final model selected from the experiment described in Point 1, Sec. \ref{sec:Internal dataset evaluation} evaluates its segmentation performance and reliability on both the Internal and SAROS datasets. We report Dice coefficients, MRAE, and $R^2$ between the model predictions and manual annotations across body regions, including L3, T12–L4, and all slices within the selected volumes (chest, abdomen, and pelvis), thereby addressing Point 3. Notably, $R^2$ is computed only for skeletal muscle and SAT due to the limited number of volumetric annotations in the Internal Test set and the availability of only skeletal muscle and SAT labels in the SAROS dataset.

Sec. \ref{sec:Analysis metric evaluation} evaluates the accuracy of automated body composition measurements by comparing them with values derived from manual annotations. Specifically, we report the mean error for muscle density, VAT/SAT ratio, muscle area/volume, and SMI in both 2D and 3D analyses. 

To enable more flexible use of muscular fat (including intra-muscular and inter-muscular components), we additionally report the Dice coefficient for muscular fat alone in Sec.\ref{sec:muscular fat segmentation}. We also include Dice scores for combined labels such as muscular fat + VAT and muscular fat + SAT, compared against manual annotations in the internal dataset.

\subsubsection{Comparison with benchmark models} \label{sec:External dataset evaluation}
To demonstrate the rationale for selecting nnU-Net ResEnc XL for our later performance comparison and body composition analysis, we compare nine selected models trained on the same Internal Training dataset and test them on both the Internal Test and SAROS datasets. The performance is evaluated by comparing the manually annotated labels with the automatically generated labels across all slices. The results are presented in Table \ref{tab:external_evaluation}, in the \textit{Models with the Same Training Set} section, which summarizes the segmentation performance on both internal and external datasets using the Dice coefficient and MRAE with bolded value for the best-performing models across datasets and metrics.

As a result, nnU-Net ResEnc XL demonstrates the best Dice performance on both the Internal Test and SAROS datasets, along with the lowest MRAE on the SAROS dataset. Therefore, we adopt the nnU-Net ResEnc XL architecture as our segmentation model for subsequent performance comparisons and body composition analysis.

With the selected nnU-Net ResEnc XL, we further compare our model with the chosen external benchmarks proposed by Hou et al. \citep{hou2024enhanced} and TotalSegmentator model \citep{wasserthal2023totalsegmentator}. Performance is evaluated by the Dice coefficient between our segmented mask and the publicly available skeletal muscle and SAT annotations on SAROS dataset \citep{koitka2023saros, clark2013cancer}. To compare our method with the chosen benchmarks, we follow the instructions provided in \citep{hou2024enhanced}, constraining the analysis to the abdomen section, specifically L1–L5 and T9–T12 following the instructions provided \citep{hou2024enhanced}. The performance results are illustrated in Table \ref{tab:external_evaluation}. As a result, our model outperform the enhanced segmentation model \citep{hou2024enhanced} by 2.10\% for skeletal muscle and 8.6\% for SAT. Additionally, it surpasses the TotalSegmentator \citep{wasserthal2023totalsegmentator} by 6.5\% for skeletal muscle and 11.6\% for SAT. Notably, due to iiicense restrictions, our evaluation dataset is a large subset of theirs, with 650 commercially licensed volumes used in our study compared to 900 volumes in theirs. However, due to the considerable amount of data and data overlap, it is still representative of the original dataset, ensuring the confidence of our advancements.

\begin{table*}[t]
\centering
\resizebox{\textwidth}{!}{%
\begin{tabular}{c|c|c|c|c|c|c|c|c}
\toprule
\multicolumn{9}{c}{\textbf{Models with the Same Training Set}} \\\toprule
\multicolumn{9}{c}{Internal Dataset} \\
\midrule
 & \multicolumn{2}{c|}{Skeletal Muscle} & \multicolumn{2}{c|}{SAT} & \multicolumn{2}{c|}{VAT} & \multicolumn{2}{c}{Average} \\
 \midrule
 & Dice $\uparrow$ (\%) & MRAE $\downarrow$ (\%) & Dice $\uparrow$ (\%) & MRAE $\downarrow$ (\%) & Dice $\uparrow$ (\%) & MRAE $\downarrow$ (\%) & Dice $\uparrow$ (\%) & MRAE $\downarrow$ (\%) \\
\midrule
MexNetX-S & 88.23 $\pm$ 4.47 & 4.14 $\pm$ 2.48 & 91.62 $\pm$ 6.36 & 5.06 $\pm$ 6.99 & 81.35 $\pm$ 9.63 & 9.81 $\pm$ 12.79 & 87.07 $\pm$ 6.82 & 6.34 $\pm$ 7.42 \\\midrule
MexNetX-M & 88.52 $\pm$ 3.45 & 5.70 $\pm$ 4.65 & 92.10 $\pm$ 4.56 & 6.75 $\pm$ 5.64 & 83.17 $\pm$ 6.50 & 6.77 $\pm$ 5.01 & 87.93 $\pm$ 4.84 & 6.41 $\pm$ 5.10 \\\midrule
MexNetX-L & 87.94 $\pm$ 4.10 & 4.53 $\pm$ 3.87 & 91.65 $\pm$ 5.32 & 5.68 $\pm$ 6.34 & 82.80 $\pm$ 6.40 & 9.36 $\pm$ 8.76 & 87.46 $\pm$ 5.27 & 6.52 $\pm$ 6.32 \\\midrule
SAM-Med2D fine-tuning & 82.01 $\pm$ 3.83 & 14.97 $\pm$ 5.88 & 88.35 $\pm$ 5.91 & 10.03 $\pm$ 6.24 & 66.22 $\pm$ 9.60 & 32.83 $\pm$ 17.92 & 78.86 $\pm$ 6.45 & 19.28 $\pm$ 10.01 \\\midrule
SAM fine-tuning & 80.56 $\pm$ 5.35 & 6.64 $\pm$ 5.49 & 86.42 $\pm$ 6.23 & 12.92 $\pm$ 7.36 & 79.07 $\pm$ 7.90 & 7.42 $\pm$ 6.99 & 82.02 $\pm$ 6.49 & 8.99 $\pm$ 6.61 \\\midrule
MedSAM fine-tuning & 77.68 $\pm$ 7.21 & 6.67 $\pm$ 6.42 & 84.39 $\pm$ 6.82 & 11.63 $\pm$ 6.99 & 72.80 $\pm$ 11.74 & 16.15 $\pm$ 9.75 & 78.29 $\pm$ 8.59 & 11.48 $\pm$ 7.72 \\\midrule
nnU-Net ResEnc M & 89.83 $\pm$ 3.18 & \textbf{3.51 $\pm$ 3.23} & 93.48 $\pm$ 4.16 & 3.92 $\pm$ 4.27 & 84.20 $\pm$ 6.65 & \textbf{4.44 $\pm$ 3.69} & 89.17 $\pm$ 4.68 & \textbf{3.96 $\pm$ 3.73} \\\midrule
nnU-Net ResEnc L & 89.71 $\pm$ 3.26 & 4.05 $\pm$ 3.55 & 93.81 $\pm$ 4.16 & 3.07 $\pm$ 4.06 & 84.30 $\pm$ 6.95 & 4.96 $\pm$ 4.19 & 89.27 $\pm$ 4.79 & 4.03 $\pm$ 3.93 \\\midrule\midrule
nnU-Net ResEnc XL & \textbf{91.85 $\pm$ 3.37} & 4.13 $\pm$ 4.10 & \textbf{94.06 $\pm$ 4.25} & \textbf{3.06 $\pm$ 3.97} & 
\textbf{89.45 $\pm$ 7.09} & 5.91 $\pm$ 5.51 & \textbf{91.79 $\pm$ 4.90} & 4.37 $\pm$ 4.53 \\
\midrule
\multicolumn{9}{c}{External Dataset} \\
\midrule
 & \multicolumn{2}{c|}{Skeletal Muscle} & \multicolumn{2}{c|}{SAT} & \multicolumn{2}{c|}{VAT} & \multicolumn{2}{c}{Average} \\
\midrule
 & Dice $\uparrow$ (\%) & MRAE $\downarrow$ (\%) & Dice $\uparrow$ (\%) & MRAE $\downarrow$ (\%) & Dice $\uparrow$ (\%) & MRAE $\downarrow$ (\%) & Dice $\uparrow$ (\%) & MRAE $\downarrow$ (\%) \\
\midrule
MexNetX-S & 89.64 $\pm$ 2.85 & 7.20 $\pm$ 3.83 & 89.24 $\pm$ 5.33 & 4.15 $\pm$ 3.88 & - & - & 89.44 $\pm$ 4.23 & 5.68 $\pm$ 3.86\\\midrule
MexNetX-M & 89.68 $\pm$ 2.66 & \textbf{6.91 $\pm$ 3.96} & 88.97 $\pm$ 5.42 & 4.55 $\pm$ 3.93 & - & - & 89.33 $\pm$ 4.22 & 5.73 $\pm$ 3.95\\\midrule
MexNetX-L & 89.39 $\pm$ 2.71 & 8.00 $\pm$ 3.85 & 89.18 $\pm$ 5.14 & 3.87 $\pm$ 3.98 & - & - & 89.29 $\pm$ 3.93 & 5.94 $\pm$ 3.92 \\\midrule
SAM-Med2D fine-tuning & 74.16 $\pm$ 7.33 & 25.03 $\pm$ 8.37 & 82.35 $\pm$ 11.58 & 8.43 $\pm$ 10.40 & - & - & 78.26 $\pm$ 9.64 & 16.73 $\pm$ 9.52 \\\midrule
SAM fine-tuning & 79.86 $\pm$ 9.09 & 12.26 $\pm$ 14.79 & 87.10 $\pm$ 10.71 & 20.34 $\pm$ 19.09 & - & - & 83.48 $\pm$ 9.90 & 16.30 $\pm$ 16.94 \\\midrule
MedSAM fine-tuning & 77.58 $\pm$ 8.10 & 12.38 $\pm$ 12.77 & 86.29 $\pm$ 9.24 & 23.37 $\pm$ 16.13 & - & - & 81.94 $\pm$ 8.69 & 17.88 $\pm$ 14.55 \\\midrule
nnU-Net ResEnc M & 89.50 $\pm$ 2.67 & 7.74 $\pm$ 3.74 & 89.61 $\pm$ 4.83 & 3.72 $\pm$ 3.68 & - & - & 89.56 $\pm$ 3.88 & 5.73 $\pm$ 3.71 \\\midrule
nnU-Net ResEnc L & \textbf{89.72 $\pm$ 2.65} & 7.45 $\pm$ 3.68 & 89.51 $\pm$ 4.89 & \textbf{3.53 $\pm$ 3.51} & - & - & 89.62 $\pm$ 3.98 & 5.49 $\pm$ 3.61 \\\midrule\midrule
nnU-Net ResEnc XL & 89.68 $\pm$ 2.75 & 7.35 $\pm$ 3.80 & \textbf{90.27 $\pm$ 4.83} & 3.60 $\pm$ 3.39 & - & - & \textbf{89.98 $\pm$ 3.79} & \textbf{5.48 $\pm$ 3.60} \\
\toprule
\multicolumn{9}{c}{\textbf{Models with the Different Training Set}} \\\toprule
\multicolumn{9}{c}{External Dataset} \\
\midrule
 & \multicolumn{2}{c|}{Skeletal Muscle} & \multicolumn{2}{c|}{SAT} & \multicolumn{2}{c|}{VAT} & \multicolumn{2}{c}{Average} \\\midrule
& \multicolumn{2}{c|}{Dice $\uparrow$ (\%) [IQR]} & \multicolumn{2}{c|}{Dice $\uparrow$ (\%) [IQR]} & \multicolumn{2}{c|}{Dice $\uparrow$ (\%) [IQR]} & \multicolumn{2}{c|}{Dice $\uparrow$ (\%)} \\
\midrule
TotalSegmentator \citep{wasserthal2023totalsegmentator} & \multicolumn{2}{c|}{83.2 $\pm$ 4.6 [80.5, 86.4]} & \multicolumn{2}{c|}{80.8 $\pm$ 10.4 [76.7, 87.7]} & \multicolumn{2}{c|}{-} & \multicolumn{2}{c}{82.0 $\pm$ 7.5} \\
\midrule
Enhanced Segmentation \citep{hou2024enhanced} & \multicolumn{2}{c|}{87.6 $\pm$ 3.3 [85.6, 90.0]} & \multicolumn{2}{c|}{83.8 $\pm$ 10.9 [80.7, 90.5]} & \multicolumn{2}{c|}{-} & \multicolumn{2}{c}{85.7 $\pm$ 7.1} \\
\midrule\midrule
Ours (nnU-Net ResEnc XL) & \multicolumn{2}{c|}{\textbf{89.7 $\pm$ 3.2 [88.3, 91.7]}} & \multicolumn{2}{c|}{\textbf{92.4 $\pm$ 3.7 [91.0, 94.7]}} & \multicolumn{2}{c|}{-} & \multicolumn{2}{c}{\textbf{91.05 $\pm$ 3.45}} \\
\bottomrule
\end{tabular}
}
\caption{\textbf{Comparison with benchmark models:} The \textit{Models with the Same Training Set} section presents the segmentation performance of nine models using Dice and MRAE, where models are trained and tested on the same dataset. The \textit{Models with the Different Training Set} section compares the segmentation performance of our best-performing model with externally trained models, reported using Dice scores. The scores are presented as mean, standard deviation, and interquartile range (IQR). Bolded values indicate the best-performing model among those evaluated.}
\label{tab:external_evaluation}
\end{table*}

\subsubsection{Internal segmentation performance} \label{sec:Internal dataset evaluation}
\begin{figure*}[htbp]
    \centering
    \begin{subfigure}[b]{0.31\textwidth}
        \includegraphics[width=\textwidth]{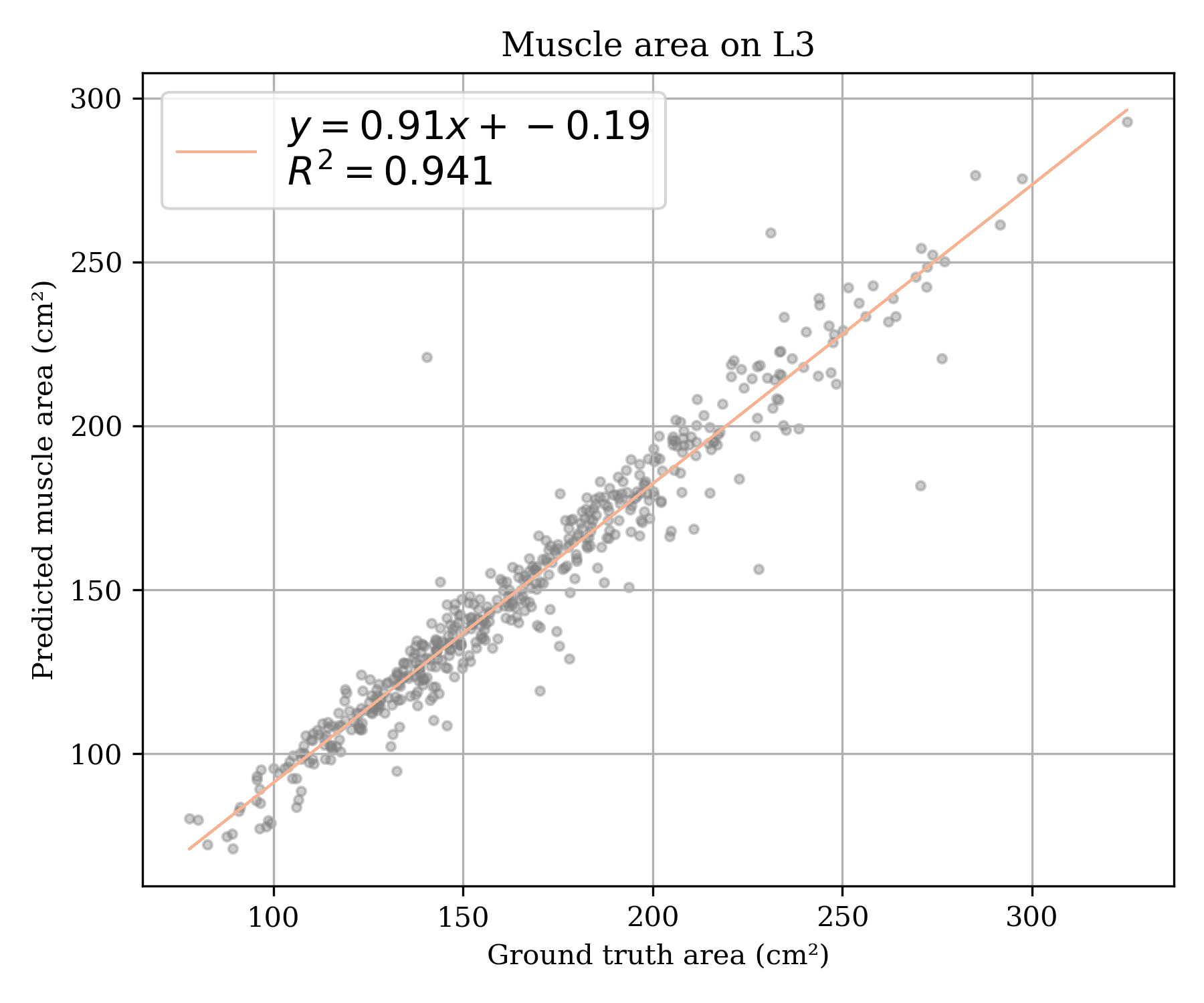}
    \end{subfigure}
    \begin{subfigure}[b]{0.31\textwidth}
        \includegraphics[width=\textwidth]{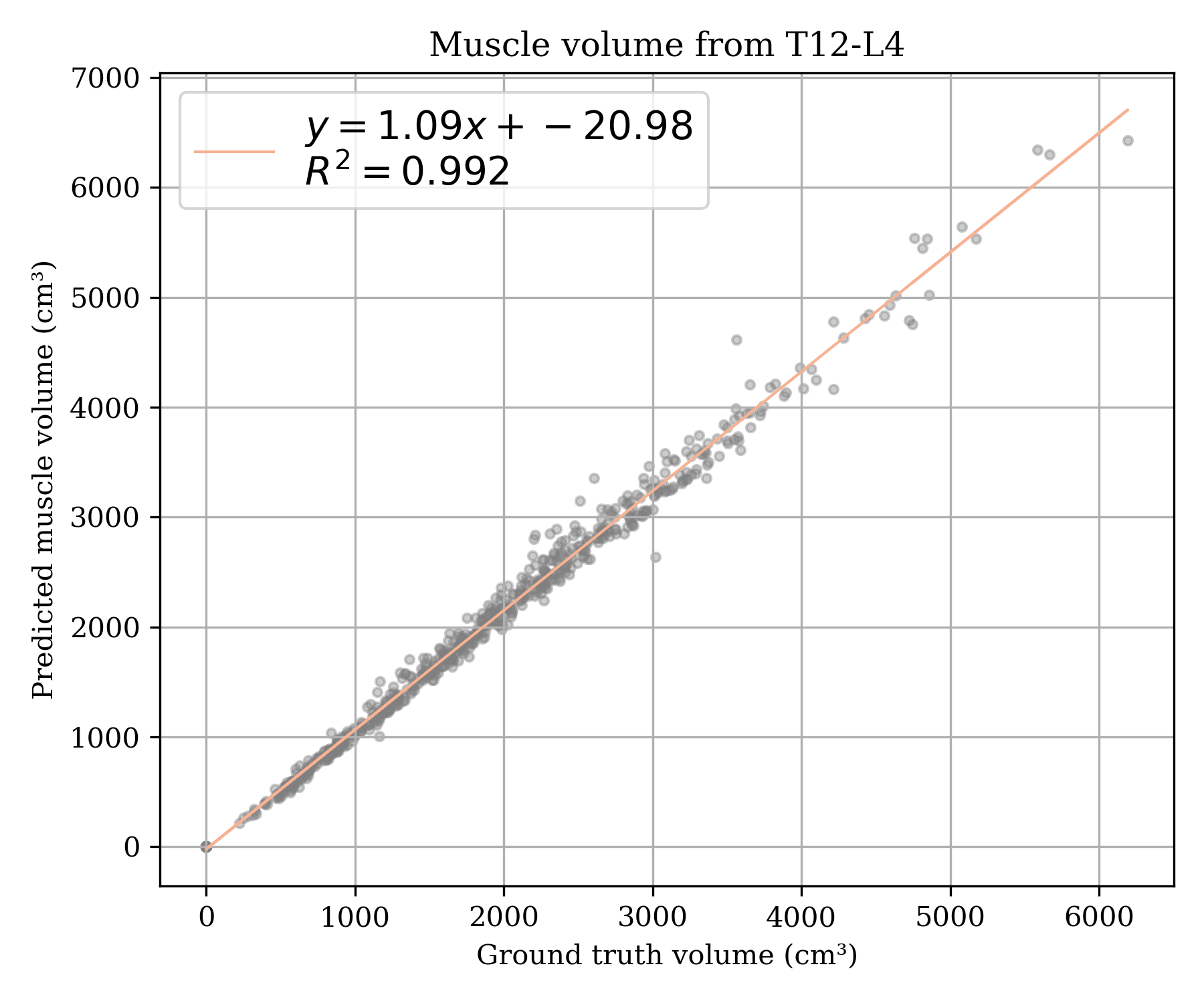}
    \end{subfigure}
    \begin{subfigure}[b]{0.31\textwidth}
        \includegraphics[width=\textwidth]{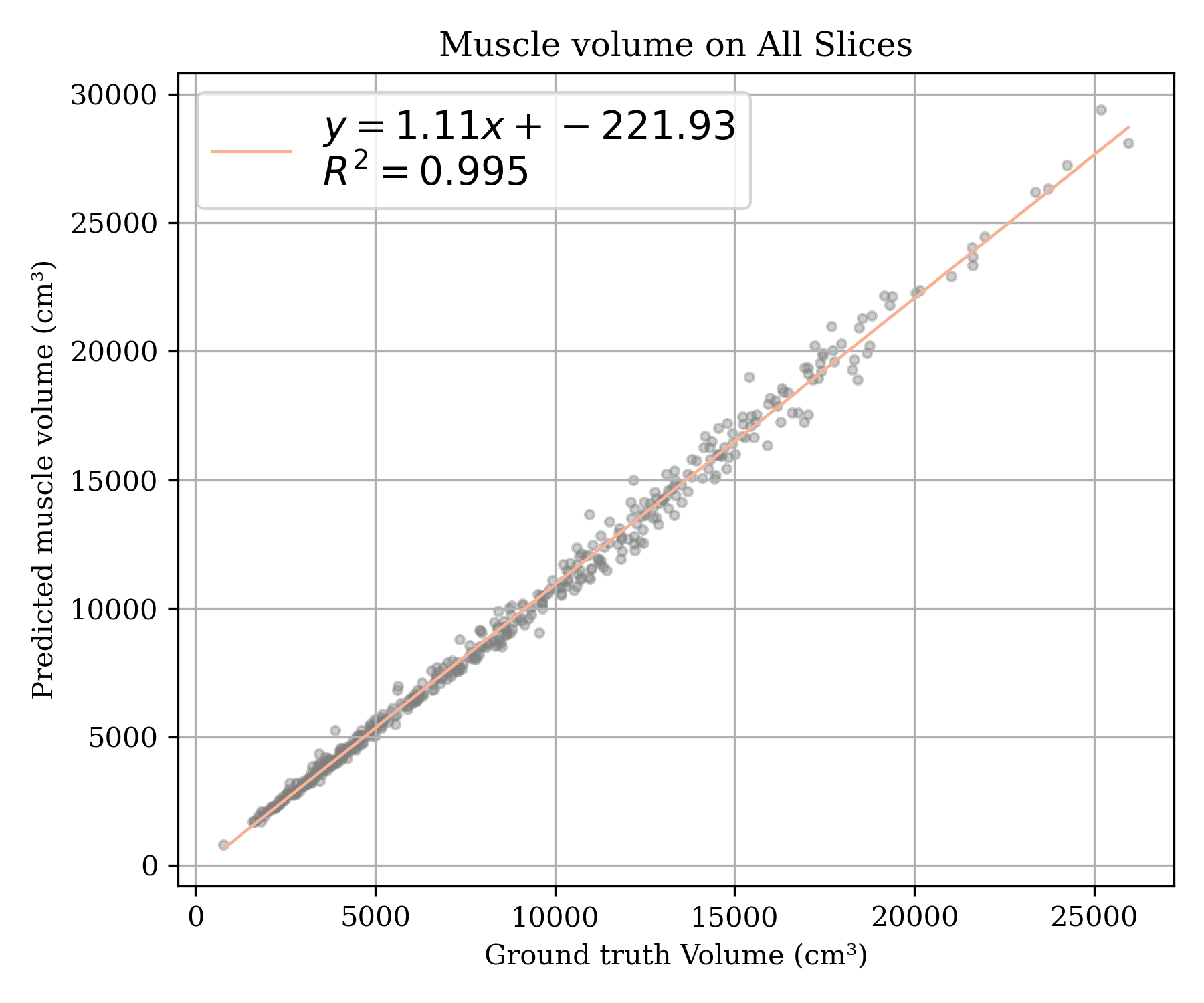}
    \end{subfigure}

    \begin{subfigure}[b]{0.31\textwidth}
        \includegraphics[width=\textwidth]{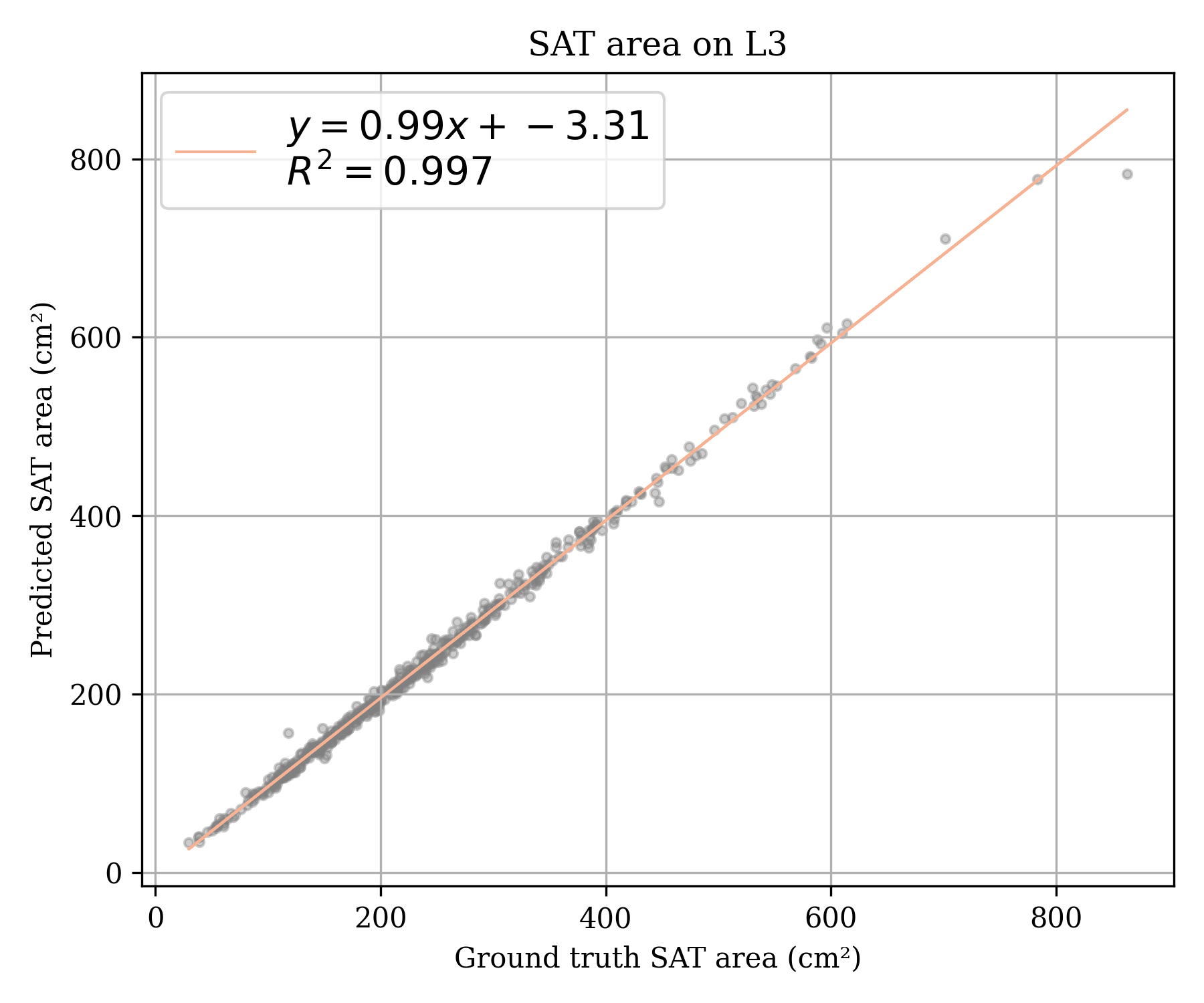}
    \end{subfigure}
    \begin{subfigure}[b]{0.31\textwidth}
        \includegraphics[width=\textwidth]{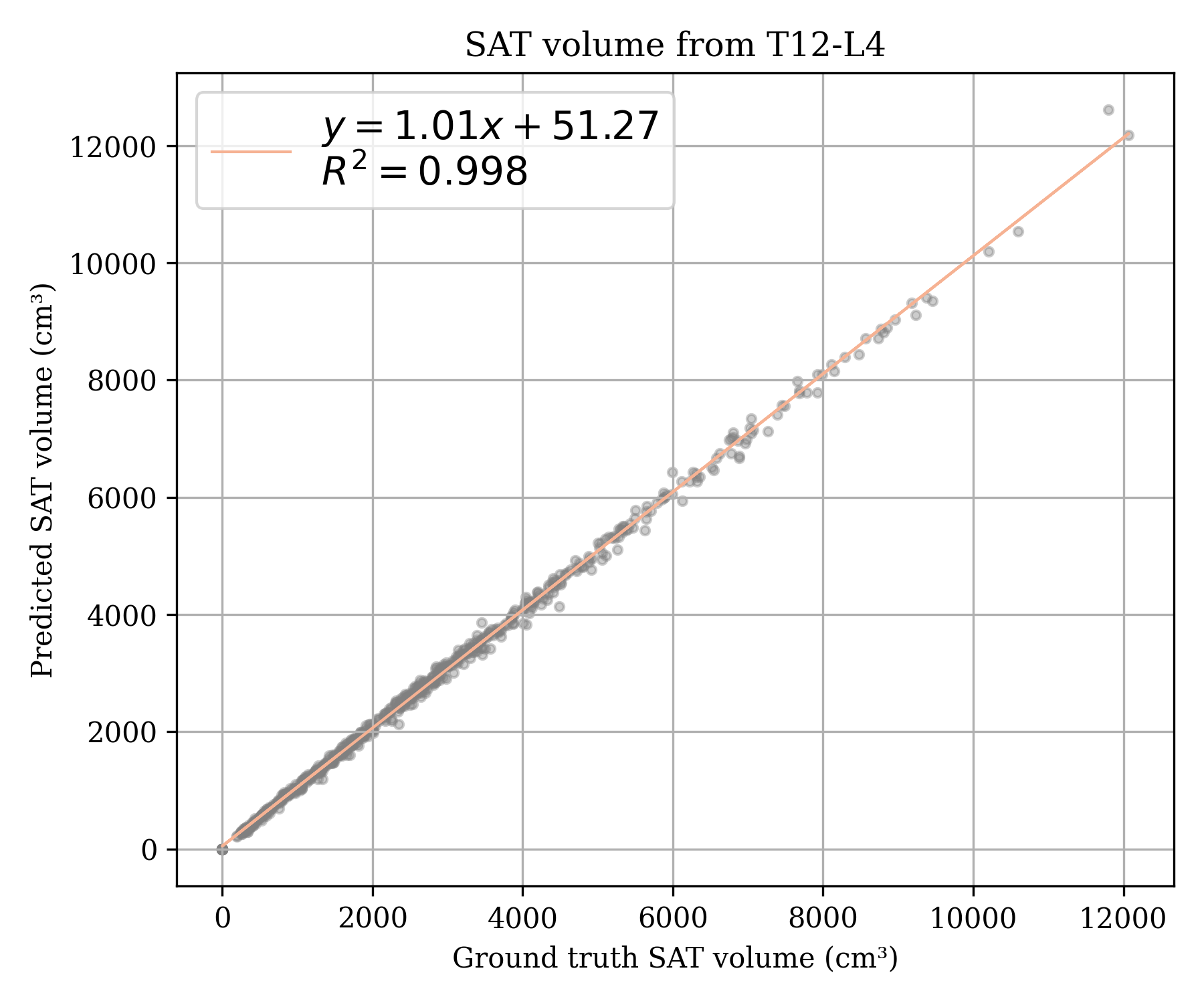}
    \end{subfigure}
    \begin{subfigure}[b]{0.31\textwidth}
        \includegraphics[width=\textwidth]{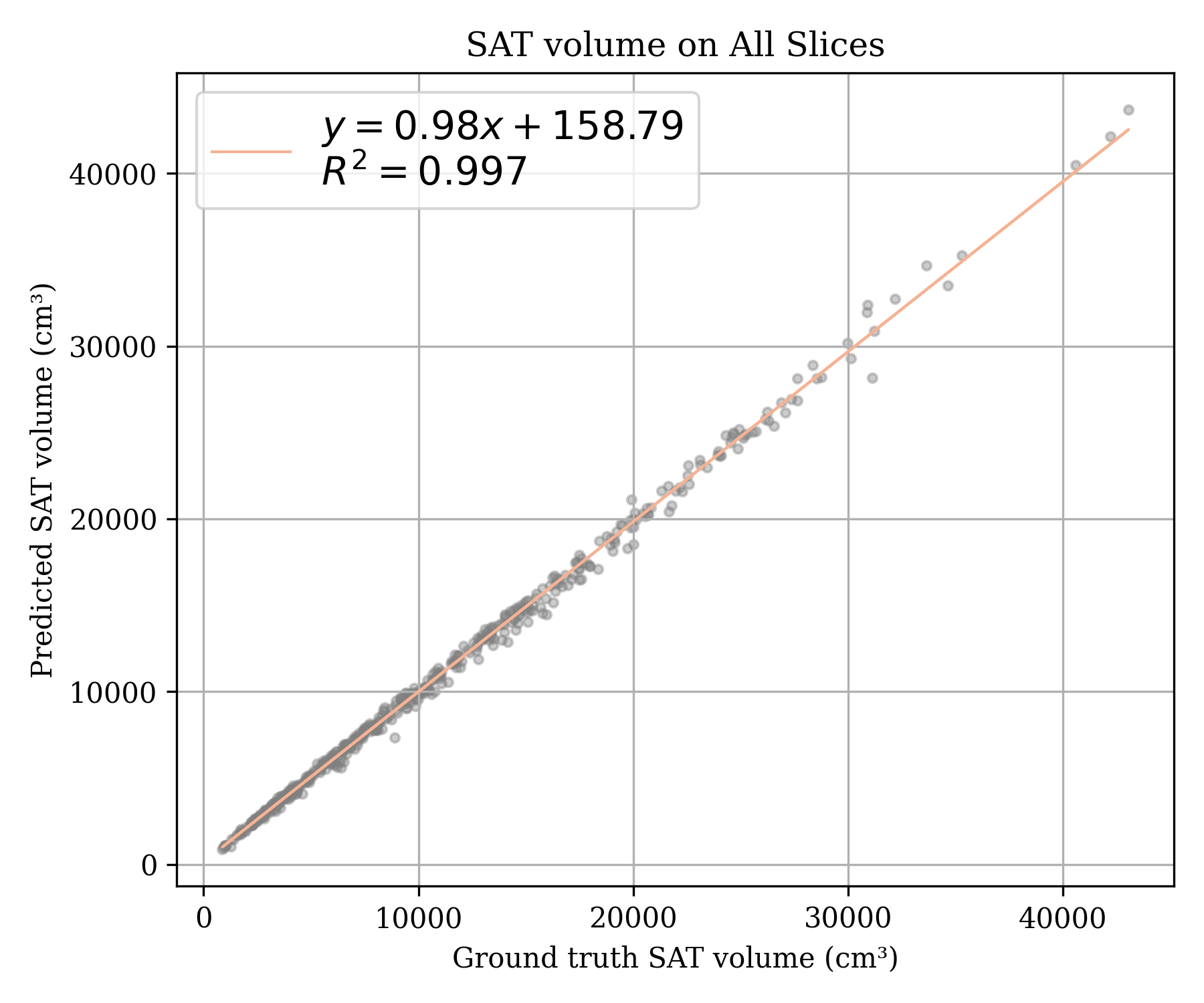}
    \end{subfigure}

    \caption{
        \textbf{$\mathbf{R^2}$ correlation plots:} for muscle (top row) and SAT (bottom row) across three evaluation ranges: L3 slice, T12--L4 range, and the All Slices setting.
    }
    \label{fig:volume_correlation_plots}
\end{figure*}

For our internal evaluation, without comparisons to other methods, we utilize the Dice coefficient, MRAE, and $R^2$ to compare our auto-segmented labels given by previously selected nnU-Net ResEnc XL with manual annotations for skeletal muscle, subcutaneous adipose tissue (SAT), and visceral adipose tissue (VAT). Table \ref{tab:internal_evaluation} summarizes the segmentation performance for both Internal Test and SAROS datasets, using the Dice coefficient and MRAE (Sec. \ref{Method:Segmentation metrics}). Fig. \ref{fig:volume_correlation_plots} demonstrates the $R^2$ correlation for skeletal muscle and SAT. 

As the result demonstrated by Table \ref{tab:internal_evaluation}, the model performs competitively across all regions, with the highest average Dice coefficient observed at the L3 slice (93.38\%) and the lowest average MRAE at all-slices (4.37\%) in the internal dataset. Among the tissues, the model achieves the consistent highest segmentation accuracy for SAT, which consistently shows superior Dice scores and lower MRAEs compared to skeletal muscle and VAT. The external dataset follows a similar trend, with the best Dice coefficient observed at L3 (92.55\%), the lowest MRAE at the all-slices setting ( 5.48\%), and the highest accuracy consistently for SAT. For $R^2$ correlation, Fig. \ref{fig:volume_correlation_plots} exhibit a strong correlation on both skeletal muscle and SAT on all body regions with $R^2$ values over 0.941 for muscle and 0.997 for SAT.

Notably, there is a label inconsistency between the annotations in our internal dataset and those in the SAROS dataset. Specifically, the SAROS annotation includes skin as part of the SAT label. To address this discrepancy, we applied a simple post-processing step to our model by dilating our SAT segmentation to include the skin. The detailed process for this post-processing is described in Appendix \ref{sec:Post-processing for label inconsistencies}. However, the post-processing step only mimics the inclusion of skin in the segmentation, which still leaves a gap between the two segmentation approaches.

\begin{table*}[t]
\centering
\resizebox{\textwidth}{!}{%
\begin{tabular}{c|c|c|c|c|c|c|c|c}
\toprule
\multicolumn{9}{c}{Internal Dataset} \\
\midrule
 & \multicolumn{2}{c|}{Skeletal Muscle} & \multicolumn{2}{c|}{SAT} & \multicolumn{2}{c|}{VAT} & \multicolumn{2}{c}{Average} \\
 \midrule
 & Dice $\uparrow$ (\%) & MRAE $\downarrow$ (\%) & Dice $\uparrow$ (\%) & MRAE $\downarrow$ (\%) & Dice $\uparrow$ (\%) & MRAE $\downarrow$ (\%) & Dice $\uparrow$ (\%) & MRAE $\downarrow$ (\%) \\
\midrule
L3 & 92.53 $\pm$ 4.48 & 6.90 $\pm$ 8.05 & 94.72 $\pm$ 5.21 & 3.20 $\pm$ 3.74 & 92.90 $\pm$ 5.04 & 5.50 $\pm$ 6.12 & 93.38 $\pm$ 4.92 & 5.20 $\pm$ 6.22 \\
\midrule
T12-L4 & 92.08 $\pm$ 4.74 & 4.86 $\pm$ 5.69 & 94.03 $\pm$ 6.88 & 4.43 $\pm$ 7.05 & 92.15 $\pm$ 5.90 & 5.63 $\pm$ 8.22 & 92.75 $\pm$ 5.84 & 4.97 $\pm$ 6.99 \\
\midrule
All Slices & 91.85 $\pm$ 3.37 & 4.13 $\pm$ 4.10 & 94.06 $\pm$ 4.25 & 3.06 $\pm$ 3.97 & 89.45 $\pm$ 7.09 & 5.91 $\pm$ 5.51 & 91.79 $\pm$ 4.90 & 4.37 $\pm$ 4.53 \\
\toprule
\multicolumn{9}{c}{External Dataset} \\
\midrule
 & \multicolumn{2}{c|}{Skeletal Muscle} & \multicolumn{2}{c|}{SAT} & \multicolumn{2}{c|}{VAT} & \multicolumn{2}{c}{Average} \\
\midrule
 & Dice $\uparrow$ (\%) & MRAE $\downarrow$ (\%) & Dice $\uparrow$ (\%) & MRAE $\downarrow$ (\%) & Dice $\uparrow$ (\%) & MRAE $\downarrow$ (\%) & Dice $\uparrow$ (\%) & MRAE $\downarrow$ (\%) \\
\midrule
L3 & 92.22 $\pm$ 3.27 & 9.20 $\pm$ 5.31 & 92.88 $\pm$ 3.62 & 3.29 $\pm$ 2.82 & - & - & 92.55 $\pm$ 3.45 & 6.75 $\pm$ 4.25 \\
\midrule
T12-L4 & 91.03 $\pm$ 3.77 & 6.61 $\pm$ 4.24 & 92.59 $\pm$ 4.35 & 4.39 $\pm$ 3.32 & - & - & 91.81 $\pm$ 4.06 & 5.50 $\pm$ 3.78 \\
\midrule
All Slices & 89.68 $\pm$ 2.75 & 7.35 $\pm$ 3.80 & 90.27 $\pm$ 4.83 & 3.60 $\pm$ 3.39 & - & - & 89.98 $\pm$ 3.79 & 5.48 $\pm$ 3.60 \\
\bottomrule
\end{tabular}
}
\caption{ \textbf{Internal segmentation Performance:} Segmentation performance for skeletal muscle, subcutaneous adipose tissue (SAT), and visceral adipose tissue (VAT) across the internal and external datasets, reported using Dice scores ($\uparrow$) and MRAE ($\downarrow$). Results are provided for L3, T12-L4, and all slices, highlighting the model's superior performance at the L3 slice and on SAT compared to skeletal muscle and VAT. The "Average" column provides the mean Dice score and MRAE across the reported tissues. VAT performance is unavailable due to the absence of VAT annotation in the SAROS dataset.}
\label{tab:internal_evaluation}
\end{table*}

\subsubsection{Analysis metric evaluation} \label{sec:Analysis metric evaluation}
For the metric evaluation, we utilize our auto-segmentation to measure four selected body composition metrics: muscle density, VAT/SAT ratio, muscle volume, and SMI, in both 2D and 3D settings on our internal test dataset. These results are compared with the body composition metrics derived from manual annotations. The MRAE result is presented in Table \ref{tab:metric_analysis}.

The model demonstrates the best performance in measuring muscle density among four body compositions, with an MRAE lower than 5\% on both internal and external datasets for both 2D and 3D measurements. Across all body composition metrics for both datasets, the model achieves an MRAE lower than 10\%, showcasing its robustness. 

We also provide the full evaluation for all nine selected segmentation models in Appendix \ref{Comprehensive analysis metric evaluation} to offer users additional insights into model performance, helping inform model selection for specific clinical or research applications.

\begin{table*}[t]
\centering
\resizebox{\textwidth}{!}{%
\begin{tabular}{c|c|c|c|c}
\toprule
\multicolumn{5}{c}{Internal Dataset} \\
\midrule
& Muscle density (\% among range -29 to +150 HU) & VAT/SAT ratio (\%) & Skeletal Muscle area/volume (\%) & SMI (\%) \\
\midrule
2D & 1.43 $\pm$ 1.19 & 5.43 $\pm$ 5.17 & 6.90 $\pm$ 8.05 & 6.90 $\pm$ 8.05 \\
\midrule
3D & 2.18 $\pm$ 2.14 & 5.09 $\pm$ 4.74 & 4.86 $\pm$ 5.69 & - \\
\toprule
\multicolumn{5}{c}{External Dataset} \\
\midrule
& Muscle density (\% among range -29 to +150 HU) & VAT/SAT ratio (\%) & Skeletal Muscle area/volume (\%) & SMI (\%) \\
\midrule
2D & 4.47 $\pm$ 2.44 & - & 9.20 $\pm$ 5.31 & - \\
\midrule
3D & 4.71 $\pm$ 2.22 & - & 6.61 $\pm$ 4.24 & - \\
\bottomrule
\end{tabular}%
}
\caption{\textbf{Analysis metric evaluation performance:} The performance of our segmentation model on both internal and external datasets, evaluated by comparing the four body composition metrics automatically calculated by our model with the ground truth measured from manual annotations. The evaluation is based on MRAE ($\downarrow$).}
\label{tab:metric_analysis}
\end{table*}

\subsubsection{Muscular fat segmentation} \label{sec:muscular fat segmentation}
During the literature review, we observed inconsistencies in how muscular fat (both intra-muscular and inter-muscular fat) is classified in research. While some studies include muscular fat as part of skeletal muscle measurements \citep{hou2024enhanced, van2018percentiles}, others classify it under VAT \citep{camus2014prognostic, wirtz2021ct, connelly2013volumetric}, and a smaller subset considers it part of SAT \citep{ozturk2020relationship, magudia2021population}. Consequently, we attempted to segment muscular fat as a separate label in our segmentation model.

In this section, we present the Dice coefficient performance for muscular fat segmented independently, muscular fat included as part of SAT, muscular fat included as part of VAT, and muscular fat included as part of muscle. Notably, for all skeletal muscle evaluations discussed in the previous sections, we follow the methodology adopted in prior studies, where muscular fat is evaluated as part of muscle segmentation (Sec. \ref{sec:Internal dataset evaluation}\ref{sec:External dataset evaluation}). Although the segmentation of muscular fat itself demonstrates a relatively low Dice coefficient (56.27 $\pm$ 10.33\%) compared to manual annotations on our internal dataset, incorporating muscular fat into other labels—specifically muscle, SAT, and VAT, as is common in body composition measurements—results in high Dice coefficients across all slices (91.85 $\pm$ 3.37\%, 92.35 $\pm$ 4.6\%, and 85.19 $\pm$ 6.73\%, respectively).

\subsection{Corner Cases}
\begin{figure*}[htbp]
    \centering
    \includegraphics[width=0.9\textwidth]{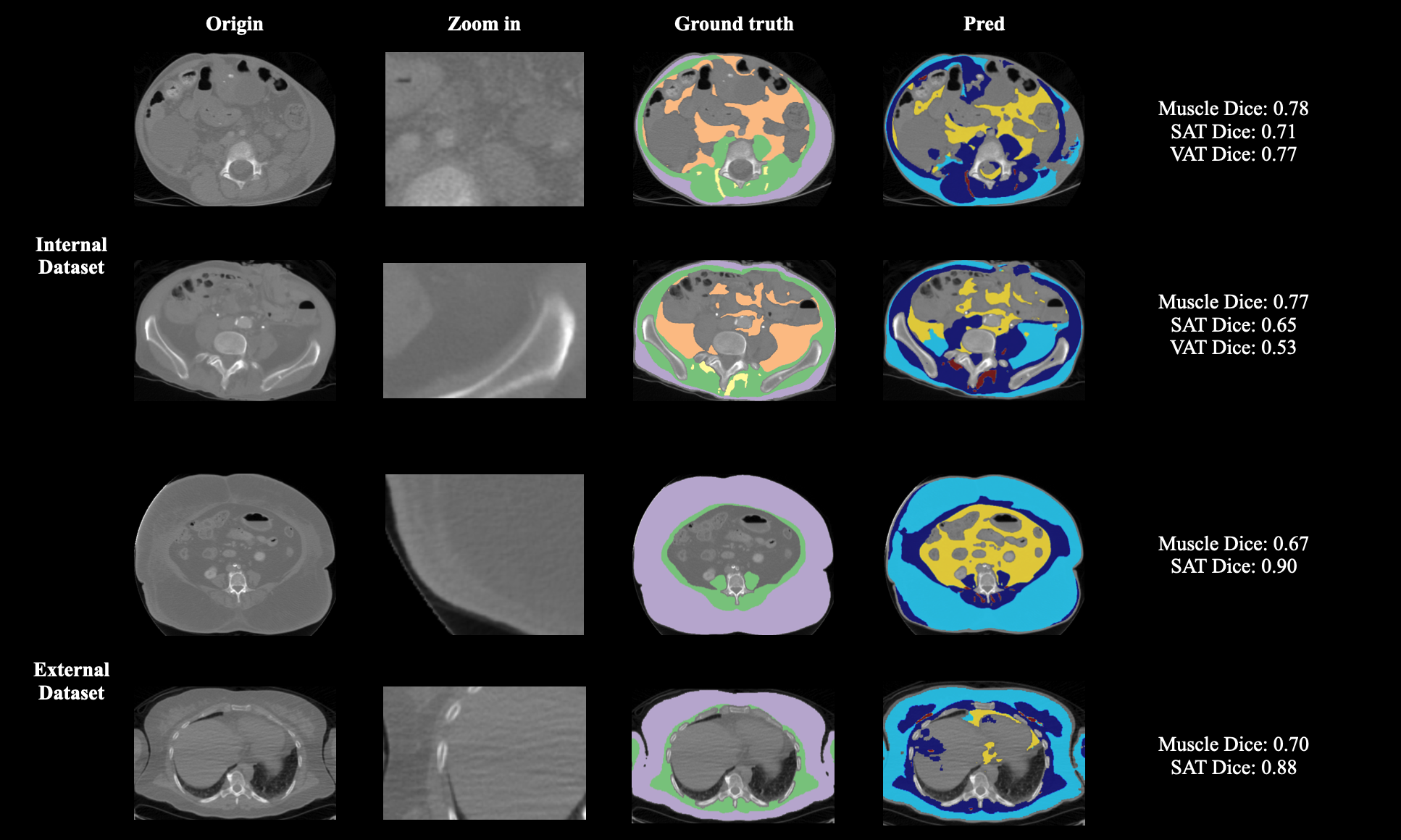}
    \caption{
    \textbf{Representative failure cases of our segmentation model:} Each row shows one case, with the top two from the internal dataset and the bottom two from the external dataset. From left to right: original CT slice, zoomed-in view, ground truth segmentation, model prediction, and corresponding Dice scores. In the Ground Truth, \textit{green} shows skeletal muscle, \textit{purple} SAT, \textit{orange} VAT, and \textit{yellow} muscular fat. In the Pred, \textit{dark blue} shows skeletal muscle, \textit{light blue} SAT, \textit{yellow} VAT, and \textit{maroon} muscular fat.
    }
    \label{fig:failure_cases}
\end{figure*}

In this section, we analyze the corner cases of our model to further provide insights for user application and future improvement. The corner cases include four slices carefully selected from the worst-performing volumes—two from the internal test set and two from the SAROS dataset—representing common scenarios where the model fails. As shown in Fig. \ref{fig:failure_cases}, the model demonstrates relatively poor performance when CT images contain clear noise, as seen in the first, third, and fourth rows. This becomes more evident when zooming into the images provided in the figure's second column. Artifacts caused by motion can also lead to model failure, as shown in the second row. The selected examples in Fig. \ref{fig:failure_cases} also highlight several common types of mistakes made by the model—for example, misclassifying VAT as SAT in the second row, misclassifying thick skin as muscle in the third row, and misclassifying organs as muscle shown in the forth row.

\section{Body composition vs. Demographic Analysis} \label{sec:Body composition analysis}
In the following sections, we analyze the body composition measurements generated by our algorithm, highlighting their relationships with patients' age, sex, and racial groups. The results produced by our algorithm are compared with previous body composition findings reported in leading medical journals. 
The aim of this section is to demonstrate the accuracy of our body composition metrics calculation. While the calculated metric values may differ slightly from those reported in previous studies due to variations in population distribution, the trends shown in this analysis strongly resemble those established before.

\subsection{Body composition metrics vs. age}
\begin{figure*}[htbp]
    \centering
    \begin{subfigure}[b]{0.36\textwidth}
        \includegraphics[width=\linewidth]{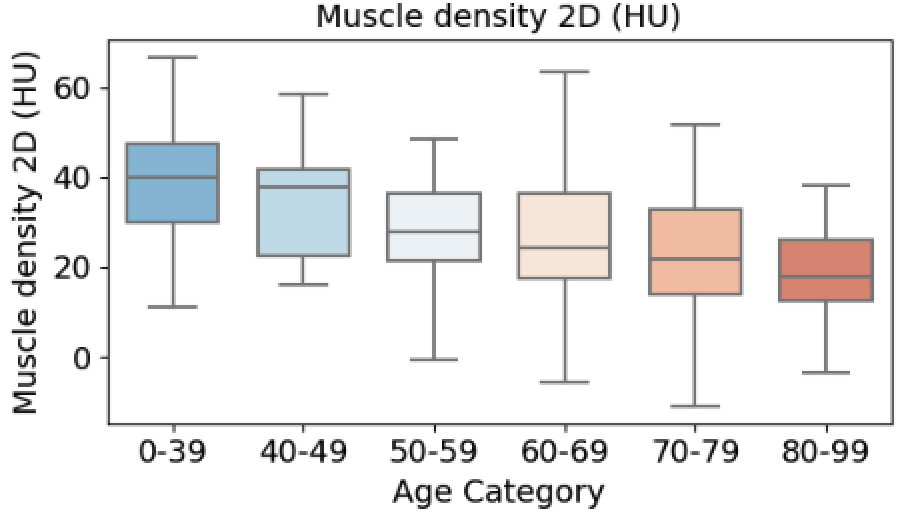}
    \end{subfigure}
    \vspace{1mm}
    \begin{subfigure}[b]{0.36\textwidth}
        \includegraphics[width=\linewidth]{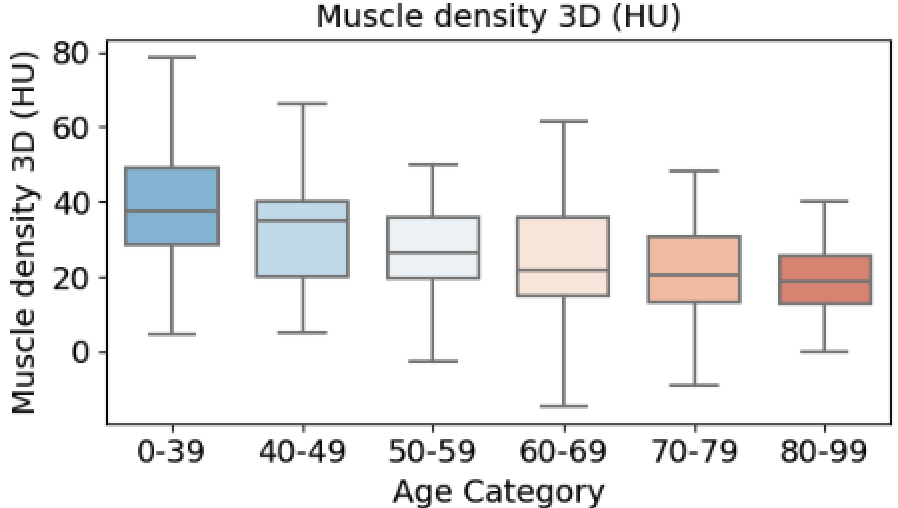}
    \end{subfigure}
    \vspace{1mm}
    \begin{subfigure}[b]{0.36\textwidth}
        \includegraphics[width=\linewidth]{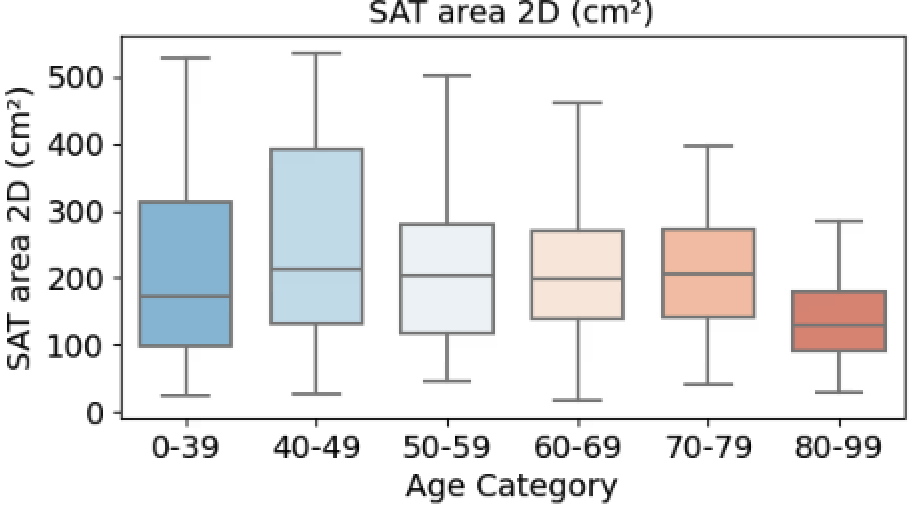}
    \end{subfigure}
    \vspace{1mm}
    \begin{subfigure}[b]{0.36\textwidth}
        \includegraphics[width=\linewidth]{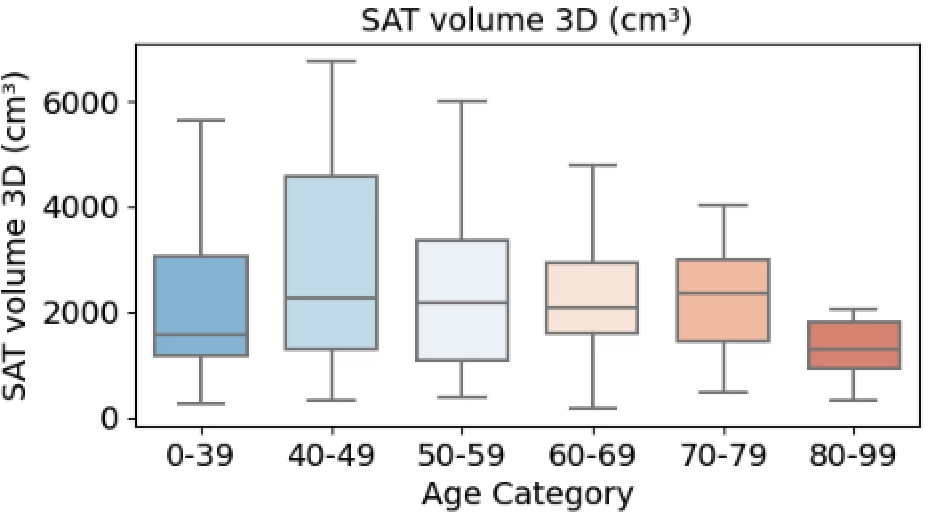}
    \end{subfigure}
    \vspace{1mm}
    \begin{subfigure}[b]{0.36\textwidth}
        \includegraphics[width=\linewidth]{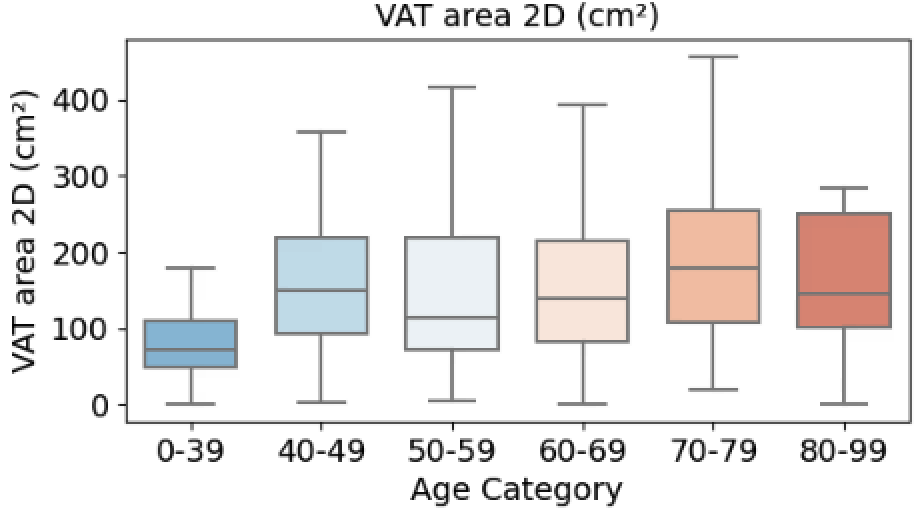}
    \end{subfigure}
    \vspace{1mm}
    \begin{subfigure}[b]{0.36\textwidth}
        \includegraphics[width=\linewidth]{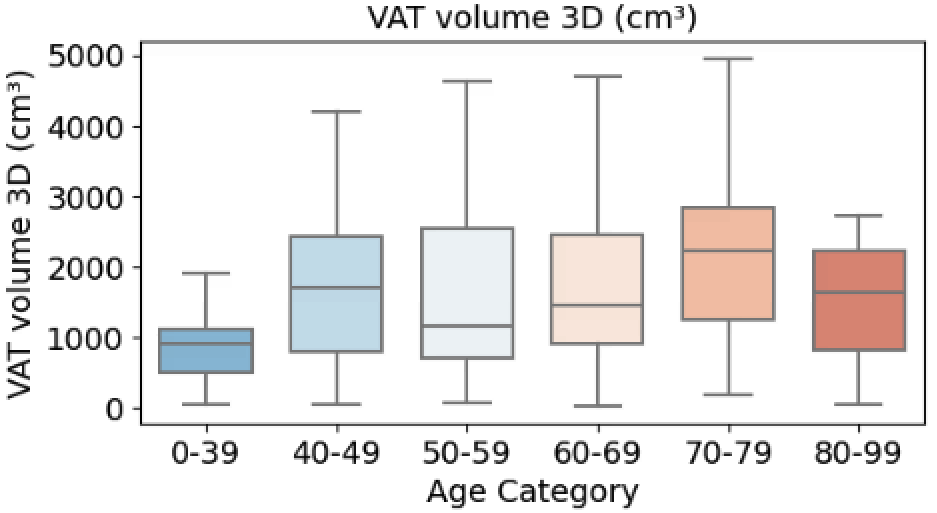}
    \end{subfigure}
    \vspace{1mm}
    \begin{subfigure}[b]{0.36\textwidth}
        \includegraphics[width=\linewidth]{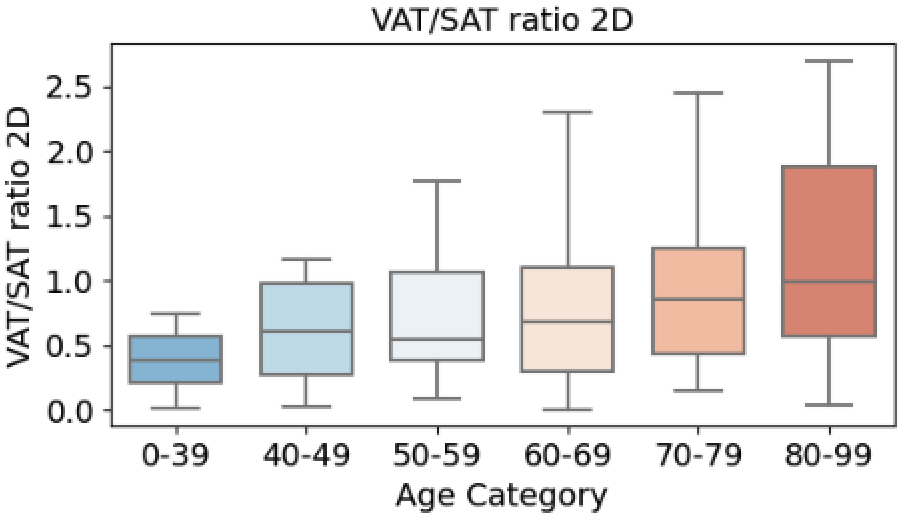}
    \end{subfigure}
    \vspace{1mm}
    \begin{subfigure}[b]{0.36\textwidth}
        \includegraphics[width=\linewidth]{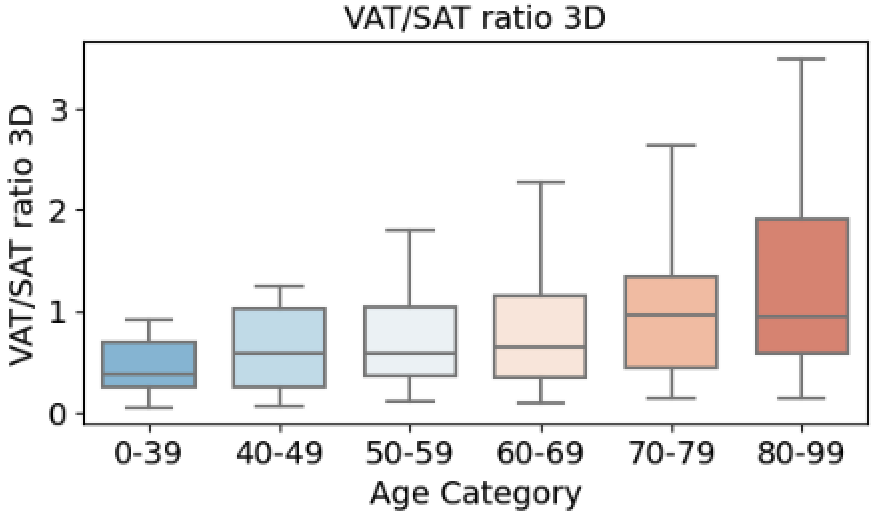}
    \end{subfigure}
    \vspace{1mm}
    \begin{subfigure}[b]{0.36\textwidth}
        \includegraphics[width=\linewidth]{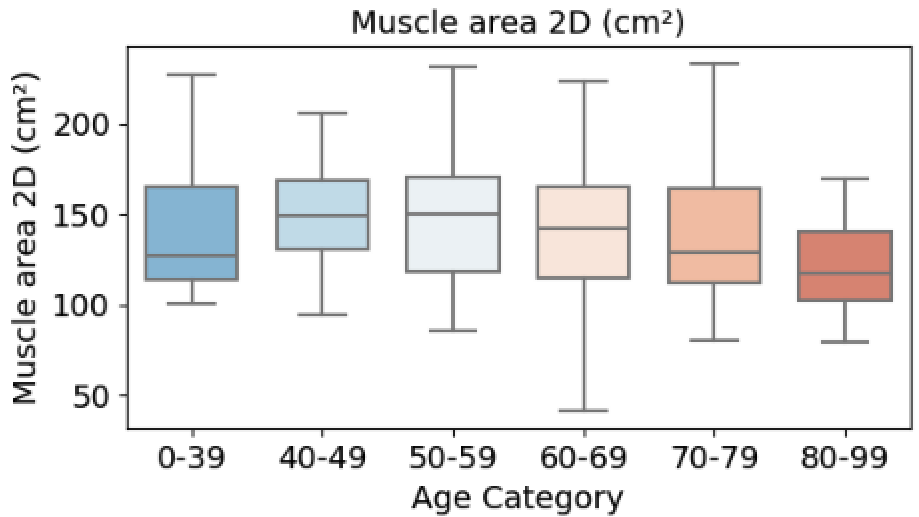}
    \end{subfigure}
    \vspace{1mm}
    \begin{subfigure}[b]{0.36\textwidth}
        \includegraphics[width=\linewidth]{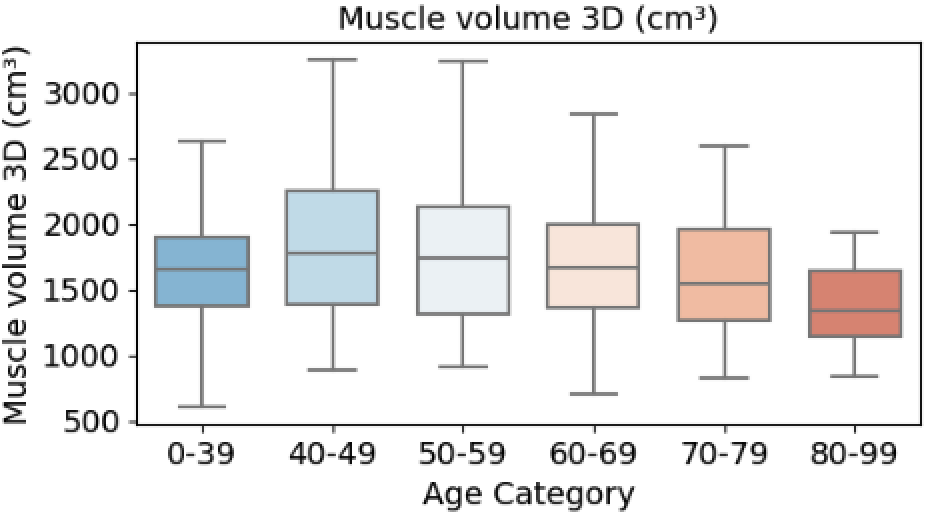}
    \end{subfigure}
    \vspace{1mm}
    \begin{subfigure}[b]{0.36\textwidth}
        \includegraphics[width=\linewidth]{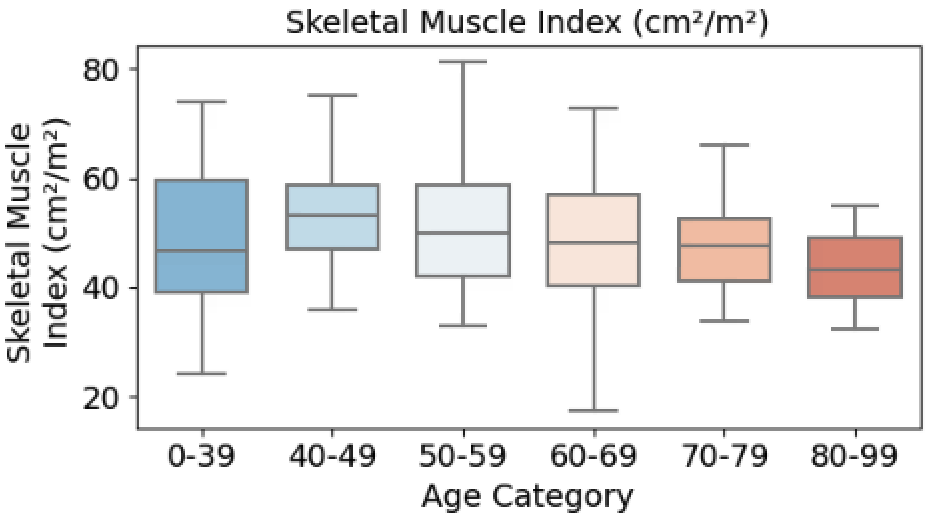}
    \end{subfigure}
    \begin{subfigure}[b]{0.36\textwidth} 
    \end{subfigure}
    \caption{Body composition metrics vs. age categories.}
    \label{fig:body composition vs. age}
\end{figure*}

To ensure a sufficient sample size for analysis, we divided the age range into six distinct groups, each containing at least 20 instances with the results shown in Fig. \ref{fig:body composition vs. age}. The observed trends in muscle area, SAT, VAT area, and SMI with increasing age closely align with findings from previous studies \citep{magudia2021population}, with both the trends and absolute measurement values showing strong consistency across age groups. Specifically, muscle density decreases with age, while VAT and the VAT/SAT ratio increase.

\subsection{Body composition metrics vs. sex}
\begin{figure*}[htbp]
    \centering
    \begin{subfigure}[b]{0.36\textwidth}
        \includegraphics[width=\linewidth]{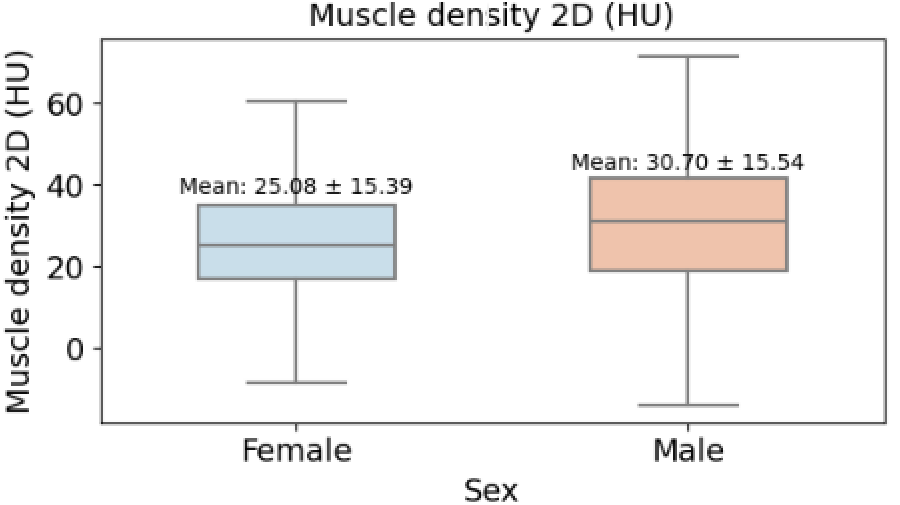}
    \end{subfigure}
    \vspace{1mm}
    \begin{subfigure}[b]{0.36\textwidth}
        \includegraphics[width=\linewidth]{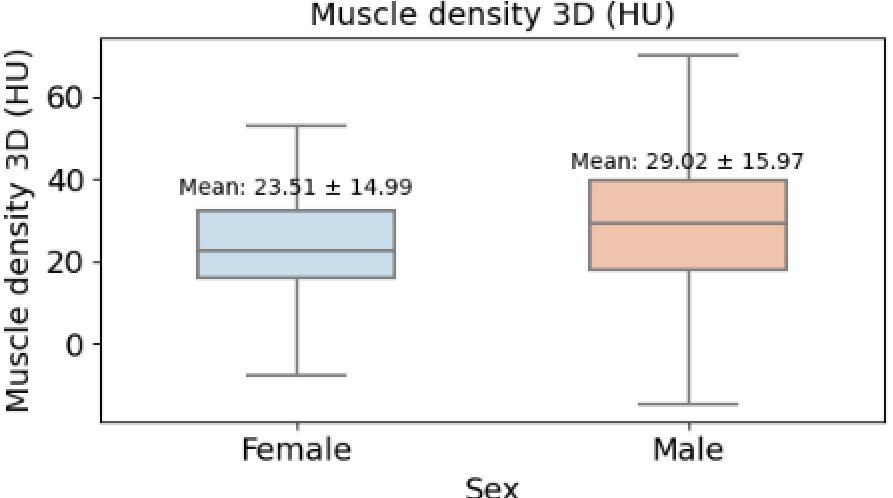}
    \end{subfigure}
    \vspace{1mm}
    \begin{subfigure}[b]{0.36\textwidth}
        \includegraphics[width=\linewidth]{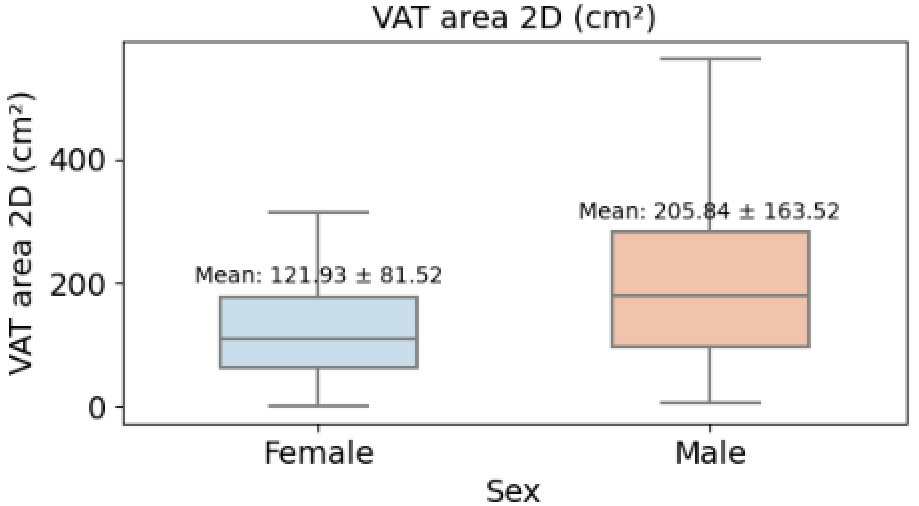}
    \end{subfigure}
    \begin{subfigure}[b]{0.36\textwidth}
        \includegraphics[width=\linewidth]{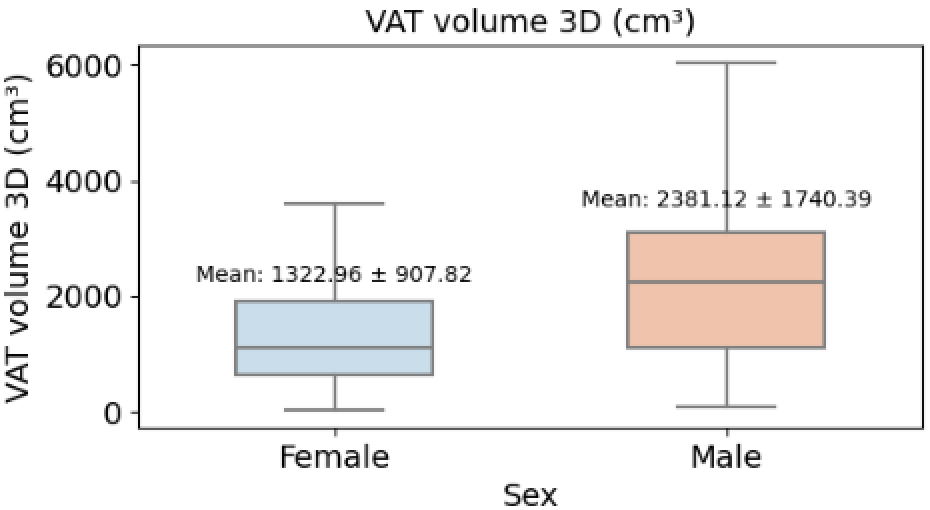}
    \end{subfigure}
    \vspace{1mm}
    \begin{subfigure}[b]{0.36\textwidth}
        \includegraphics[width=\linewidth]{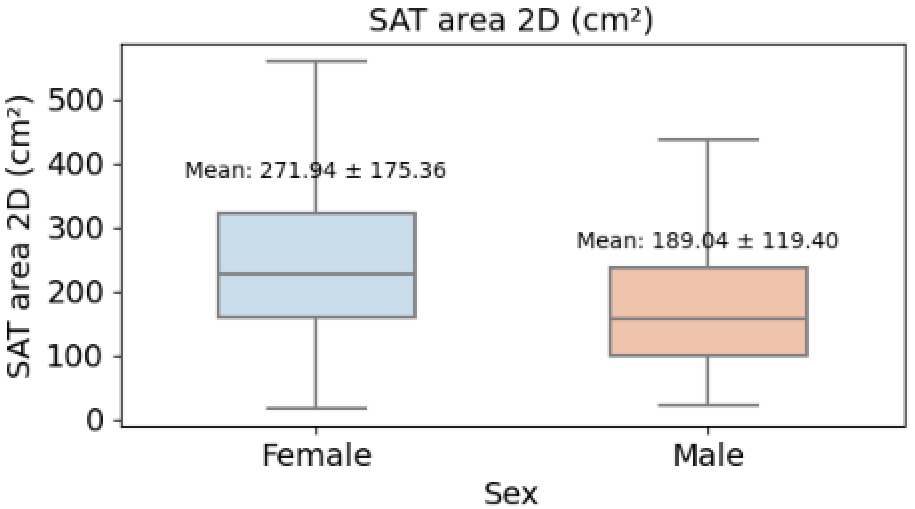}
    \end{subfigure}
    \begin{subfigure}[b]{0.36\textwidth}
        \includegraphics[width=\linewidth]{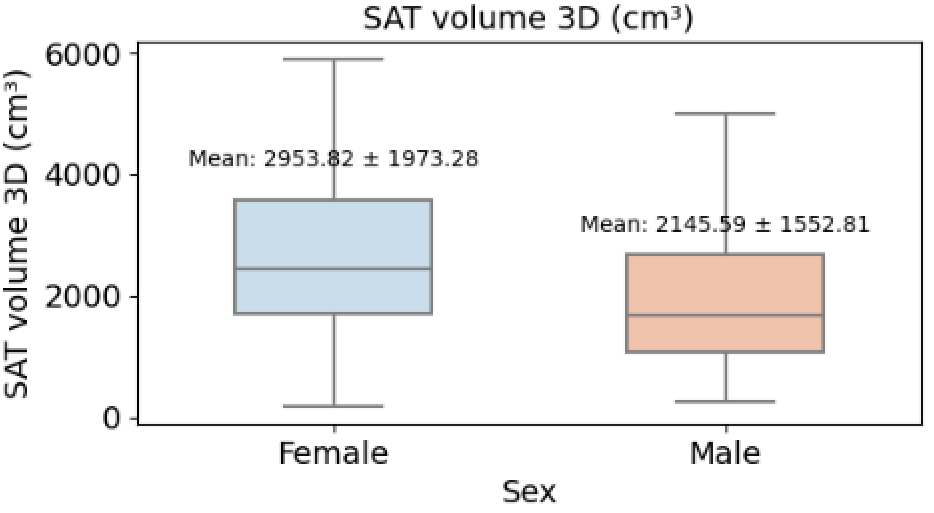}
    \end{subfigure}
    \vspace{1mm}
    \begin{subfigure}[b]{0.36\textwidth}
        \includegraphics[width=\linewidth]{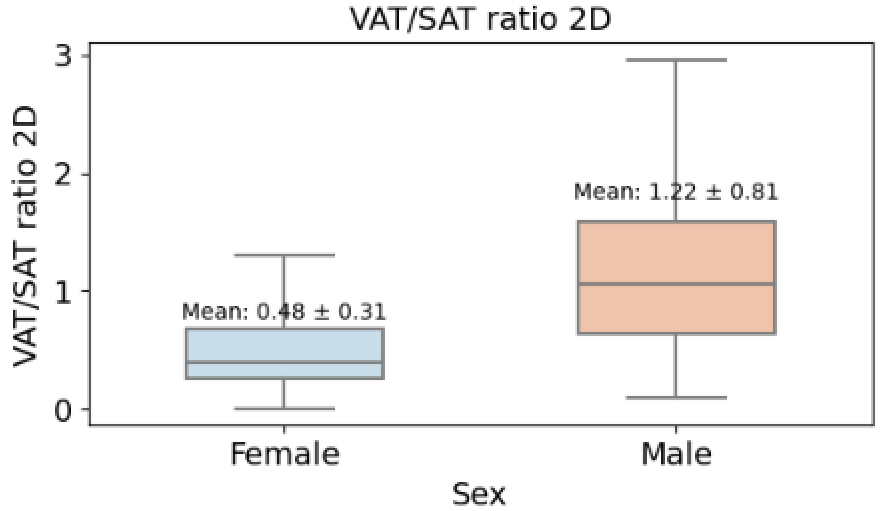}
    \end{subfigure}
    \begin{subfigure}[b]{0.36\textwidth}
        \includegraphics[width=\linewidth]{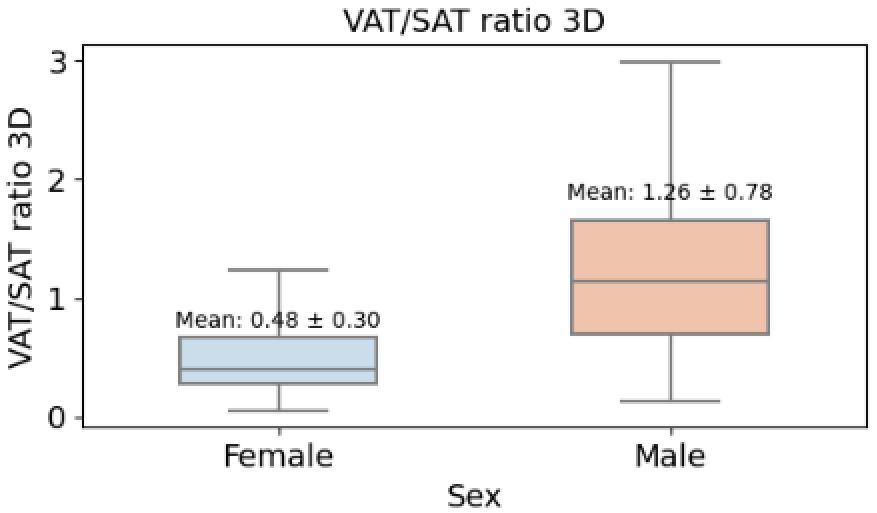}
    \end{subfigure}
    \vspace{1mm}
    \begin{subfigure}[b]{0.36\textwidth}
        \includegraphics[width=\linewidth]{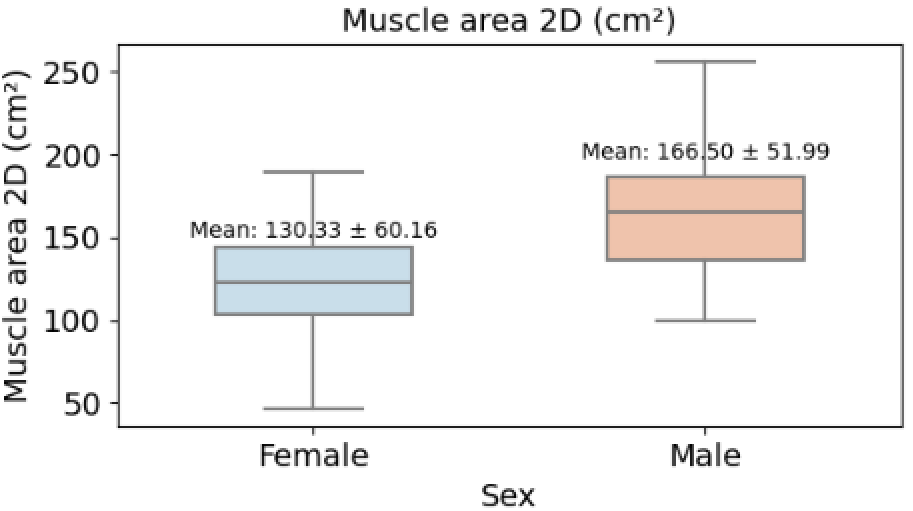}
    \end{subfigure}
    \begin{subfigure}[b]{0.36\textwidth}
        \includegraphics[width=\linewidth]{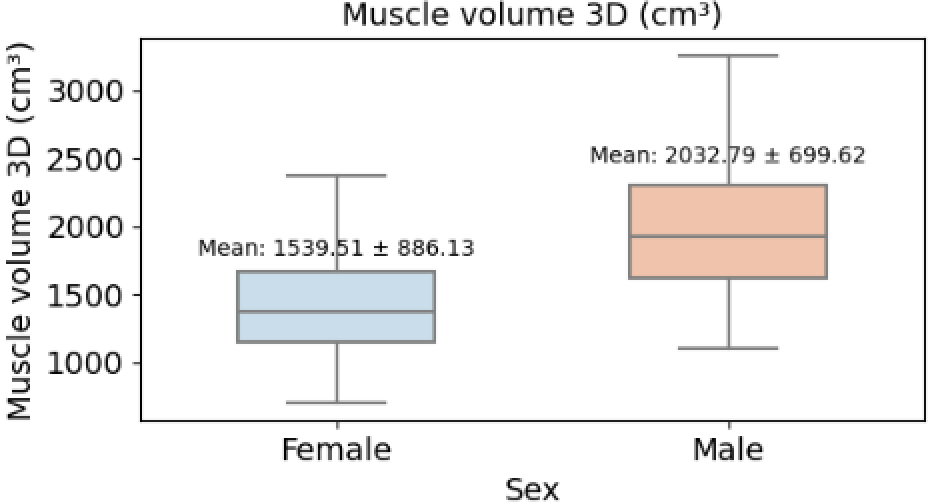}
    \end{subfigure}
    \vspace{1mm}
    \begin{subfigure}[b]{0.36\textwidth}
        \includegraphics[width=\linewidth]{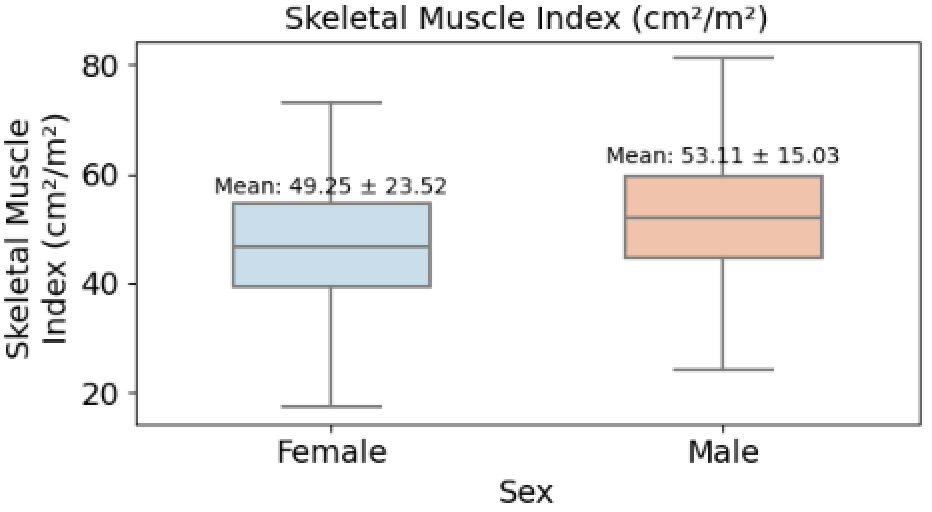}
    \end{subfigure}
    \begin{subfigure}[b]{0.36\textwidth} 
    \end{subfigure}
    \caption{Body composition metrics vs. sex.}
    \label{fig:body composition vs. sex}
\end{figure*}

The left column sub-figures in Fig. \ref{fig:body composition vs. sex} illustrate muscle density, SAT/VAT ratio, muscle area, and SMI versus gender, respectively, as measured at the L3 level. Similarly, the corresponding measurements for muscle density, SAT/VAT ratio, muscle volume as measured from T12 to L4, are shown in the right column sub-figures. The mean and standard deviation of these body composition metrics are consistent with those reported in previous studies \citep{van2018percentiles, graffy2019deep}. Our measurement also aligns with the previous findings that compared to female, male typically have a higher muscle density, VAT/SAT ratio, muscle area, SMI \citep{kammerlander2021sex, van2018percentiles, graffy2019deep}.

\subsection{Body composition metrics vs. race group}
\begin{figure*}[htbp]
    \centering
    \begin{subfigure}[b]{0.36\textwidth}
        \includegraphics[width=\linewidth]{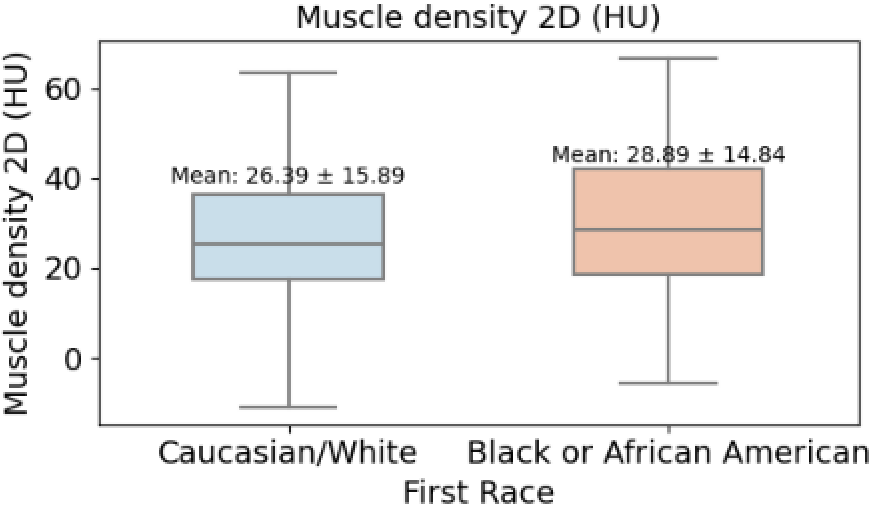}
    \end{subfigure}
    \vspace{1mm}
    \begin{subfigure}[b]{0.36\textwidth}
        \includegraphics[width=\linewidth]{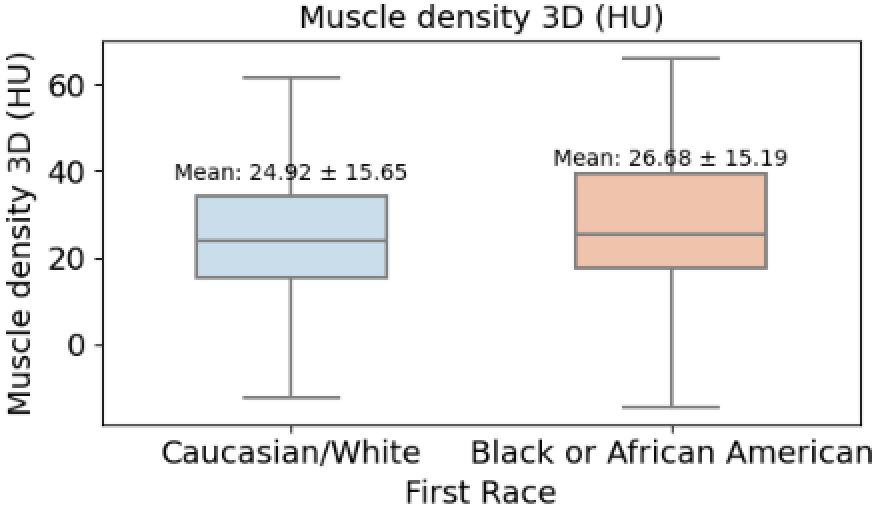}
    \end{subfigure}
    \vspace{1mm}
    \begin{subfigure}[b]{0.36\textwidth}
        \includegraphics[width=\linewidth]{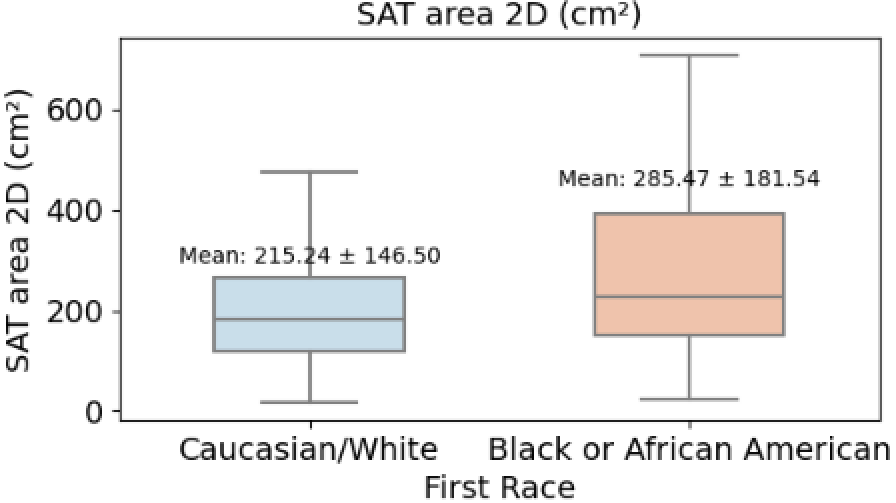}
    \end{subfigure}
    \begin{subfigure}[b]{0.36\textwidth}
        \includegraphics[width=\linewidth]{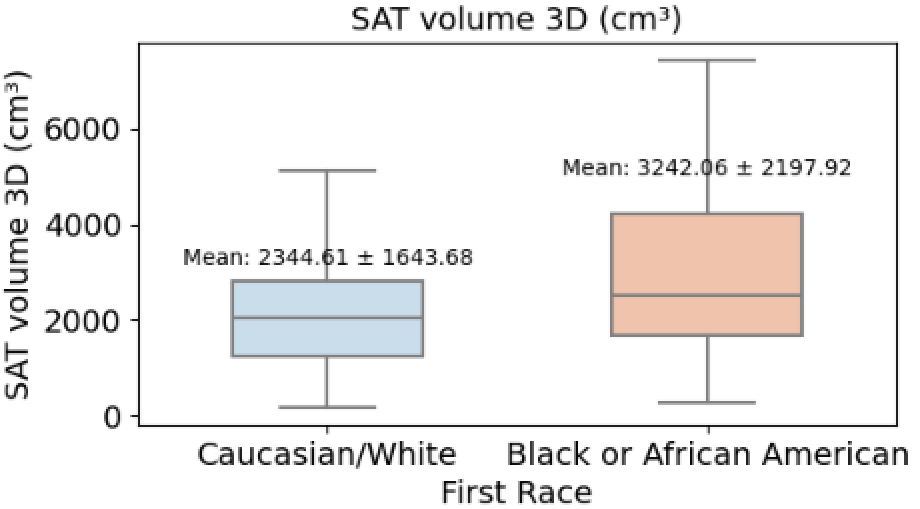}
    \end{subfigure}
    \vspace{1mm}
    \begin{subfigure}[b]{0.36\textwidth}
        \includegraphics[width=\linewidth]{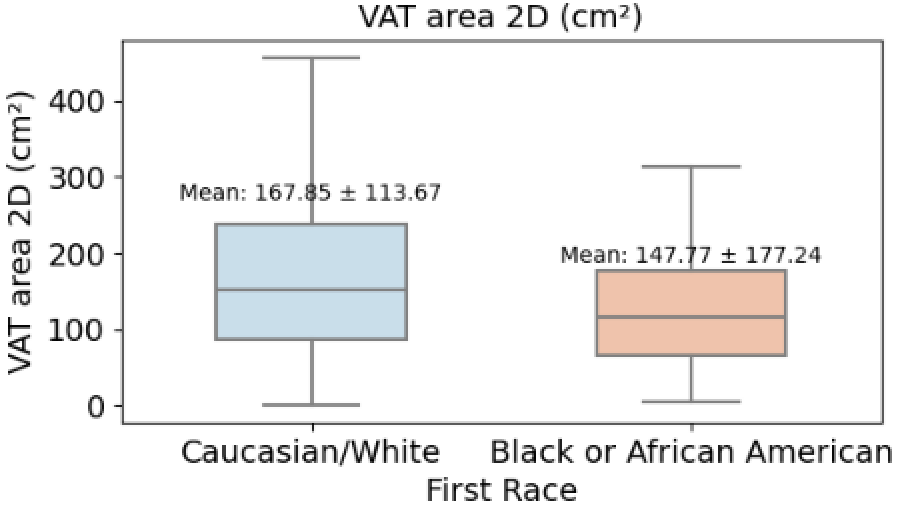}
    \end{subfigure}
    \begin{subfigure}[b]{0.36\textwidth}
        \includegraphics[width=\linewidth]{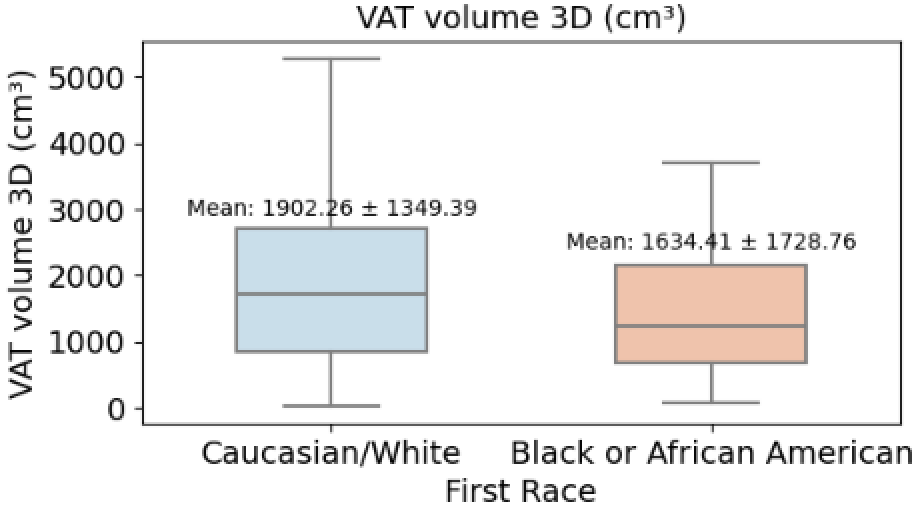}
    \end{subfigure}
    \vspace{1mm}
    \begin{subfigure}[b]{0.36\textwidth}
        \includegraphics[width=\linewidth]{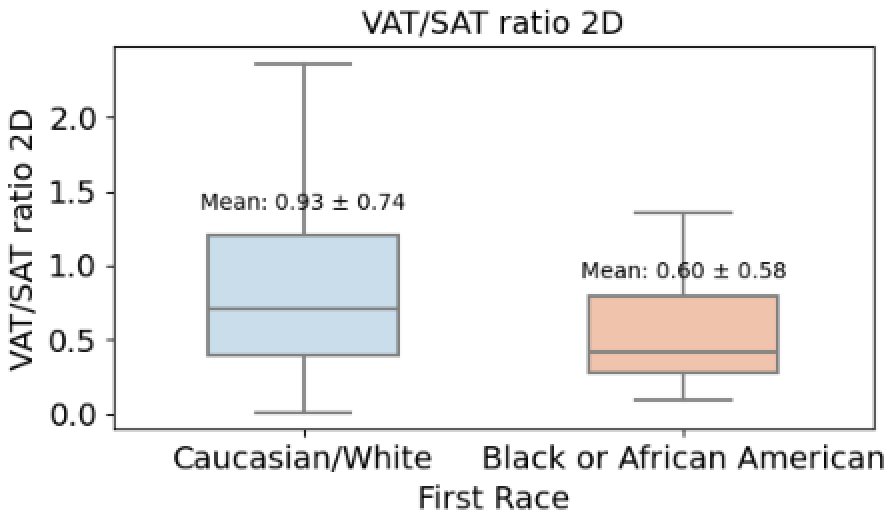}
    \end{subfigure}
    \begin{subfigure}[b]{0.36\textwidth}
        \includegraphics[width=\linewidth]{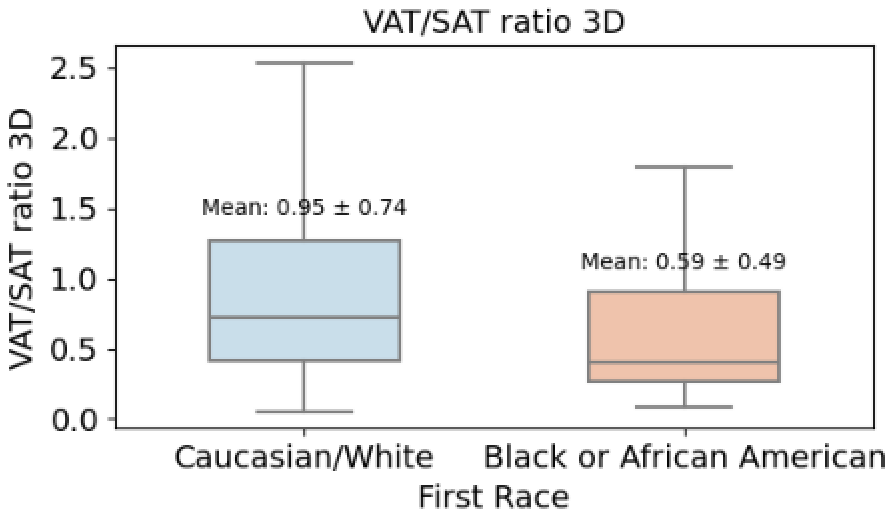}
    \end{subfigure}
    \vspace{1mm}
    \begin{subfigure}[b]{0.36\textwidth}
        \includegraphics[width=\linewidth]{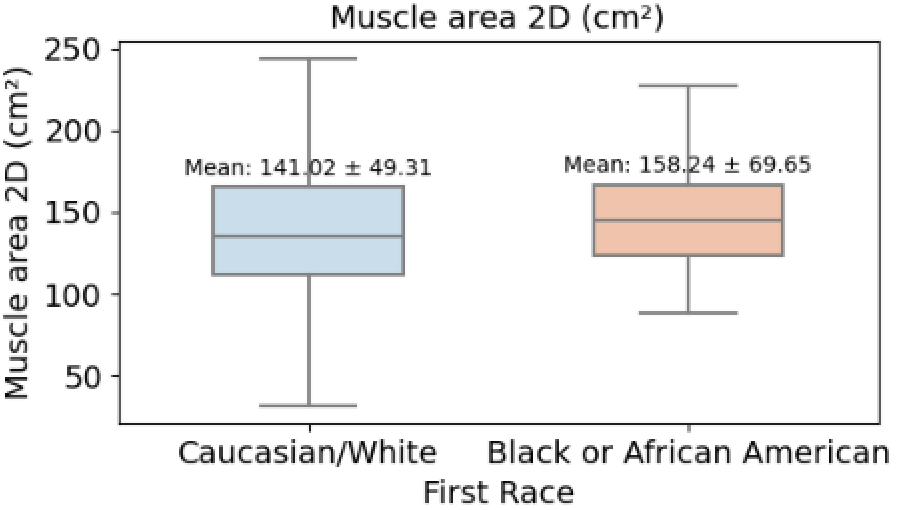}
    \end{subfigure}
    \begin{subfigure}[b]{0.36\textwidth}
        \includegraphics[width=\linewidth]{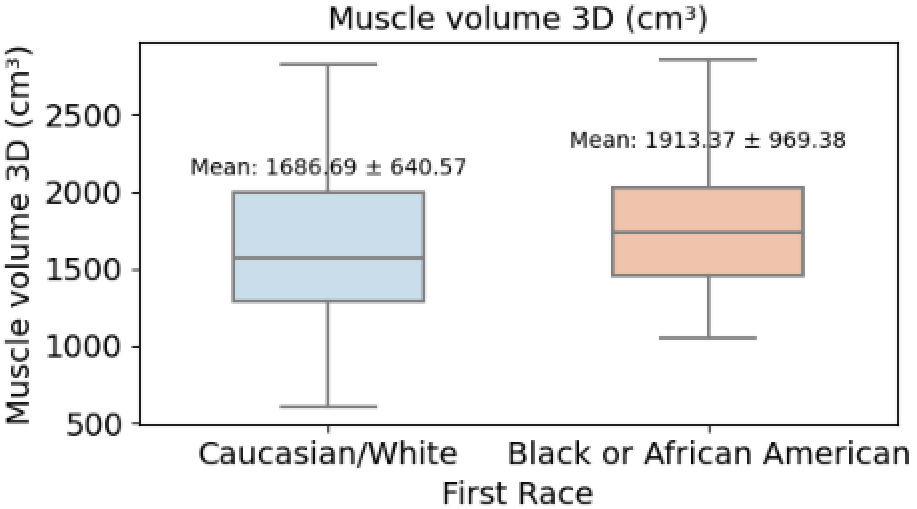}
    \end{subfigure}
    \vspace{1mm}
    \begin{subfigure}[b]{0.36\textwidth}
        \includegraphics[width=\linewidth]{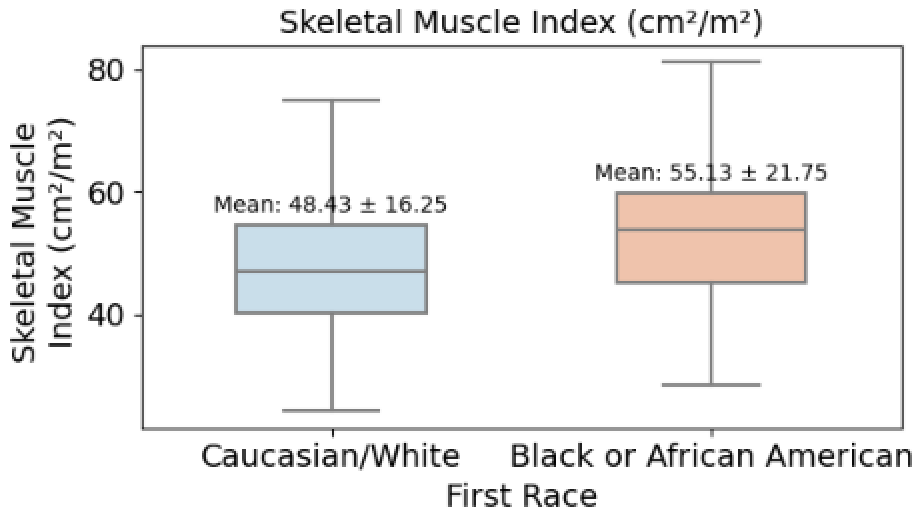}
    \end{subfigure}
    \begin{subfigure}[b]{0.36\textwidth} 
    \end{subfigure}
    \caption{Body composition metrics vs. race.}
    \label{fig:body composition vs. race}
\end{figure*}

For the race groups, to ensure sufficient data size for analysis, we conducted the analysis based only on two race groups: Caucasian/White and Black or African American. The relationship between race groups and multiple body composition metrics is demonstrated in Fig. \ref{fig:body composition vs. race}. Few previous studies have exclusively analyzed body composition across different races, limiting our ability for direct comparisons. However, several studies have examined the combined impact of both sex and race. For example, \citep{magudia2021population} demonstrates that Black or African American individuals, on average, have larger muscle areas and higher SMI for both males and females. Similarly, \citep{beasley2009body} reports an average abdominal visceral fat area of 152.0 for White individuals and 129.9 for Black individuals, as well as an average abdominal subcutaneous fat area of 266.0 for White individuals and 312.1 for Black individuals. Although our study and \citep{beasley2009body} have different population distributions, with the latter being limited to healthy elderly adults, the measurement differences between the two studies for all metrics are within 10\%.

\section{Discussion and future work}
To mitigate the gap in publicly available deep learning-based CT segmentation and body composition measurement models for abdominal muscle and fat, we evaluated nine publicly available segmentation network architectures and selected nnU-Net ResEnc XL to build our model. This model is able to segment skeletal muscle, subcutaneous adipose tissue (SAT), and visceral adipose tissue (VAT) across the chest, abdomen, and pelvis in axial CT images. It additionally automatically measures muscle density, visceral-to-subcutaneous fat (VAT/SAT) ratio, muscle area/volume, and SMI. All the code will be made publicly available at https://github.com/mazurowski-lab/CT-Muscle-and-Fat-Segmentation.git.

This study highlights the strong capability of our model in segmenting skeletal muscle, SAT, and VAT across the chest, abdomen, and pelvis in axial CT images. As detailed in Sec. \ref{sec:Internal dataset evaluation}, the model achieves an average Dice score of \textbf{91.79 ± 4.90\%} across all slices and all four labels in the internal dataset. In the external dataset, the average Dice score is \textbf{89.98 ± 3.79\%}. The model also demonstrates high agreement with manual measurements, with $R^2$ values exceeding \textbf{0.99} for both skeletal muscle and SAT.

The model demonstrates even better performance when segmenting the L3 slice and the T12–L4 region. For the L3 slice, it achieves an average Dice score of 93.38 ± 4.92\% in the internal dataset and 92.55 ± 3.45\%  in the external dataset. For the T12–L4 region, the average Dice score is 92.75 ± 5.84\% on internal dataset and 91.81 ± 4.06\% on external dataset. The model also achieves similarly high agreement in body composition measurements at L3 and T12–L4, with $R^2$ values consistently exceeding 0.95 for skeletal muscle and 0.99 for SAT.

When compared to previous methods, our model shows significant improvements, outperforming the recently published in-house segmentation model \citep{hou2024enhanced} by 2.10\% for skeletal muscle and 8.6\% for SAT. Additionally, it surpasses the TotalSegmentator \citep{wasserthal2023totalsegmentator} by 6.5\% for skeletal muscle and 11.6\% for SAT. These evaluations are based on manual annotations from the publicly available SAROS dataset. A detailed comparison with benchmark models is provided in Sec. \ref{sec:External dataset evaluation}.

Apart from segmentation performance, our model also demonstrates high accuracy in measuring commonly used body composition metrics, including muscle density, visceral-to-subcutaneous fat ratio, muscle area/volume, and SMI in both 2D and 3D settings. The average MRAE for all metrics is below \textbf{10\%}. As detailed in Sec. \ref{sec:Analysis metric evaluation}, the model achieves its best performance in measuring muscle density, with an MRAE of less than \textbf{5\%} compared to manual annotations across internal and external datasets in both 2D and 3D settings.

Furthermore, utilizing our model, we performed body composition metrics analysis across different age, sex, and race groups on 371 randomly selected patients from Duke Hospital. The results demonstrate clear differences in muscle density, adipose tissue distribution, and SMI among patients of different ages, sexes, and races. \textbf{With increasing age}, there was a noticeable decline in muscle density and SMI, coupled with an increase in visceral adipose tissue (VAT) and the VAT/SAT ratio, indicating age-related muscle loss and fat redistribution. \textbf{Sex-based comparisons} revealed that males generally had higher muscle density, muscle volume, and SMI, while females exhibited higher subcutaneous adipose tissue (SAT) levels. Additionally, \textbf{race-based analysis} showed that Black or African American individuals had higher muscle mass and SAT but lower VAT levels and VAT/SAT ratios compared to Caucasian/White individuals. All the findings also align with results from previous studies, highlighting the robustness of our model for both segmentation and body composition measurement.

Despite the promising results, this study has several limitations that present opportunities for further development and improvement. First, the scope of the study is restricted to three body regions and relies solely on axial views. To enhance the generalizability and clinical utility of our approach, future work will focus on expanding the analysis to additional body regions, such as the hip, leg, and shoulder, which are also commonly assessed in body composition studies.

While our model demonstrates reasonable performance on sagittal and coronal views by stacking segmented axial slices and extracting intersections across different planes, as illustrated in Fig. \ref{fig2:3D_qualitative_evaluation}, this approach has inherent limitations. For example, since axial slices are segmented independently, which may lead to inconsistencies between adjacent slices. We recognize the potential benefits of directly incorporating sagittal and coronal views into the training and evaluation pipeline, which may improve segmentation accuracy and consistency across all anatomical planes.

Secondly, although the segmentation model includes a muscular fat label, its performance is comparatively lower than that of the other three labels. This discrepancy is primarily due to variability in annotation granularity among different annotators. To enhance annotation consistency in future versions, we will establish clear annotation standards. Specifically, we will define muscular fat regions by applying a Hounsfield Unit (HU) threshold between -220 and -50 for fat tissue, as suggested by Chougule et al. \citep{chougule2018clinical}, and retain only contiguous fat regions comprising more than six pixels.


\acks{This work was supported through a partnership between the Duke Departments of Surgery and Radiology and the Duke Spark Initiative for AI in Medical Imaging.}

%
\ethics{The research protocol was approved by the Duke Health System Institutional Review Board (IRB) with ethical standards for research and manuscript preparation, adhering to all relevant laws and regulations concerning the treatment of human subjects and animals.}

\coi{We declare we don't have conflicts of interest.}

\data{The external dataset utilized for model evaluation and analysis in this study is publicly accessible \citep{koitka2023saros, clark2013cancer}. However, the internal dataset is currently unavailable, as de-identifying the data requires an extensive institutional review process. Readers interested in evaluating the accuracy of the method can access and utilize the publicly available datasets, which are readily accessible and user-friendly. The code for this study is also publicly available at https://github.com/mazurowski-lab/CT-Muscle-and-Fat-Segmentation.git.}

\bibliography{refs}


\clearpage
\appendix
\section{CT parameters}\label{appendix:CT parameters}
Table \ref{tab:ct_parameters_summary} summarizes the detailed acquisition parameters of the CT scans used in the internal dataset.

\begin{table*}[ht]
\centering
\begin{tabular}{l|c|c|c}
\hline
\textbf{CT Parameter} & \textbf{Internal Training} & \textbf{Internal Test} & \textbf{Demographic Analysis} \\
\hline\midrule
Tube Voltage (kVp) & 113.15 $\pm$ 13.43 & 114.14 $\pm$ 9.66 & 114.82 $\pm$ 11.75 \\\midrule
Tube Current (mA) & 413.87 ± 250.51 & 313.97 ± 163.45 &  413.83 $\pm$ 258.10 \\\midrule
Exposure (mAs) & 193.02 ± 139.50 & 51.62 ± 63.23 & 104.22 $\pm$ 130.56 \\\midrule
Slice Thickness (mm) & 1.57 $\pm$ 1.32 & 2.19 $\pm$ 1.08 & 2.07 $\pm$ 1.71 \\\midrule
Manufacturer & GE, SIEMENS & GE, SIEMENS & GE, SIEMENS \\\midrule
\hline
\end{tabular}
\caption{Summary of CT Acquisition Parameters Across Datasets}
\label{tab:ct_parameters_summary}
\end{table*}

\section{Data collections from SAROS} \label{sec:Data collections from SAROS}
While only a subset of the collections within SAROS is provided with a commercially permitted license (specifically CC BY 3.0 and CC BY 4.0), we exclusively utilized this subset in external evaluation to ensure maximum flexibility for our model. In this section, we provide a detailed list of the dataset collections used in this study, shown in Table \ref{tab:SAROS_data_collection}, including the collection name, scan region (Abdomen, Thorax, Whole-body) assigned by SAROS \citep{koitka2023saros, clark2013cancer}, and their license type.

\begin{table*}[t]
\centering
\resizebox{\textwidth}{!}{%
\begin{tabular}{c|c|c|c|c|c}
\toprule
Collection & Number of studies & Abdomen & Thorax & Whole-body & License \\
\toprule
ACRIN-FLT-Breast$^{1,2}$ & 32 & 0 & 0 & 32 & CC BY 3.0 \\
\midrule
ACRIN-NSCLC-FDG-PET$^{3,4}$ & 129 & 0 & 78 & 51 & CC BY 3.0 \\
\midrule
Anti-PD-1\_Lung$^{5}$ & 12 & 0 & 0 & 12 & CC BY 3.0 \\
\midrule
C4KC-KiTS$^{6,7}$ & 175 & 175 & 0 & 0 & CC BY 3.0 \\
\midrule
CPTAC-CM$^{8}$ & 1 & 0 & 0 & 1 & CC BY 3.0 \\
\midrule
CPTAC-LSCC$^{9}$ & 3 & 0 & 0 & 3 & CC BY 3.0 \\
\midrule
CPTAC-LUAD$^{10}$ & 1 & 0 & 0 & 1 & CC BY 3.0 \\
\midrule
CPTAC-PDA$^{11}$ & 8 & 0 & 0 & 8 & CC BY 3.0 \\
\midrule
CPTAC-UCEC$^{12}$ & 26 & 25 & 0 & 1 & CC BY 3.0 \\
\midrule
LIDC-IDRI$^{13,14}$ & 133 & 0 & 133 & 0 & CC BY 3.0 \\
\midrule
NSCLC Radiogenomics$^{15,16,17,18}$ & 7 & 0 & 0 & 7 & CC BY 3.0 \\
\midrule
Pancreas CT$^{19,20}$ & 58 & 58 & 0 & 0 & CC BY 3.0 \\
\midrule
Soft-tissue-Sarcoma$^{21,22}$ & 6 & 0 & 0 & 6 & CC BY 3.0 \\
\midrule
TCGA-LIHC$^{23}$ & 33 & 32 & 0 & 1 & CC BY 3.0 \\
\midrule
TCGA-LUAD$^{24}$ & 2 & 0 & 0 & 2 & CC BY 3.0 \\
\midrule
TCGA-LUSC$^{25}$ & 3 & 0 & 0 & 3 & CC BY 3.0 \\
\midrule
TCGA-STAD$^{26}$ & 2 & 2 & 0 & 0 & CC BY 3.0 \\
\midrule
TCGA-UCEC$^{27}$ & 1 & 0 & 0 & 1 & CC BY 3.0 \\
\midrule
COVID-19-NY-SBU$^{28}$ & 1 & 0 & 0 & 1 & CC BY 4.0 \\
\midrule
Lung-PET-CT-Dx$^{29}$ & 17 & 0 & 15 & 2 & CC BY 4.0 \\
\bottomrule
In total & 650 & 292 & 226 & 132 & - \\
\bottomrule
\end{tabular}
}
\caption{\textbf{Dataset collections from SAROS. References:} 
$^{1}$\citep{kostakoglu2015phase}, $^{2}$\citep{Kinahan2017}, 
$^{3}$\citep{machtay2013prediction}, $^{4}$\citep{Kinahan2019}, 
$^{5}$\citep{Madhavi2019}, $^{6}$\citep{heller2021state}, $^{7}$\citep{Heller2019}, 
$^{8}$\citep{CPTAC2018}, $^{9}$\citep{CPTAC2018LSCC}, $^{10}$\citep{CPTAC2018LUAD}, 
$^{11}$\citep{CPTAC2018_v2}, $^{12}$\citep{CPTAC2019UCEC}, 
$^{13}$\citep{armato2011lung}, $^{14}$\citep{Armato2015}, 
$^{15}$\citep{Napel2014}, $^{16}$\citep{Bakr2017}, $^{17}$\citep{bakr2018radiogenomic}, $^{18}$\citep{gevaert2012non}, 
$^{19}$\citep{Roth2016}, $^{20}$\citep{roth2015deeporgan}, 
$^{21}$\citep{vallieres2015radiomics}, $^{22}$\citep{Vallieres2015}, 
$^{23}$\citep{Erickson2016}, $^{24}$\citep{Albertina2016}, 
$^{25}$\citep{Kirk2016}, $^{26}$\citep{Lucchesi2016}, 
$^{27}$\citep{Erickson2016UCEC}, $^{28}$\citep{Saltz2021}, $^{29}$\citep{Li2020}.}
\label{tab:SAROS_data_collection}
\end{table*}

\section{Summary of body composition metrics}\label{sec:Summary of Body Composition Metrics}
Table \ref{tab:body_composition_summary} summarizes detailed information for four body composition metrics used in this study: muscle density, VAT/SAT ratio, muscle area/volume, and SMI. The table includes their calculation methods, clinical significance, and associations with diseases, and is intended to help readers quickly understand the relevance and interpretation of each metric.

\begin{table*}[h]
\centering
\begin{tabular}{|p{3.5cm}|p{4.5cm}|p{3.5cm}|p{3.5cm}|}
\hline
\textbf{Metric} & \textbf{Calculation Method} & \textbf{Clinical Significance} & \textbf{Associated Diseases} \\
\hline
Muscle Density & Mean Hounsfield Units (HU) of segmented skeletal muscle on CT & Indicator of muscle quality and intramuscular fat infiltration & Sarcopenia, frailty, cancer prognosis \\
\hline
VAT/SAT Ratio & Volume of visceral adipose tissue (VAT) divided by volume of subcutaneous adipose tissue (SAT) & Reflects abdominal fat distribution, metabolic risk & Type 2 diabetes, cardiovascular disease, metabolic syndrome \\
\hline
Muscle Area/Volume & Cross-sectional area or total volume of skeletal muscle from CT slices or 3D volume & Estimates muscle mass, useful for nutritional and physical health assessment & Sarcopenia, cachexia, post-operative complications \\
\hline
Skeletal Muscle Index (SMI) & Skeletal muscle area normalized by height squared (cm\textsuperscript{2}/m\textsuperscript{2}) & Standardized metric for diagnosing sarcopenia & Cancer-related cachexia, decreased treatment tolerance \\
\hline
\end{tabular}
\caption{\textbf{Summary of the four body composition metrics:}-muscle density, VAT/SAT ratio, muscle area/volume, and skeletal muscle index (SMI)—including their calculation methods, clinical significance, and known associations with disease.}
\label{tab:body_composition_summary}
\end{table*}

\section{Vertebrae detection qualitative evaluation}\label{appendix:vertebra}
\begin{figure*}[ht]
    \centering
    \includegraphics[width=\textwidth]{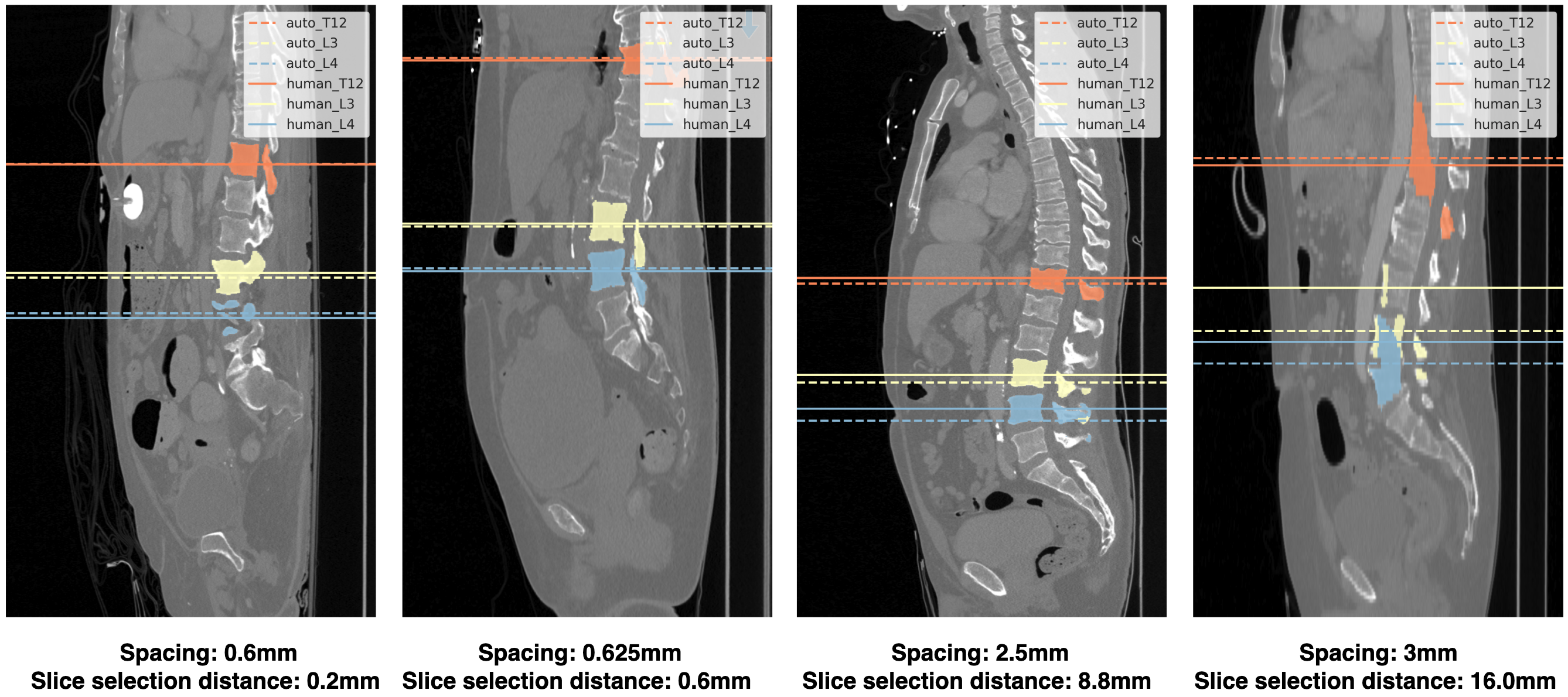}  
    \caption{\textbf{Qualitative evaluation for vertebrae detection given by TotalSegmentator:} Randomly selected four volumes in the sagittal view, with the automatically segmented vertebrae, the automatically selected slices (dashed lines), the manually selected slices (solid lines), and the slice distance between the automated and manual selections.}
    \label{fig:slice_selection_analysis_qual}
\end{figure*}

This section demonstrates the qualitatively assessment of TotalSegmentator's vertebrae detection ability. We randomly selected four representative cases and visualized them in the sagittal view, as shown in Fig. \ref{fig:slice_selection_analysis_qual}. Each example displays the automatically segmented vertebrae and compares the slice locations selected by the automated method (dashed lines) with those selected manually (solid lines) for T12, L3, and L4. The CT scans vary in slice spacing, ranging from 0.6$mm$ to 3.0$mm$.

\section{Comprehensive qualitative visualizations of segmentation performance} \label{appendix:comprehensive_vis}
To provide a more in-depth understanding of our model’s performance across diverse cases, we include here the comprehensive qualitative visualization for both 2D and 3D segmentation result.

\begin{figure*}[t]
    \centering
    \begin{subfigure}[b]{0.9\textwidth}
        \centering
        \includegraphics[width=\linewidth]{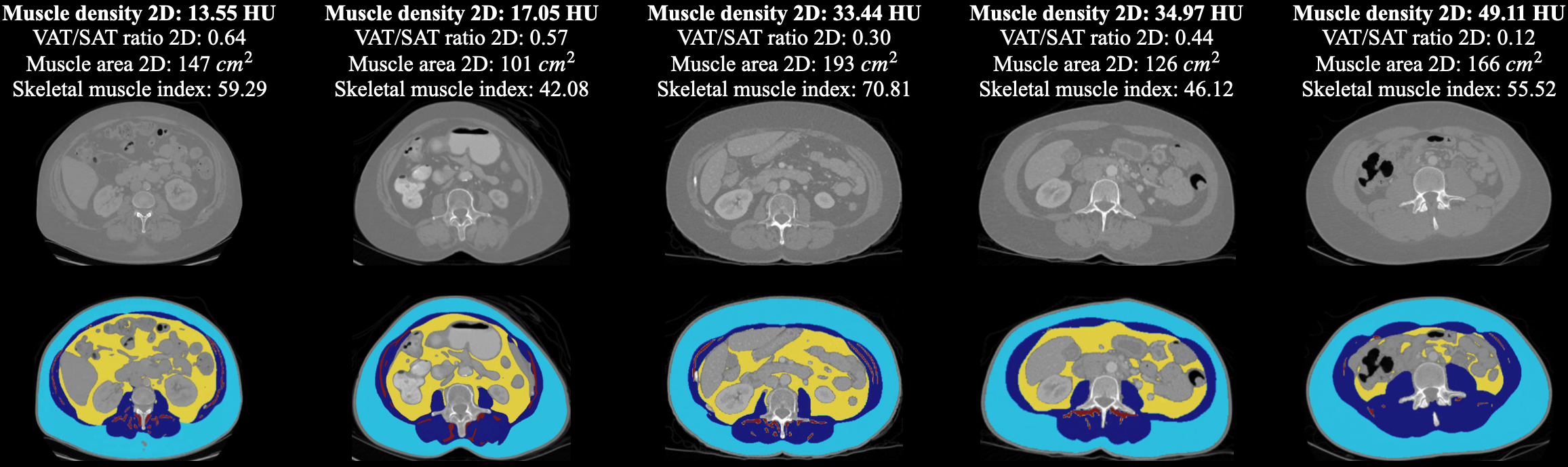}
    \end{subfigure}
     \vspace{0.05cm}
    
    \begin{subfigure}[b]{0.9\textwidth}
        \centering
        \includegraphics[width=\linewidth]{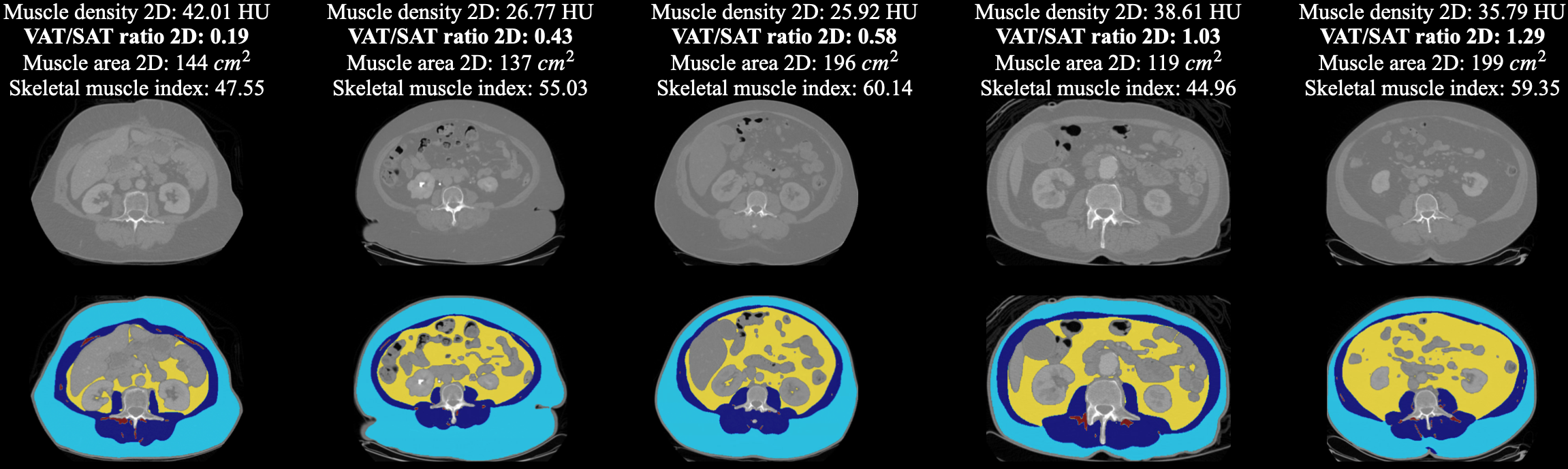}
    \end{subfigure}
     \vspace{0.05cm}
    
    \begin{subfigure}[b]{0.9\textwidth}
        \centering
        \includegraphics[width=\linewidth]{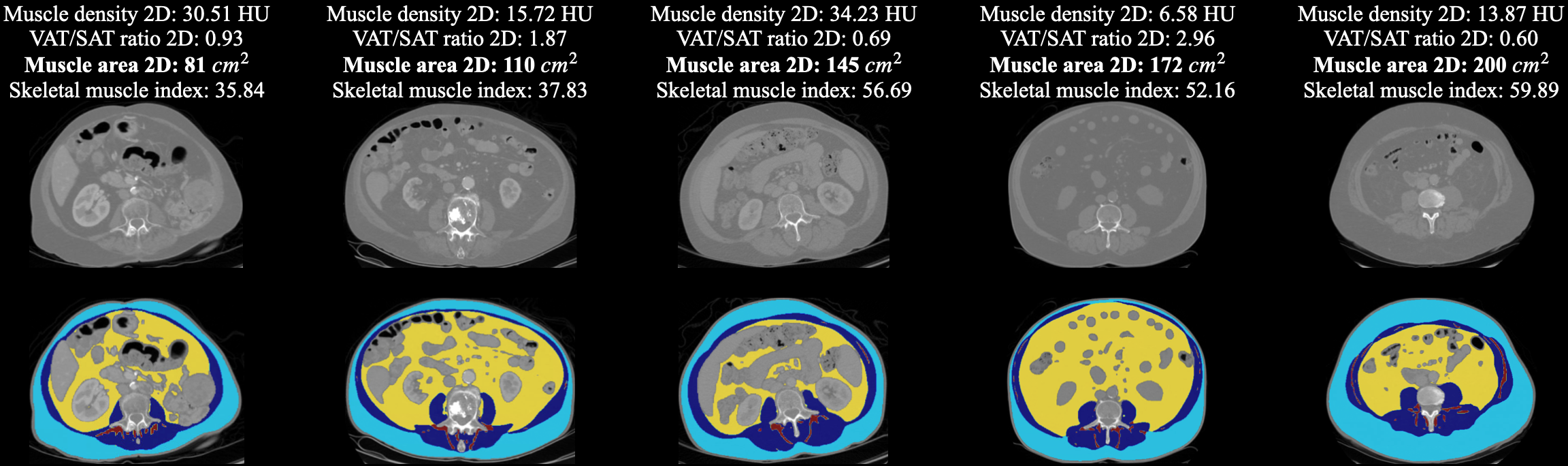}
    \end{subfigure}
     \vspace{0.05cm}
    
    \begin{subfigure}[b]{0.9\textwidth}
        \centering
        \includegraphics[width=\linewidth]{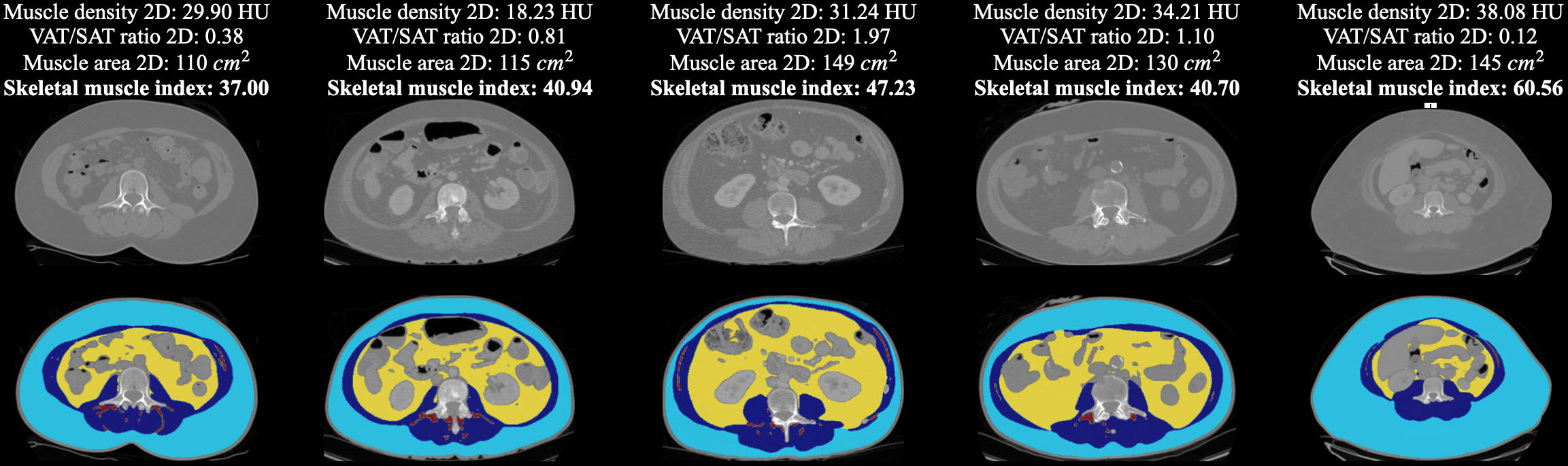}
    \end{subfigure}
    \vspace{0.05cm}
    
    \caption{\textbf{Comprehensive qualitative evaluation of our segmentation model:} Figure shows segmentation results of the abdominal L3 slice. Each row represents a specific body composition metric (in bold), with five patients arranged from left to right in categories: Low, Moderately Low, Moderate, Moderately High, and High. For example, in the first row (muscle density), the first patient has low muscle density, and the fifth has high muscle density. The second, third, and fourth rows show the VAT/SAT ratio, muscle area/volume, and SMI, respectively, following the same left-to-right order. In the segmentation, \textit{dark blue} shows skeletal muscle, \textit{light blue} SAT, \textit{yellow} VAT, and \textit{maroon} muscular fat.}
    \label{fig1:2D_qualitative_evaluation}
\end{figure*}

Fig. \ref{fig1:2D_qualitative_evaluation} presents the L3 segmentation results and their corresponding body composition metrics (muscle density, VAT/SAT ratio, muscle area/volume, and SMI) for selected patients. The samples are categorized based on their body composition metric values into five groups: Low, Moderately Low, Moderate, Moderately High, and High, with cut-off points set at the 20th, 40th, 60th, and 80th percentiles of the entire population. For each body composition metric, one sample is randomly selected from each category for visualization. Each column in Fig. \ref{fig1:2D_qualitative_evaluation} corresponds to a specific category, with patients arranged from low to high values across the columns, ensuring a consistent representation of the metric's progression. Each row, in turn, highlights a specific body composition metric. The first row illustrates muscle density, the second depicts the VAT/SAT ratio, the third represents muscle area/volume, and the fourth corresponds to the SMI. 

\begin{figure*}[ht]
    \centering
    \begin{subfigure}[b]{0.65\textwidth}
        \centering
        \includegraphics[width=\textwidth]{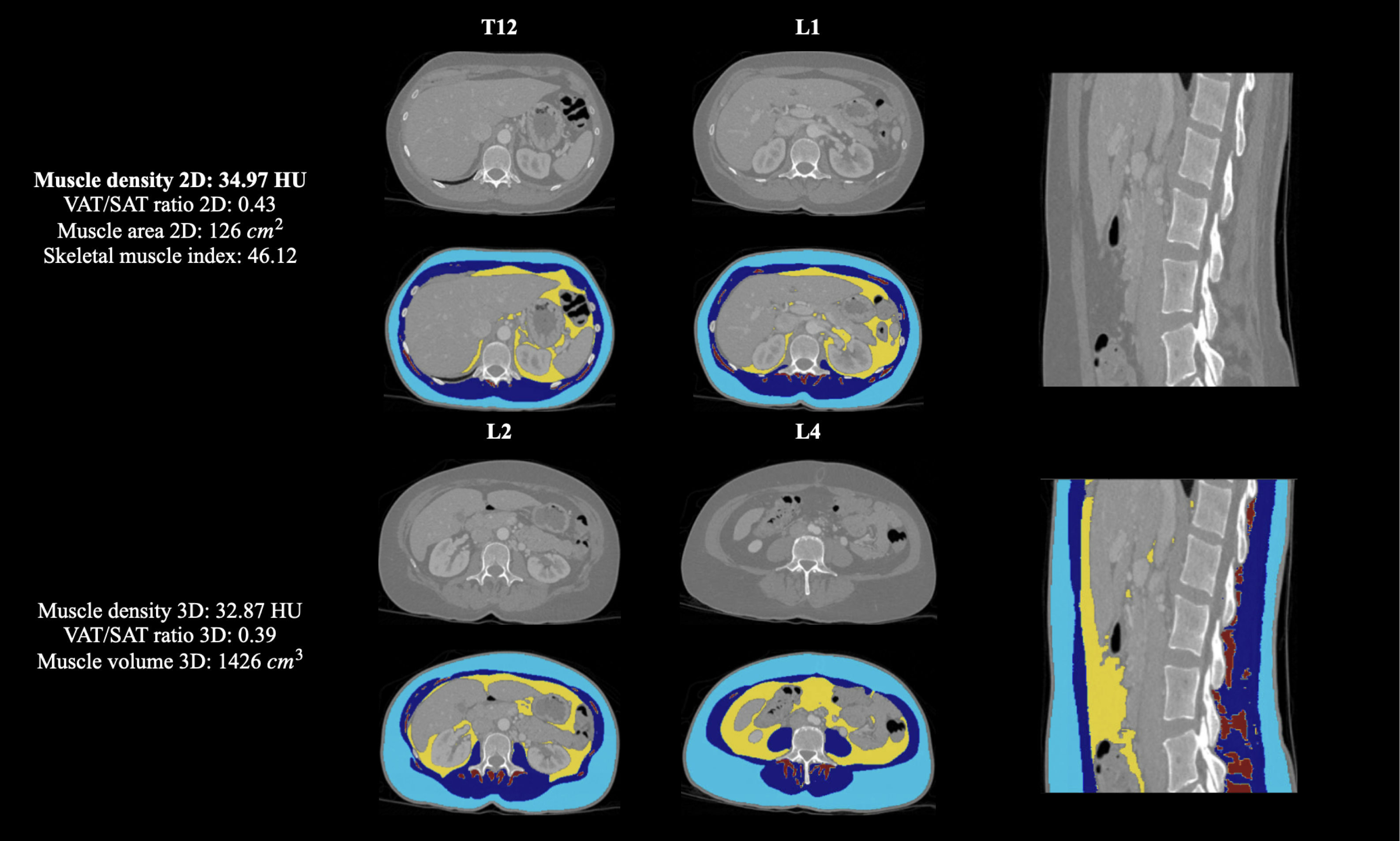} 
    \end{subfigure}
    \begin{subfigure}[b]{0.65\textwidth}
        \centering
        \includegraphics[width=\textwidth]{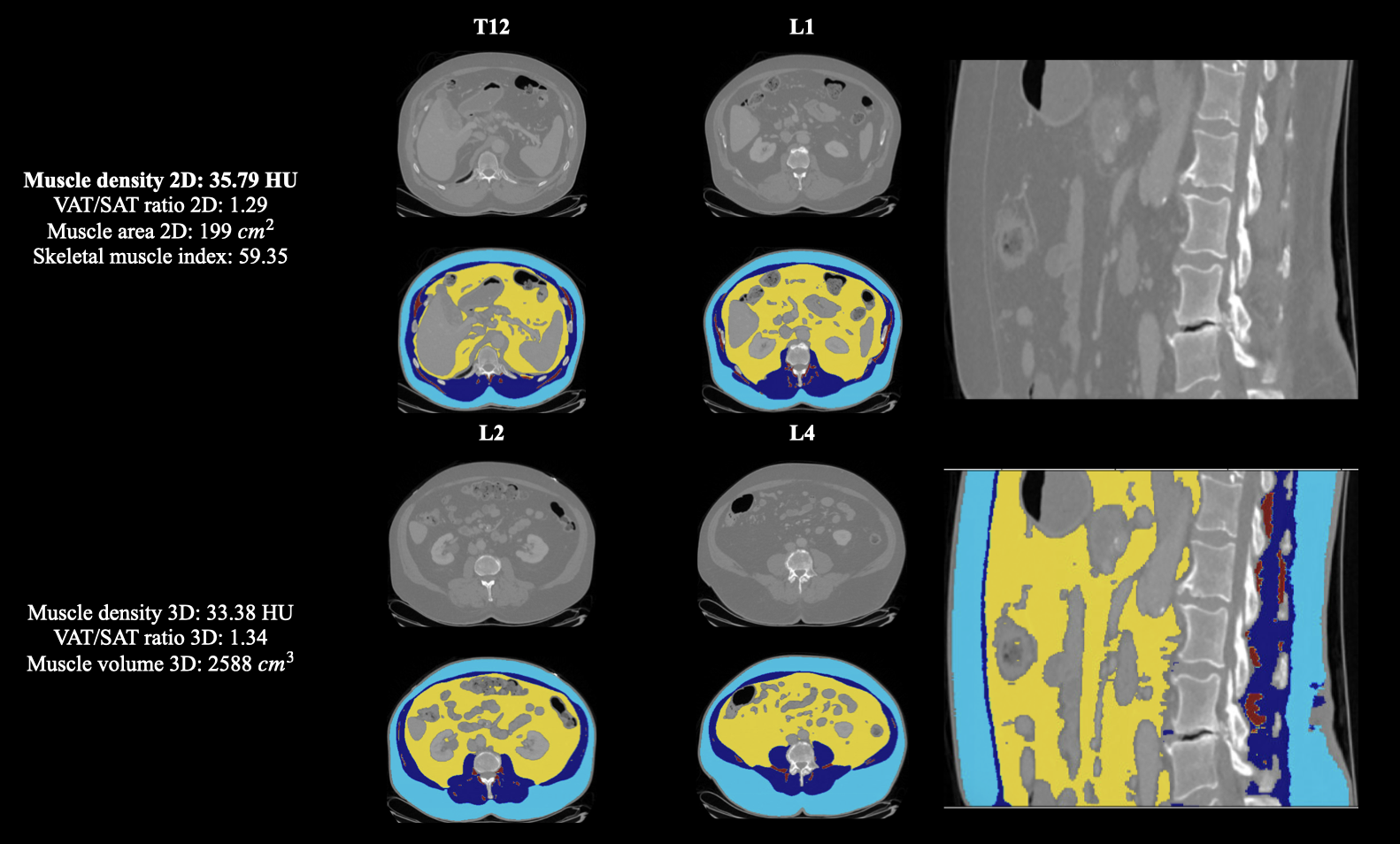} 
    \end{subfigure}
    
    \begin{subfigure}[b]{0.65\textwidth}
        \centering
        \includegraphics[width=\textwidth]{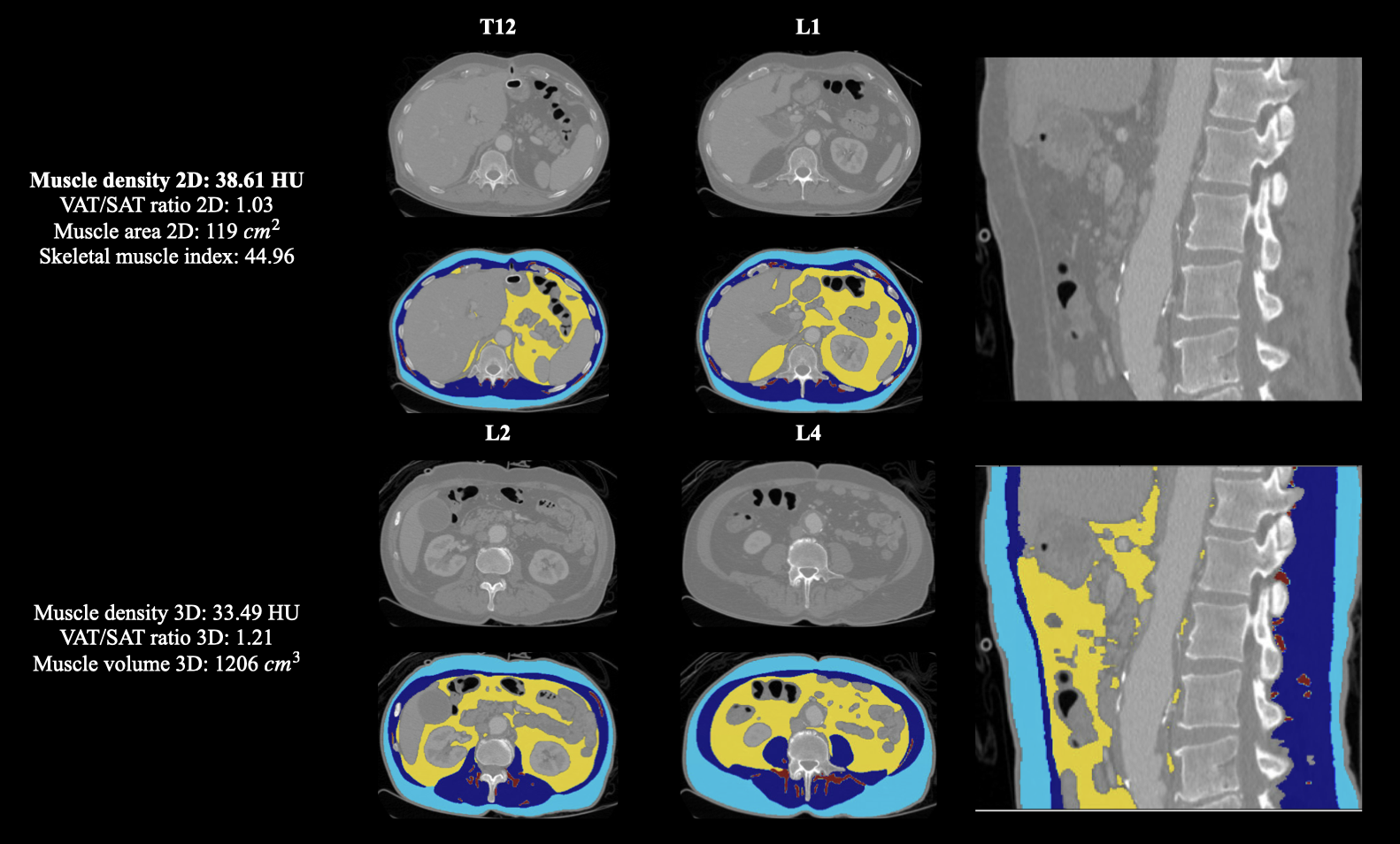} 
    \end{subfigure}
    \caption{\textbf{Qualitative evaluation of our segmentation model for 3D setting:} Four slices corresponding to T12, L1, L2, and L4 are displayed on the left, while the stacked slices from T12 to L4 are visualized in the sagittal view on the right side of the figure. The visualization volumes are ordered by increasing of 2D muscle density (in bold). In the segmentation, \textit{dark blue} shows skeletal muscle, \textit{light blue} SAT, \textit{yellow} VAT, and \textit{maroon} muscular fat.}
    \label{fig2:3D_qualitative_evaluation}
    \label{fig:main}
\end{figure*}

In figure \ref{fig2:3D_qualitative_evaluation}, three volumes are randomly selected from the Demographic Analysis dataset to demonstrate the algorithm's performance for 3D body composition measurement. For each volume, four slices corresponding to T12, L1, L2, and L4 are displayed on the left, along with their automatically generated segmentation masks. L3 segmentation is not included in this figure, as it is fully demonstrated in Fig. \ref{fig1:2D_qualitative_evaluation}. Additionally, the stacked slices from T12 to L4 are visualized in the sagittal view, shown on the right side of the figure.

\section{Body composition metrics relationship} \label{sec:Body composition metrics relationship}
\begin{figure*}[t]
    \centering
    \captionsetup{font=footnotesize} 

    \begin{subfigure}[b]{0.31\textwidth} 
        \centering
        \includegraphics[width=\textwidth]{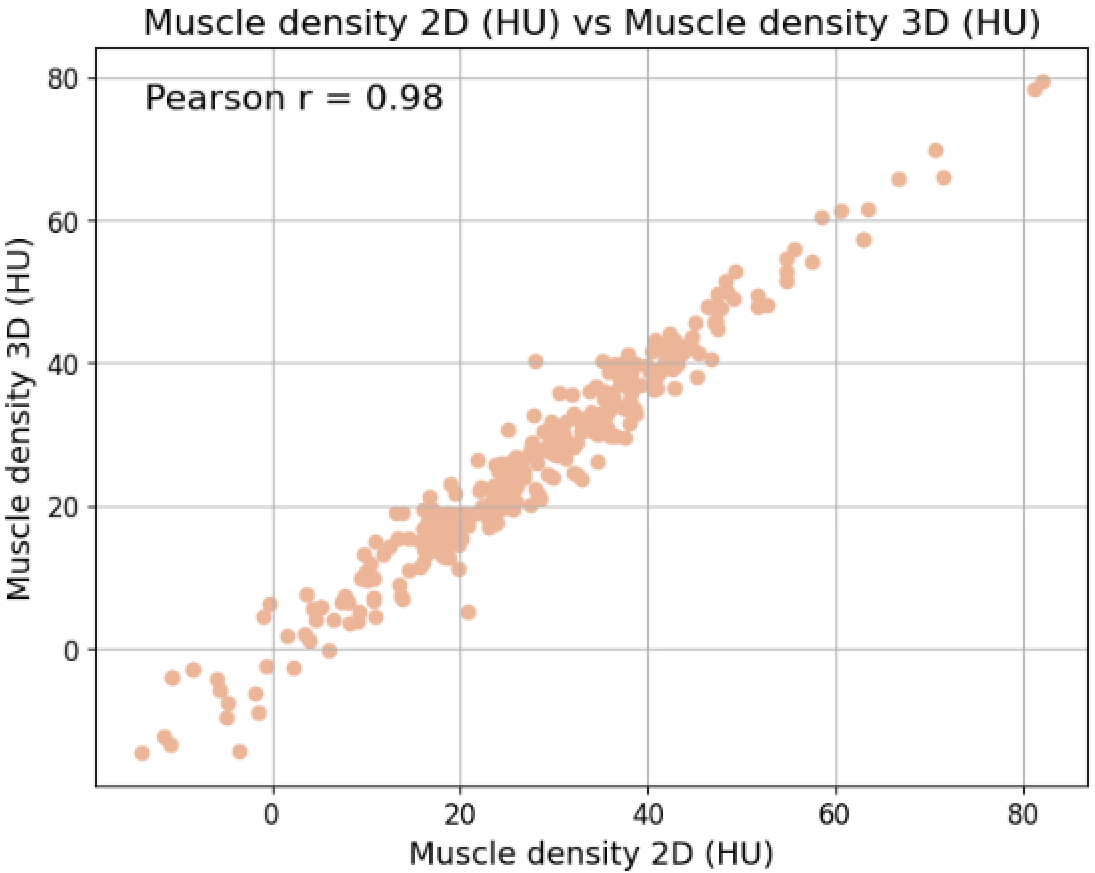}
    \end{subfigure}
    \begin{subfigure}[b]{0.31\textwidth}
        \centering
        \includegraphics[width=\textwidth]{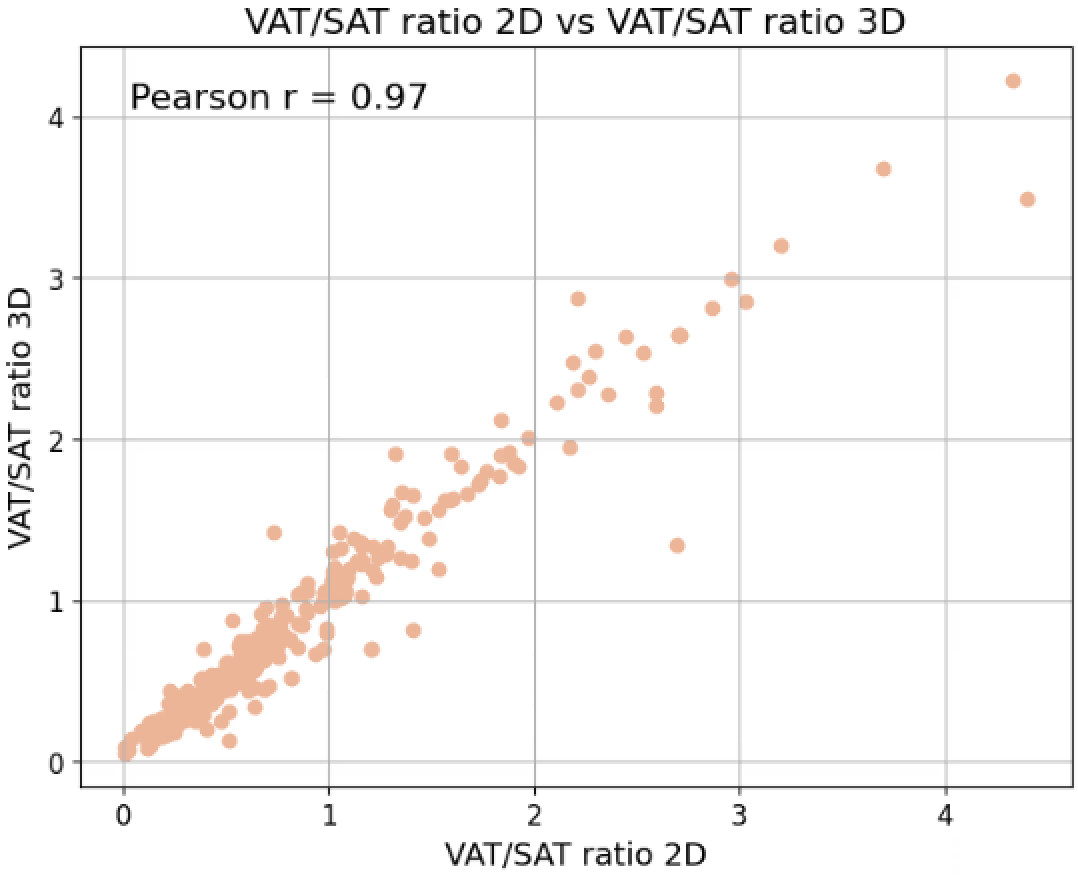}
    \end{subfigure}
    \begin{subfigure}[b]{0.31\textwidth}
        \centering
        \includegraphics[width=\textwidth]{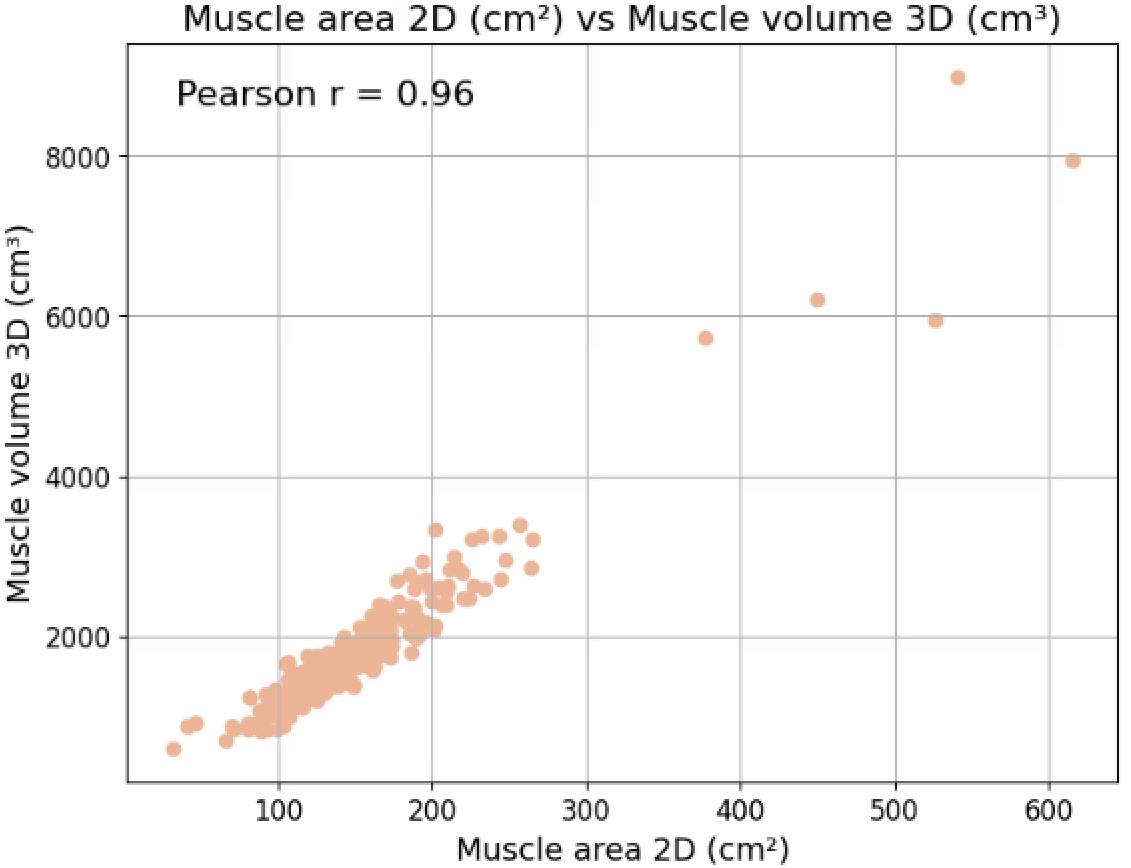}
    \end{subfigure}
    \caption{\textbf{Scatter plots illustrating the relationships between 2D and 3D settings of body composition metrics:} Muscle density 2D (HU), VAT/SAT ratio 2D, and Muscle area 2D (cm²). Each subplot represents a specific metric pair, displaying the distribution of data points alongside the calculated Pearson correlation coefficient (Pearson r) to quantify the strength and direction of their linear relationship.}
    \label{fig:self-correlation}
\end{figure*}

\begin{figure*}[t]
    \centering
    \begin{subfigure}[b]{0.49\textwidth}
        \includegraphics[width=\linewidth]{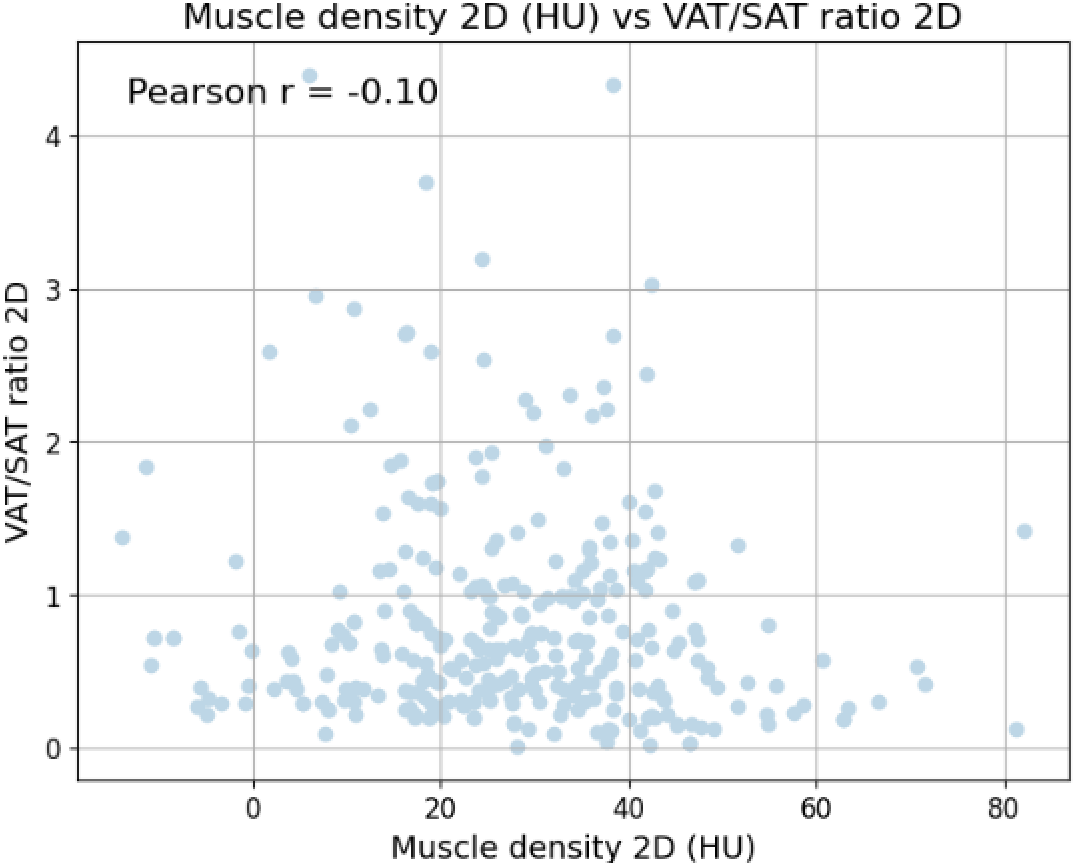}
    \end{subfigure}
    \vspace{1mm}
    \begin{subfigure}[b]{0.49\textwidth}
        \includegraphics[width=\linewidth]{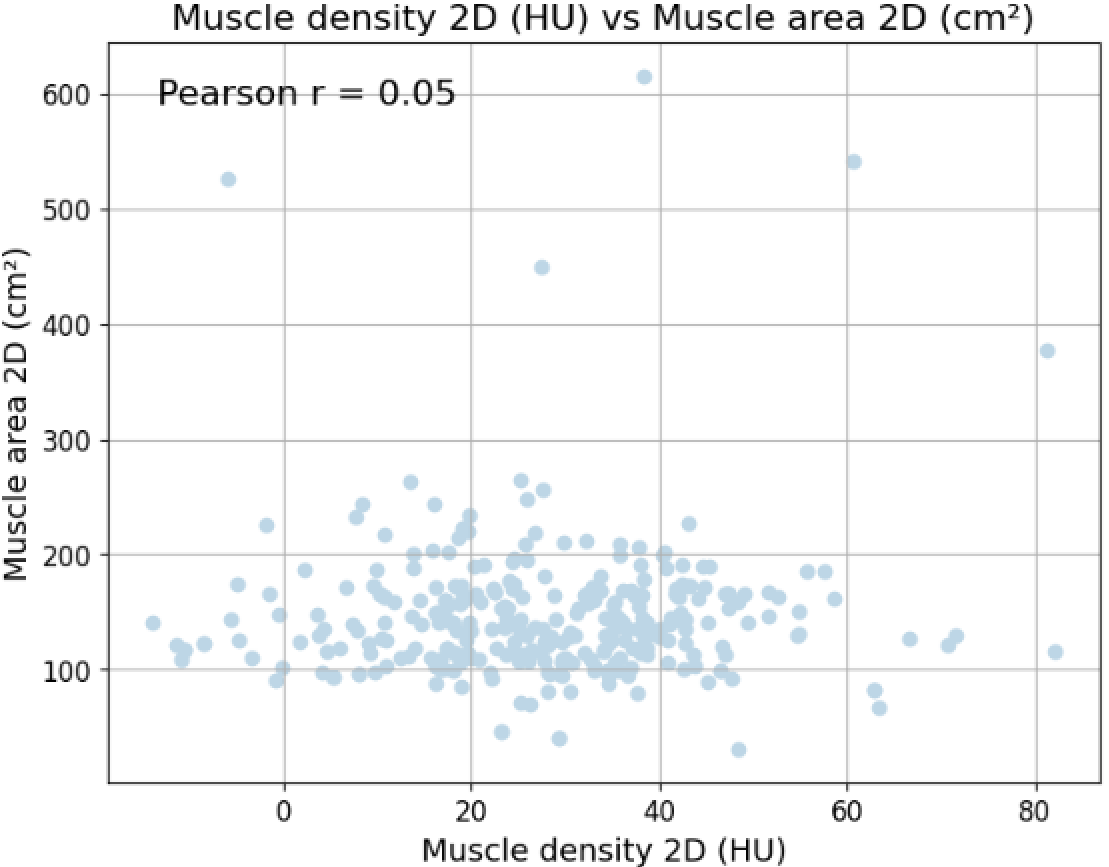}
    \end{subfigure}
    \vspace{1mm}
    \begin{subfigure}[b]{0.49\textwidth}
        \includegraphics[width=\linewidth]{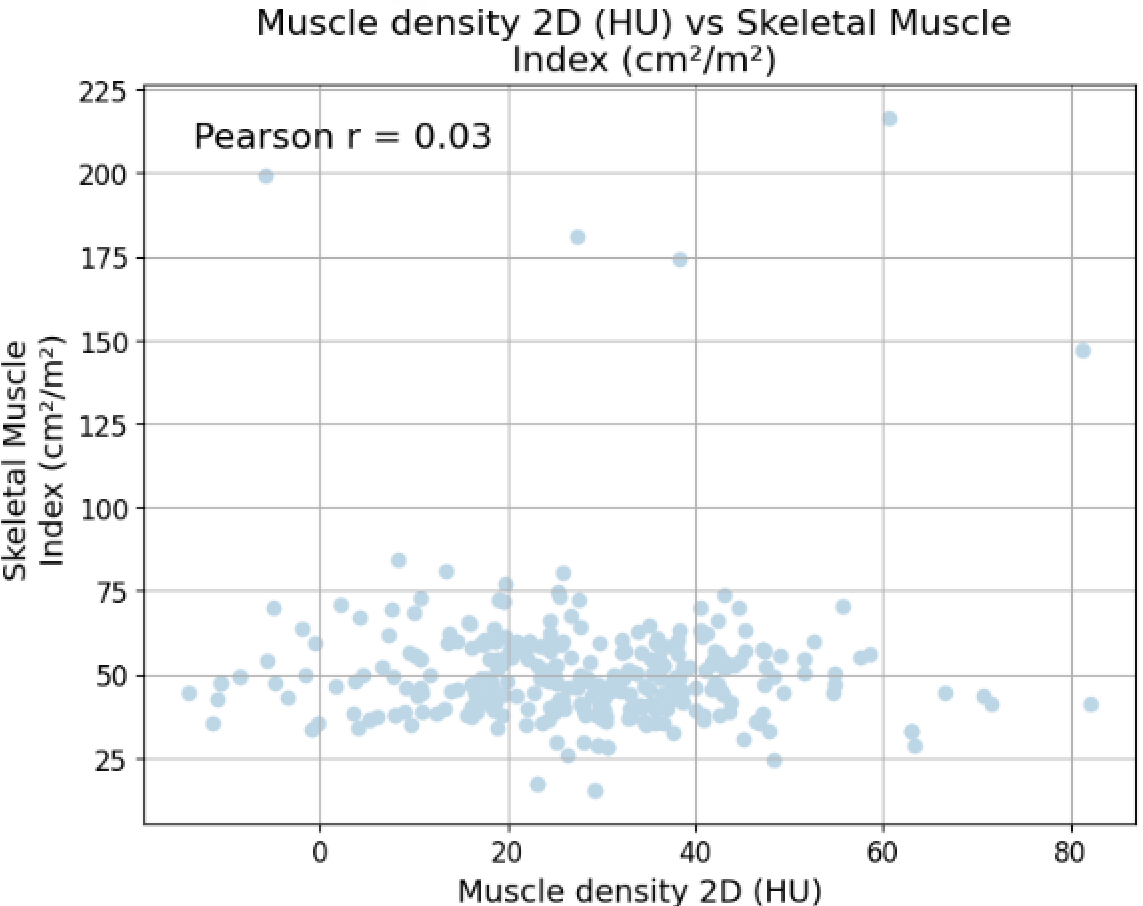}
    \end{subfigure}
    \vspace{1mm}
    \begin{subfigure}[b]{0.49\textwidth}
        \includegraphics[width=\linewidth]{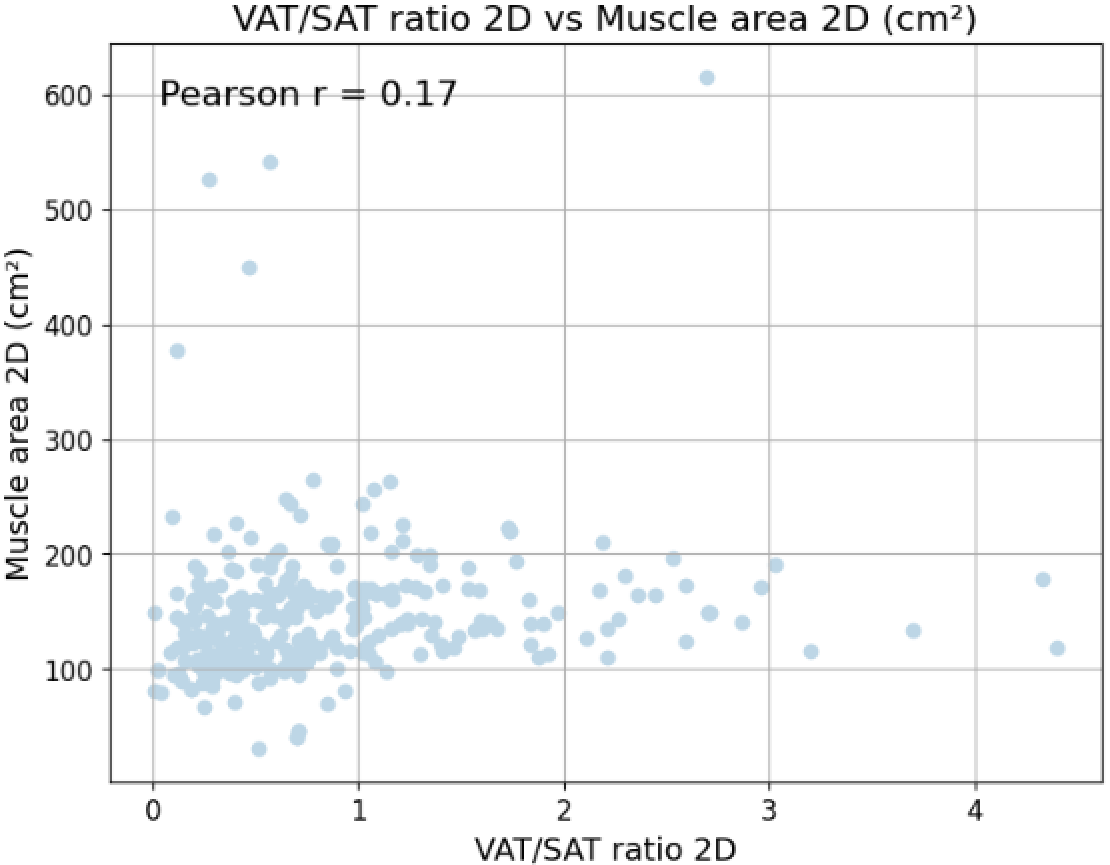}
    \end{subfigure}
    \vspace{1mm}
    \begin{subfigure}[b]{0.49\textwidth}
        \includegraphics[width=\linewidth]{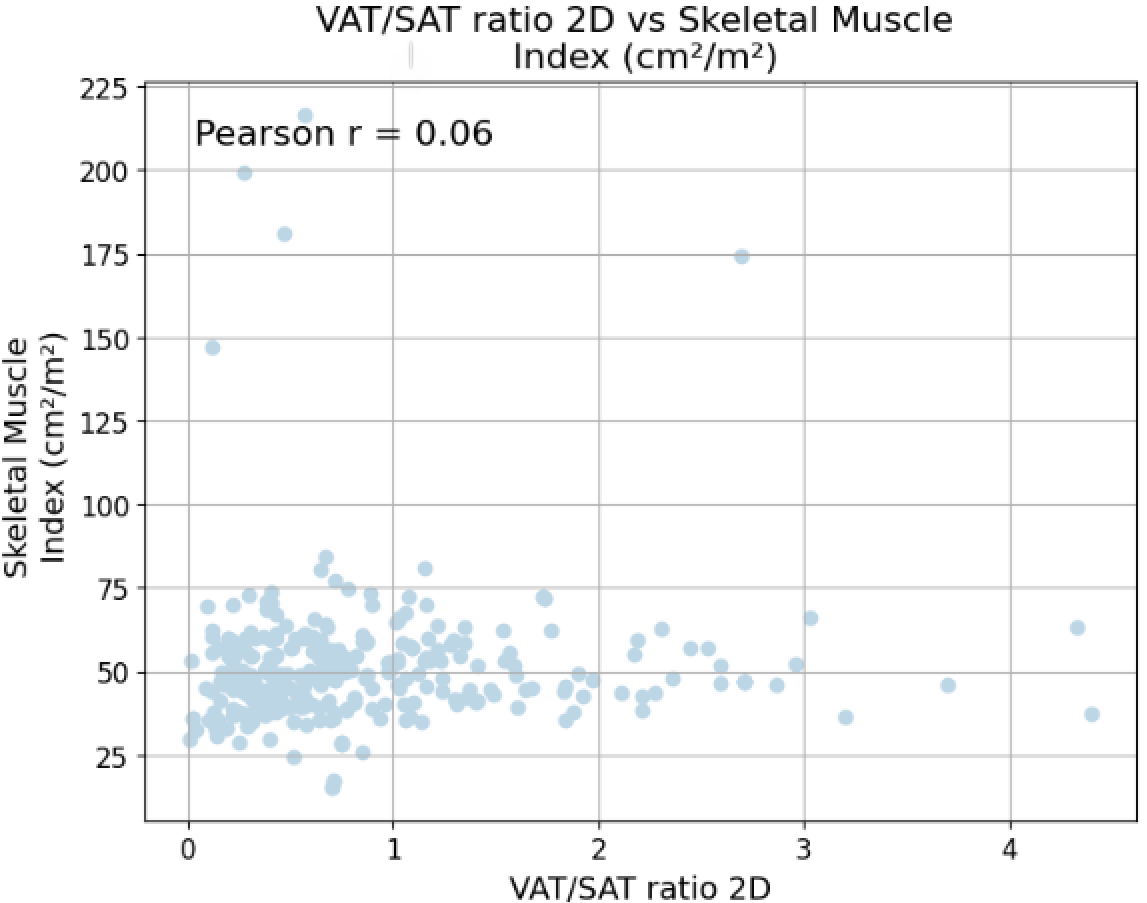}
    \end{subfigure}
    \vspace{1mm}
    \begin{subfigure}[b]{0.49\textwidth}
        \includegraphics[width=\linewidth]{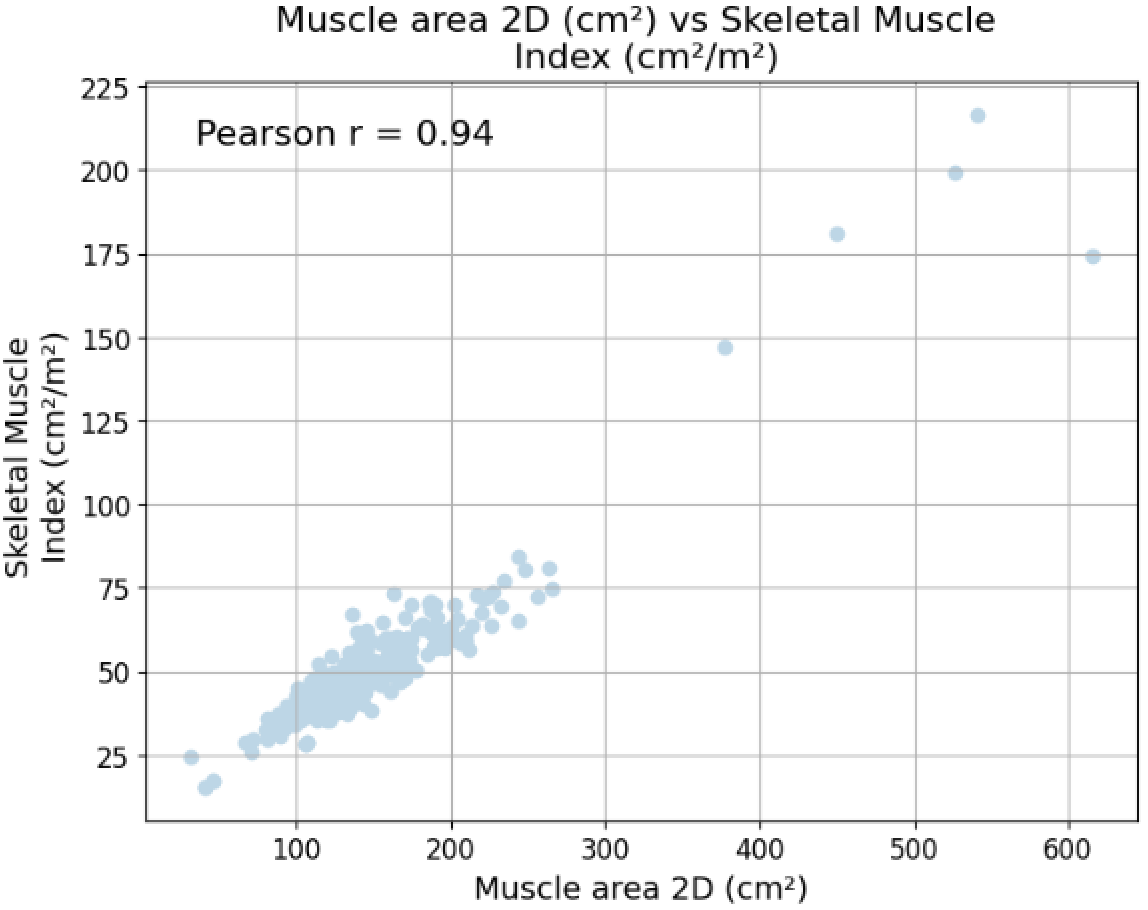}
    \end{subfigure}
    \vspace{1mm}
    \caption{\textbf{Scatter plots illustrating the relationships between pairs of body composition metrics:} Muscle density 2D (HU), VAT/SAT ratio 2D, Muscle area 2D (cm²), and Skeletal Muscle Index (cm/m²). Each subplot represents a specific metric pair, displaying the distribution of data points alongside the calculated Pearson correlation coefficient (Pearson r) to quantify the strength and direction of their linear relationship.}
    \label{fig:correlation}
\end{figure*}

Pearson correlation coefficient (r) is the most common method of measuring a linear correlation between two variables, with its definition shown in Eq. \eqref{r}, where $Cov(X, Y)$ represents the covariance of $X$ and $Y$, and $\sigma_X$, $\sigma_Y$ are standard deviations of $X$ and $Y$ respectively. This section demonstrates the correlation between four selected body composition metrics: muscle density, VAT/SAT ratio, muscle area, and SMI based both on Pearson correlation coefficient and scatter plots. The 2D and 3D measurements of the same metrics are highly correlated, as shown in Fig. \ref{fig:self-correlation}, with all three Pearson correlation coefficients exceeding 0.96. In this experiment, we utilize only the body composition metrics measured in 2D settings (at the L3 level). The results are shown in Fig. \ref{fig:correlation}, as we can observe, except for the relationship between muscle area and SMI (with $r$ equals to 0.94), all other pairs of metrics show insignificant or no linear correlation, with $r$ having an absolute value smaller than 0.2. Scatter plots also do not show a clear monotonic relationship.

\begin{equation} \label{r}
    r = \frac{\text{Cov}(X, Y)}{\sigma_X \sigma_Y}
\end{equation}

\section{Post-processing for label inconsistencies} \label{sec:Post-processing for label inconsistencies}
\begin{figure*}[t]
    \centering
    \captionsetup{font=footnotesize} 

    \begin{subfigure}[b]{0.19\textwidth} 
        \centering
        \includegraphics[width=\textwidth]{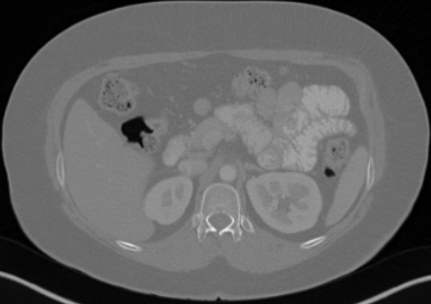}
        \includegraphics[width=\textwidth]{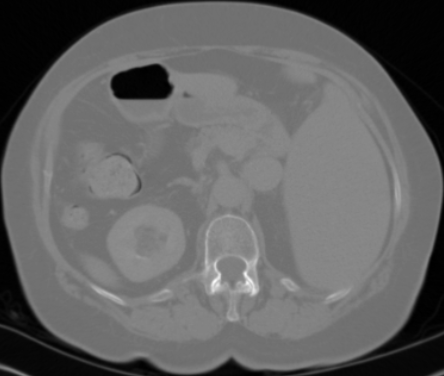}
        \caption{Raw slice}
    \end{subfigure}
    \begin{subfigure}[b]{0.19\textwidth}
        \centering
        \includegraphics[width=\textwidth]{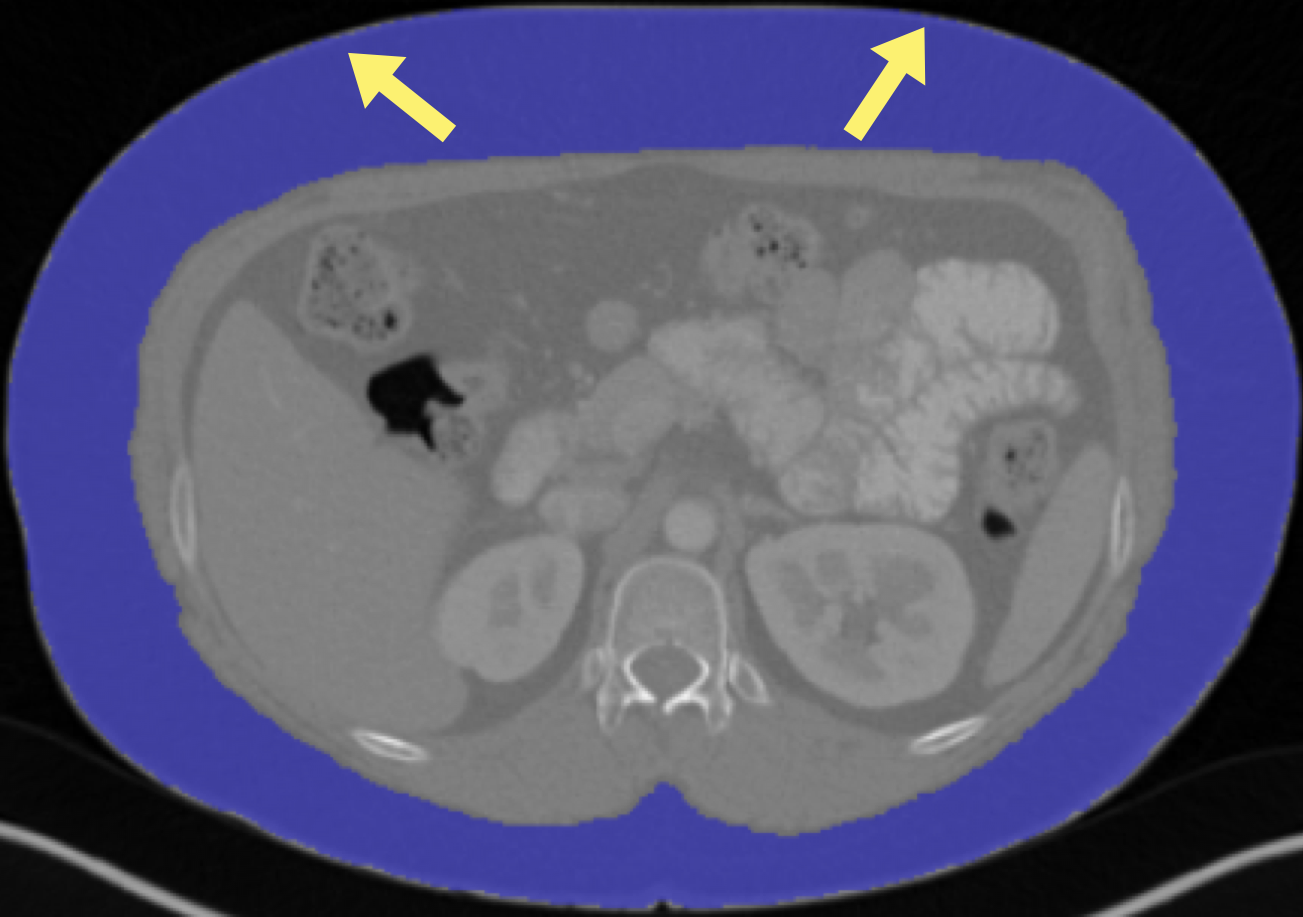}
        \includegraphics[width=\textwidth]{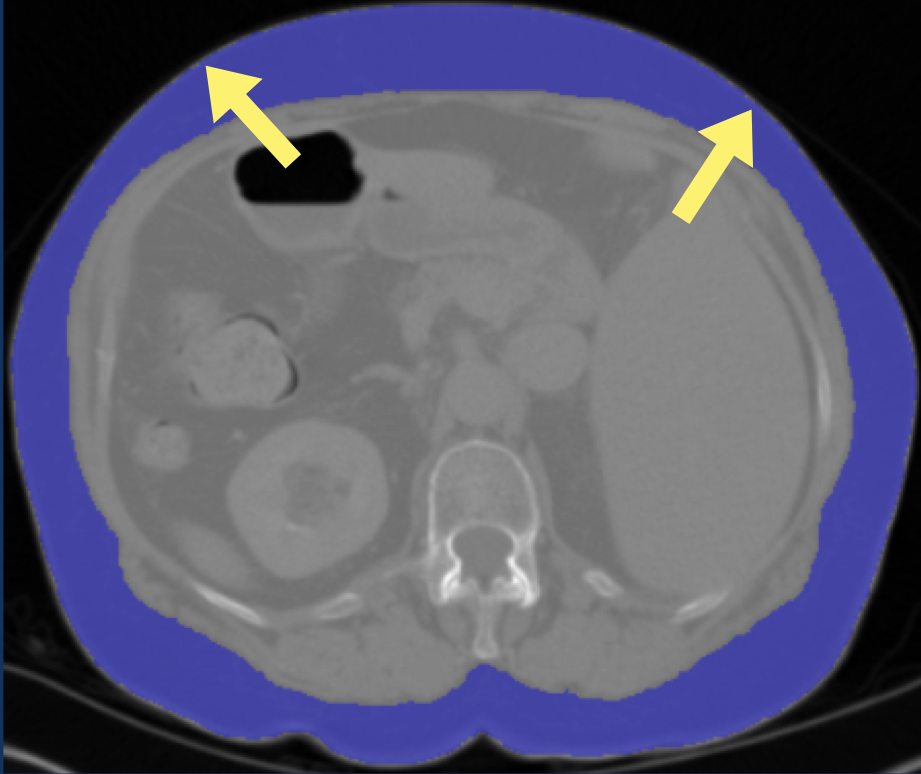}
        \caption{Ground Truth}
    \end{subfigure}
    \begin{subfigure}[b]{0.19\textwidth}
        \centering
        \includegraphics[width=\textwidth]{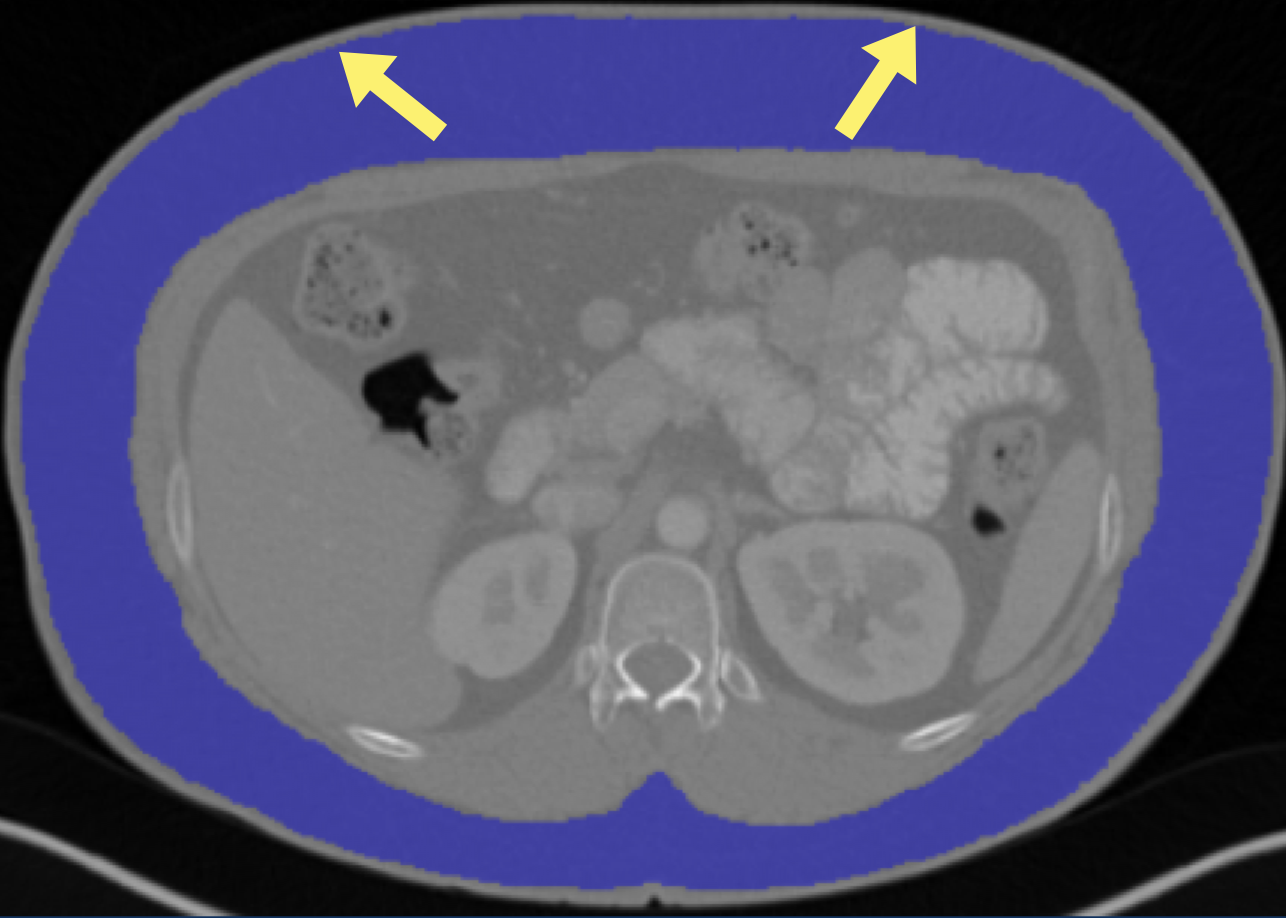}
        \includegraphics[width=\textwidth]{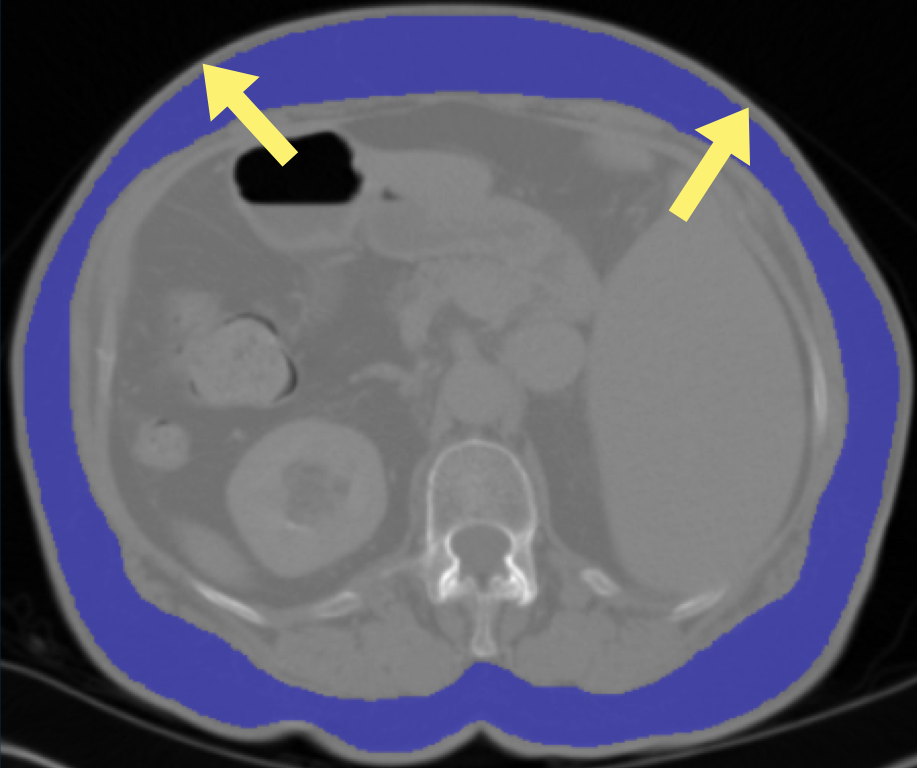}
        \caption{Our prediction} 
    \end{subfigure}
    \begin{subfigure}[b]{0.19\textwidth}
        \centering
        \includegraphics[width=\textwidth]{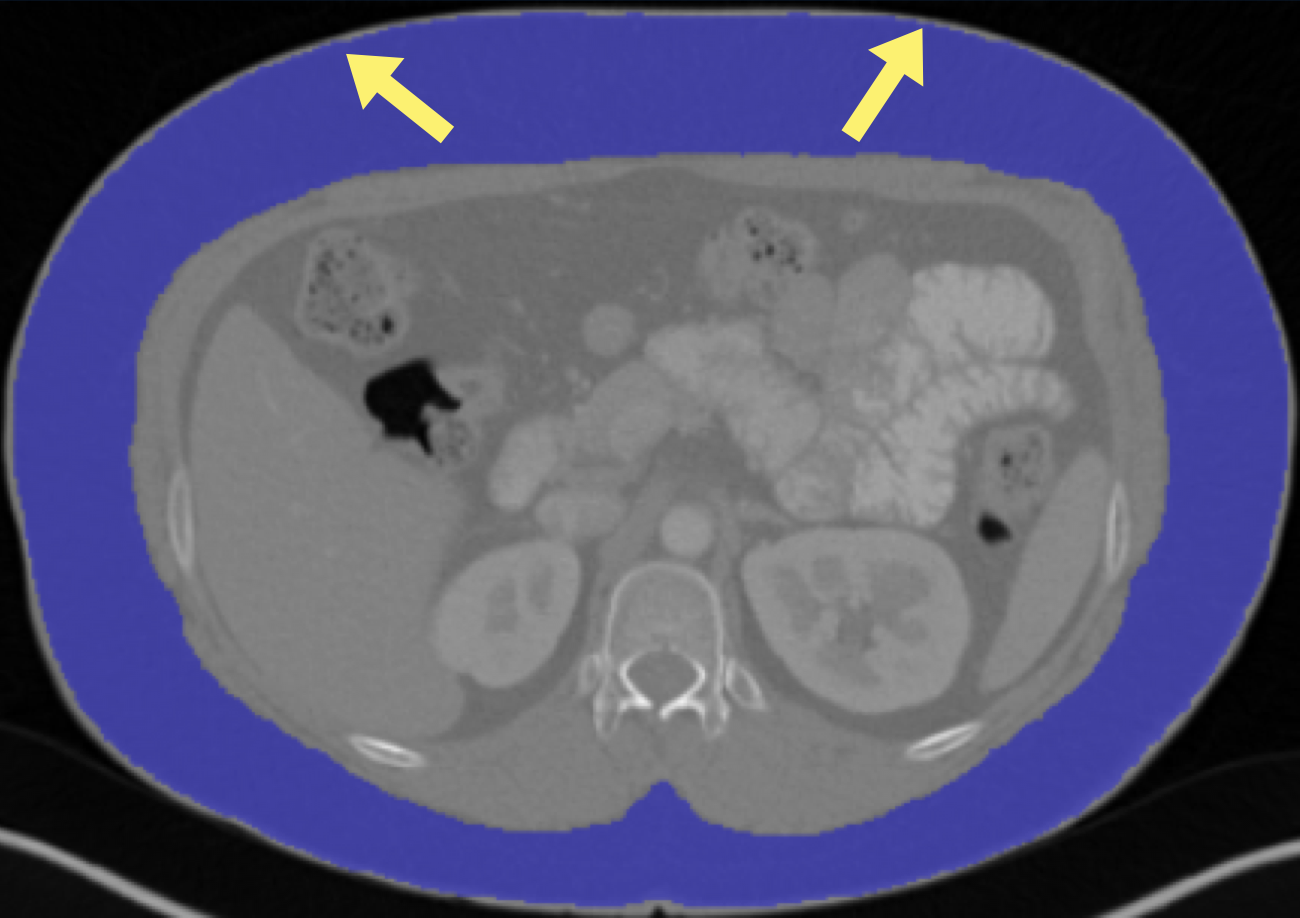}
        \includegraphics[width=\textwidth]{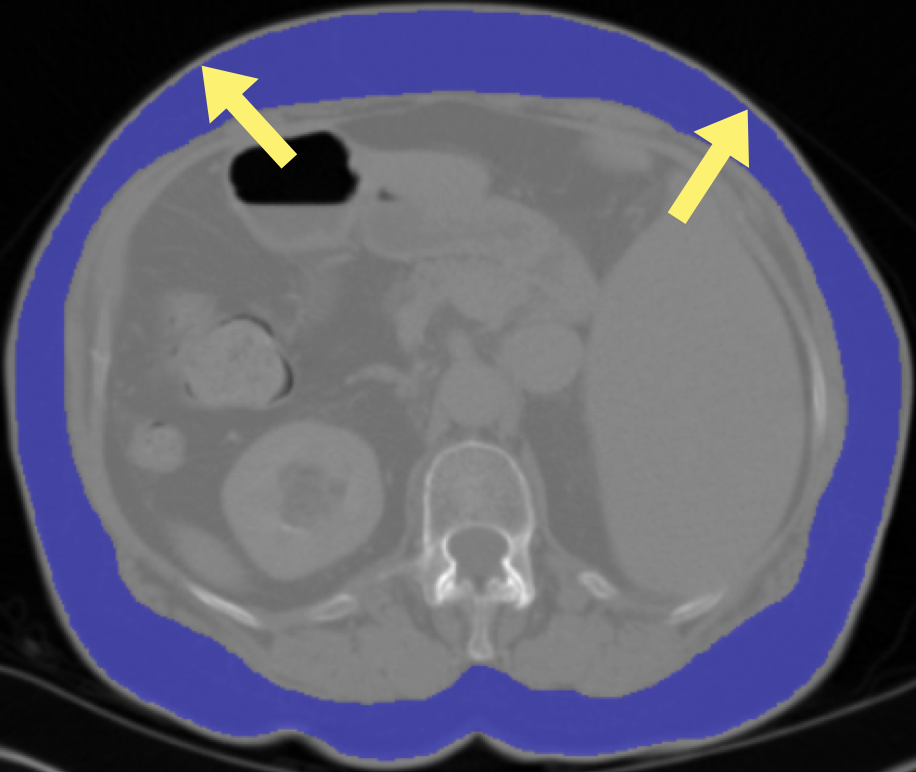}
        \caption{\makebox[2.5cm][c]{Prediction + dilation}}
    \end{subfigure}
    \begin{subfigure}[b]{0.19\textwidth}
        \centering
        \includegraphics[width=\textwidth]{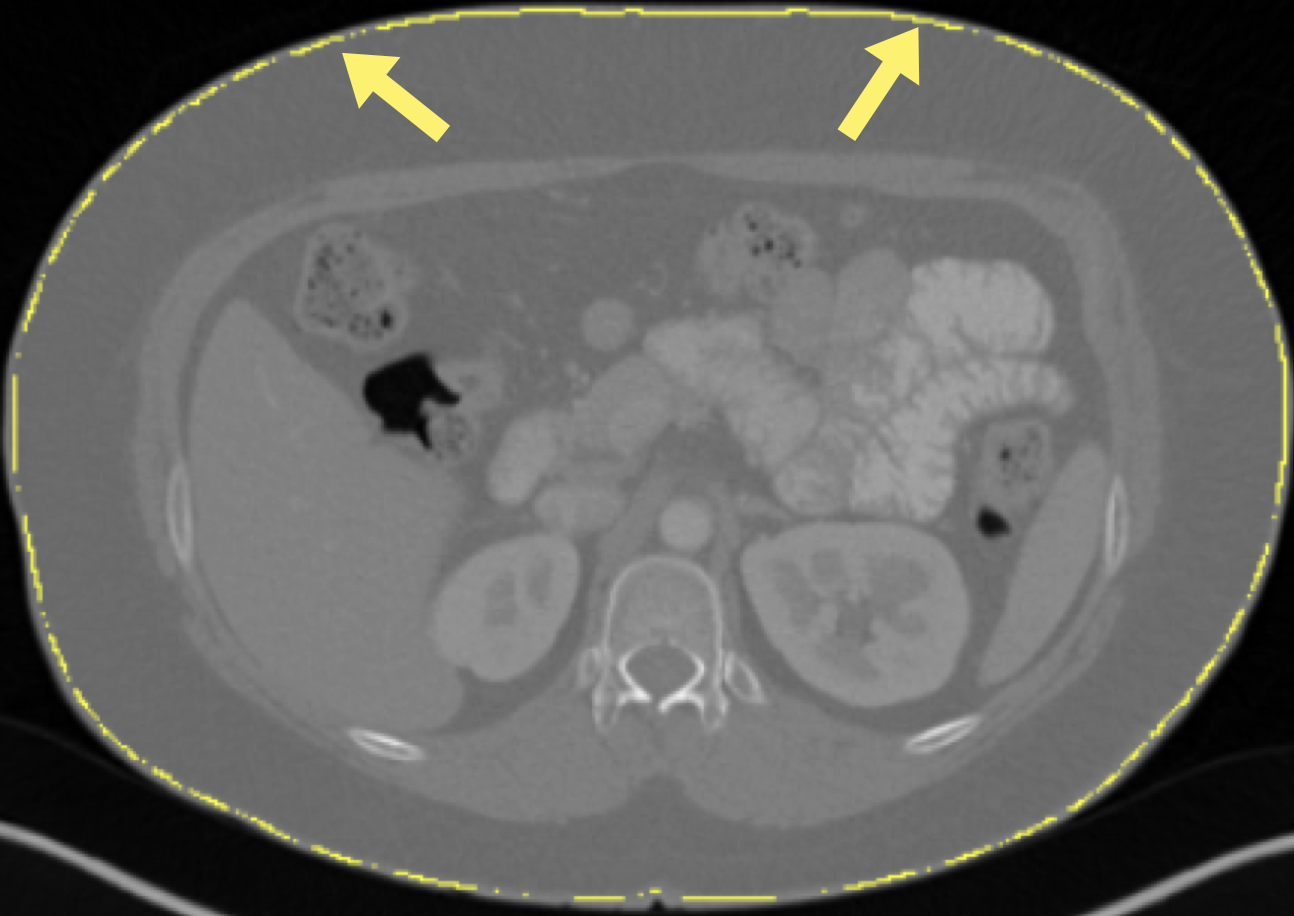}
        \includegraphics[width=\textwidth]{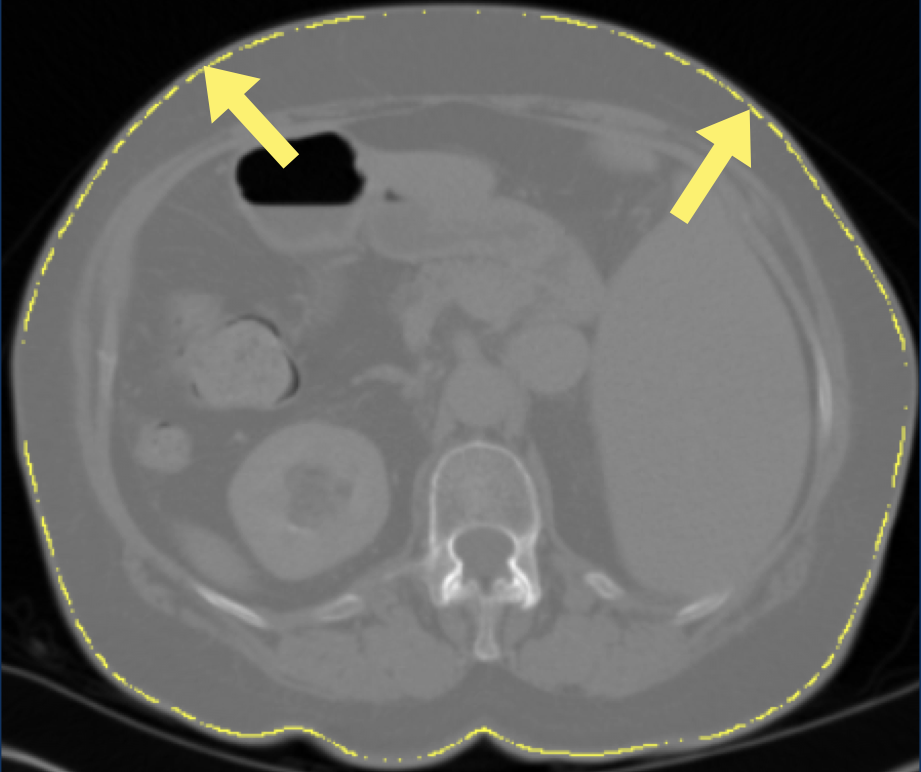}
        \caption{Dilation difference}
    \end{subfigure}

    \caption{\textbf{Dilation examples for two randomly selected slices:} The first column shows raw data without a mask. The second column displays the ground truth from the SAROS dataset. The third column is our original prediction, which excludes skin for SAT. The fourth column demonstrates our prediction after dilation, and the last column illustrates the area added by dilation (in yellow). The yellow arrow highlights the difference introduced by the dilation and blue mask shows SAT.}
    \label{fig:appendix:dilation}
\end{figure*}

Label post-processing is performed in two steps. First, a 5×5 structuring element is applied to morphologically dilate the SAT label. Second, the expanded region is constrained to ensure that (1) it does not overlap with any previously labeled areas and (2) it remains within the abdominal region. The abdominal boundary is determined by thresholding at -800 Hounsfield Units (HU), as skin typically exhibits an HU value around this level \citep{chougule2018clinical, villa2012hounsfield}. Fig. \ref{fig:appendix:dilation} demonstrates the dilation result with  on two randomly selected slices. Notably, the post-processing step merely approximates the inclusion of skin in the segmentation, leaving a remaining discrepancy between the two approaches.

\section{Comprehensive analysis metric evaluation}\label{Comprehensive analysis metric evaluation}
We provide the comprehensive metrics evaluation table here, reporting the percentage error for four key body composition metrics: muscle density, VAT/SAT ratio, skeletal muscle area/volume, and skeletal muscle index (SMI) across both internal and external datasets for \textbf{nine} publicly available network architectures mentioned in Sec. \ref{sec:method}. Each value reflects the MRAE between predictions and human annotation. 

\begin{table*}[t]
\centering
\resizebox{\textwidth}{!}{%
\begin{tabular}{c|c|c|c|c|c}
\toprule
\multicolumn{6}{c}{\textbf{Internal Dataset}} \\\midrule
\textbf{Model} & \textbf{Setting} & \textbf{Muscle density (\%)} & \textbf{VAT/SAT ratio (\%)} & \textbf{Skeletal muscle area/volume (\%)} & \textbf{SMI (\%)} \\
\midrule
\multirow{2}{*}{MexNetX-S} & 2D & 1.03 $\pm$ 0.89 & 12.21 $\pm$ 19.75 & 4.66 $\pm$ 3.26 & 4.66 $\pm$ 3.26 \\
                           & 3D & 2.27 $\pm$ 2.26 & 11.27 $\pm$ 11.25 & 3.75 $\pm$ 3.23 & - \\
\midrule
\multirow{2}{*}{MexNetX-M} & 2D & 1.26 $\pm$ 0.91 & 10.55 $\pm$ 7.76 & 5.12 $\pm$ 4.28 & 5.12 $\pm$ 4.28 \\
                           & 3D & 2.22 $\pm$ 1.57 & 8.96 $\pm$ 5.20 & 4.91 $\pm$ 5.01 & - \\
\midrule
\multirow{2}{*}{MexNetX-L} & 2D & 1.16 $\pm$ 0.77 & 13.86 $\pm$ 20.54 & 4.97 $\pm$ 3.83 & 4.97 $\pm$ 3.83 \\
                           & 3D & 2.21 $\pm$ 1.70 & 10.76 $\pm$ 8.80 & 4.46 $\pm$ 4.27 & - \\
\midrule
\multirow{2}{*}{SAM-Med2D fine-tuning} & 2D & 2.93 $\pm$ 2.45 & 21.47 $\pm$ 41.84 & 12.02 $\pm$ 8.18 & 12.02 $\pm$ 8.18 \\
                                       & 3D & 4.93 $\pm$ 3.86 & 19.16 $\pm$ 29.64 & 11.24 $\pm$ 5.87 & - \\
\midrule
\multirow{2}{*}{SAM fine-tuning} & 2D & 2.59 $\pm$ 2.75 & 36.81 $\pm$ 28.48 & 9.70 $\pm$ 8.68 & 9.70 $\pm$ 8.68 \\
                                       & 3D & 5.30 $\pm$ 7.59 & 32.68 $\pm$ 30.29 & 8.35 $\pm$ 6.60 & - \\
\midrule
\multirow{2}{*}{MedSAM fine-tuning} & 2D & 2.11 $\pm$ 2.20 & 23.97 $\pm$ 17.59 & 8.96 $\pm$ 5.62 & 8.96 $\pm$ 5.62 \\
                                       & 3D & 4.40 $\pm$ 6.99 & 19.30 $\pm$ 16.23 & 9.34 $\pm$ 10.18 & - \\
\midrule
\multirow{2}{*}{nnU-Net ResEnc M} & 2D & 1.60 $\pm$ 1.23 & 5.68 $\pm$ 5.03 & 5.20 $\pm$ 3.89 & 5.20 $\pm$ 3.89 \\
                                  & 3D & 2.22 $\pm$ 2.08 & 4.69 $\pm$ 4.87 & 4.54 $\pm$ 4.04 & - \\
\midrule
\multirow{2}{*}{nnU-Net ResEnc L} & 2D & 1.43 $\pm$ 1.09 & 5.13 $\pm$ 4.26 & 4.95 $\pm$ 3.70 & 4.95 $\pm$ 3.70 \\
                                  & 3D & 2.28 $\pm$ 1.91 & 5.22 $\pm$ 5.36 & 4.09 $\pm$ 3.81 & - \\
\midrule
\multirow{2}{*}{nnU-Net ResEnc XL} & 2D & 1.43 $\pm$ 1.19 & 5.43 $\pm$ 5.17 & 6.90 $\pm$ 8.05 & 6.90 $\pm$ 8.05 \\
                                   & 3D & 2.18 $\pm$ 2.14 & 5.09 $\pm$ 4.74 & 4.86 $\pm$ 5.69 & - \\
\midrule
\multicolumn{6}{c}{\textbf{External Dataset}} \\\midrule
\textbf{Model} & \textbf{Setting} & \textbf{Muscle density (\%)} & \textbf{VAT/SAT ratio (\%)} & \textbf{Skeletal muscle area/volume (\%)} & \textbf{SMI (\%)} \\
\midrule
\midrule
\multirow{2}{*}{MexNetX-S} & 2D & 4.35 $\pm$ 2.60 & - & 8.72 $\pm$ 5.03 & - \\
                           & 3D & 4.60 $\pm$ 2.03 & - & 6.32 $\pm$ 3.70 & - \\
\midrule
\multirow{2}{*}{MexNetX-M} & 2D & 4.49 $\pm$ 2.85 & - & 9.49 $\pm$ 5.07 & - \\
                           & 3D & 4.83 $\pm$ 2.13 & - & 6.73 $\pm$ 3.95 & - \\
\midrule
\multirow{2}{*}{MexNetX-L} & 2D & 4.57 $\pm$ 2.70 & - & 9.87 $\pm$ 5.55 & - \\
                           & 3D & 4.88 $\pm$ 1.95 & - & 7.77 $\pm$ 4.43 & - \\
\midrule
\multirow{2}{*}{SAM-Med2D fine-tuning} & 2D & 5.79 $\pm$ 6.67 & - & 23.08 $\pm$ 12.54 & - \\
                                       & 3D & 7.18 $\pm$ 6.61 & - & 25.28 $\pm$ 8.91 & - \\
\midrule
\multirow{2}{*}{SAM fine-tuning} & 2D & 8.61 $\pm$ 21.87 & - & 21.22 $\pm$ 18.72 & - \\
                                       & 3D & 8.26 $\pm$ 17.11 & - & 18.07 $\pm$ 16.82 & - \\
\midrule
\multirow{2}{*}{MedSAM fine-tuning} & 2D & 4.75 $\pm$ 9.18 & - & 23.44 $\pm$ 19.81 & - \\
                                       & 3D & 6.84 $\pm$ 6.70 & - & 19.80 $\pm$ 20.39 & - \\
\midrule
\multirow{2}{*}{nnU-Net ResEnc M} & 2D & 4.65 $\pm$ 4.10 & - & 9.49 $\pm$ 5.02 & - \\
                                  & 3D & 5.00 $\pm$ 3.78 & - & 6.96 $\pm$ 4.39 & - \\
\midrule
\multirow{2}{*}{nnU-Net ResEnc L} & 2D & 4.71 $\pm$ 6.26 & - & 9.24 $\pm$ 5.04 & - \\
                                  & 3D & 4.87 $\pm$ 3.17 & - & 6.94 $\pm$ 4.32 & - \\
\midrule
\multirow{2}{*}{nnU-Net ResEnc XL} & 2D & 4.47 $\pm$ 2.44 & - & 9.20 $\pm$ 5.31 & - \\
                                   & 3D & 4.71 $\pm$ 2.22 & - & 6.61 $\pm$ 4.24 & - \\
\bottomrule
\end{tabular}%
}
\caption{\textbf{Comprehensive analysis metric evaluation performance:} The performance of \textbf{nine} segmentation models on both internal and external datasets, evaluated by comparing the four body composition metrics automatically calculated by models with the ground truth measured from manual annotations. The evaluation is based on MRAE ($\downarrow$). Notably, the muscle density error percentage is calculated relative to the normal muscle density range, which spans from -29 to 150 Hounsfield Units (HU).}
\label{tab:metric_analysis_all}
\end{table*}

\end{document}